\documentstyle[12pt]{article}
\input epsf.tex
\newtheorem{theorem}{Theorem}[section]

\newtheorem{lemma}[theorem]{Lemma}

\renewcommand{\title}[1]{\Large\bf #1\bigskip\medskip\\}
\renewcommand{\author}[1]{\large #1\\ \smallskip}
\newcommand{\address}[1]{{\normalsize\it #1\\}\bigskip}

\newcommand{\hs}[1]{\hspace*{#1cm}}
\newcommand{\vs}[1]{\vspace*{#1cm}}

\newcommand{\Z}{\mbox{\sf Z\hspace*{-0.45em}Z}}

\def\tha#1{\vartheta_1(#1)}
\def\thd#1{\vartheta_4(#1)}
\def\la{\lambda}
\def\hl{\leaders \hbox  to 2em{---}\hfill}
\def\vl{\!\!\!\!|\!\!\!\!}
\def\ade{$A$--$D$--$E$\space}
\def\wt#1#2#3#4#5#6{#1\mbox{\small
 $\left(\matrix{#5&#4\cr#2&#3\cr}\biggm|\mbox{$#6$}\right)$}}

\def\dddots{\mathinner{\mkern1mu\raise1pt\vbox{\kern1pt\hbox{.}}
                       \mkern2mu\raise4pt\hbox{.}
                       \mkern2mu\raise7pt\hbox{.}\mkern1mu}}
\def\wf#1#2#3#4#5#6#7#8#9{#1\mbox{$\left(
   \matrix{#5&#8&\vs{-0.2}#4\cr#9&&\vs{-0.2}#7\cr a&#6&#3\cr}
           \biggm|\mbox{$#2$}\right)$}}
\def\wfa#1#2#3#4#5#6#7#8#9{#1\mbox{$\left(
   \matrix{#5&#8&\vs{-0.2}#4\cr#9&&\vs{-0.2}#7\cr d&#6&#3\cr}
           \biggm|\mbox{$#2$}\right)$}}
\def\wb#1#2#3#4#5#6#7#8#9{#1\mbox{$\left(
   \matrix{#5&#8&#4\cr#9&&#7\cr
    c^i&#6&#3\cr}\biggm|\mbox{$#2$}\right)$}}
\def\wdd#1#2#3#4#5#6#7#8#9{#1\mbox{$\left(
   \matrix{#5&#8&#4\cr#9&&#7\cr
    e^i&#6&#3\cr}\biggm|\mbox{$#2$}\right)$}}
\def\row#1#2#3#4#5#6#7#8{#1\mbox{$\left(
   \matrix{#6&#8&#5\cr #3&#7&#4\cr}\biggm|\mbox{$#2$}\right)$}}
\def\Biggm{\right}

\def\Wb#1#2#3#4#5#6#7#8#9{W^{(m,0)}_{(n,0)}\mbox{$\left(\left.
   \matrix{#5&#8&#4\cr#9&&#7\cr #1&#6&#3\cr}\Biggm|\mbox{$#2$}\right)$}}

\def\Wbml#1#2#3#4#5#6#7#8#9{W^{(m,0)}_{(l,0)}\mbox{$\left(\left.
   \matrix{#5&#8&#4\cr#9&&#7\cr #1&#6&#3\cr}\Biggm|\mbox{$#2$}\right)$}}

\def\Wbln#1#2#3#4#5#6#7#8#9{W^{(l,0)}_{(n,0)}\mbox{$\left(\left.
   \matrix{#5&#8&#4\cr#9&&#7\cr #1&#6&#3\cr}\Biggm|\mbox{$#2$}\right)$}}
\def\WF#1#2#3#4#5#6#7#8#9{W^{\!(\!n\!,m\!)}_{(\!\ol{n}\!,\ol{m}\!)}
       \mbox{$\left(\left.
   \matrix{#5&#8&#4\cr#9&&#7\cr #1&#6&#3\cr}\Biggm|\mbox{$#2$}\right)$}}
\def\WWb#1#2#3#4#5#6#7#8#9{W^{(m,\ol{m})}_{(n,\ol{n})}\mbox{$\left(\left.
   \matrix{#5&#8&#4\cr#9&&#7\cr #1&#6&#3\cr}\Biggm|\mbox{$#2$}\right)$}}
\def\WWbml#1#2#3#4#5#6#7#8#9{W^{(m,\ol{m})}_{(l,\ol{l})}\mbox{$\left(\left.
   \matrix{#5&#8&#4\cr#9&&#7\cr #1&#6&#3\cr}\Biggm|\mbox{$#2$}\right)$}}
\def\WWbln#1#2#3#4#5#6#7#8#9{W^{(l,\ol{l})}_{(n,\ol{n})}\mbox{$\left(\left.
   \matrix{#5&#8&#4\cr#9&&#7\cr #1&#6&#3\cr}\Biggm|\mbox{$#2$}\right)$}}

\def\one{\hbox{\boldmath $I$}}

\def\ba{\begin{array}}
\def\ea{\end{array}}
\def\be{\begin{eqnarray}}
\def\ee{\end{eqnarray}}
\def\no{\nonumber}
\def\-{\!-\!}
\def\+{\!+\!}
\def\={\!=\!}
\def\ol{\overline}
\def\sc{\scriptsize}
\def\scs{\scriptscriptstyle}

\def\h{\hspace*{0.5cm}}
\def\mh{\hspace*{-0.5cm}}
\def\disp{\displaystyle}

\def\ra{\raisebox{0pt}[0pt][0pt]}
\def\ma{\makebox(0,0)[lb]}
\def\T{\mbox{\boldmath $T$}}
\def\Q{\mbox{\boldmath $Q$}}

\def\({\biggl(}
\def\){\biggr)}
\def\path{p}

\newcommand{\val}[1]{\mbox{val($#1$)}}

\def\dynkin{
\setlength{\unitlength}{0.0125in}
\begin{picture}(320,30)(-10,0)
 \thicklines
 \multiput(  0,20)(50,0){3}{\circle*{8}}
 \multiput(250,20)(50,0){2}{\circle*{8}}
 \put(  0,20){\line(1,0){300}}
 \put(-2.5,4){$1$}
 \put(  47,4){$2$}
 \put(  97,4){$3$}
 \put( 236,4){$L\!-\!1$}
 \put( 296,4){$L$}
\end{picture}}

\def\dynkin2{
\setlength{\unitlength}{0.0105in}
\begin{picture}(375,120)(-60,0)
 \thicklines
 \multiput(0,0)(0,80){2}{
 \multiput(  0,30)(50,0){3}{\circle*{8}}
 \multiput(250,30)(50,0){2}{\circle*{8}}
 \put(   0,30){\line(1,0){300}}
 \put(-2.5,14){$1$}
 \put(  47,14){$2$}
 \put(  97,14){$3$}
 \put( 236,14){$L\!-\!1$}
 \put( 296,14){$L$}}
 \multiput(  0,40)(50,0){3}{\circle{20}}
 \multiput(250,40)(50,0){2}{\circle{20}}
 \put(-60,105){\Large (a)}
 \put(-60, 25){\Large (b)}
\end{picture}}

\def\sqbox#1#2#3#4{\setlength{\unitlength}{0.0105in}
  \rule[#2\unitlength]{0pt}{25\unitlength}
  \begin{picture}(20,0)(-#1,-#2)
   \put(0,0){\line(1,0){20}}
   \put(0,0){\line(0,1){20}}
   \put(20,20){\line(-1,0){20}}
   \put(20,20){\line(0,-1){20}}
   \put(7,7){\sc $#3$}
   \put(21,21){\sc $#4$}
\end{picture}}

\def\sqboxd#1#2#3{\setlength{\unitlength}{0.0105in}
  \rule[#2\unitlength]{0pt}{25\unitlength}
  \begin{picture}(40,0)(-#1,-#2)
   \put(0,0){\line(0,1){20}}
   \put(40,0){\line(0,1){20}}
   \put(0,0){\line(1,0){40}}
   \put(0,20){\line(1,0){40}}
   \put(10,7){$\cdots$}
   \put(17,-20){\sc $#3$}
\end{picture}}

\def\sqqbox#1#2#3#4#5{\setlength{\unitlength}{0.0105in}
  \rule[#2\unitlength]{0pt}{40\unitlength}
  \begin{picture}(20,0)(-#1,-#2)
   \put(0,0){\line(0,1){40}}
   \put(20,0){\line(0,1){40}}
   \put(0,0){\line(1,0){20}}
   \put(0,20){\line(1,0){20}}
   \put(0,40){\line(1,0){20}}
   \put(7,27){\sc $#3$}
   \put(7,7){\sc $#4$}
   \put(21,41){\sc $#5$}
\end{picture}}

\def\sqqboxd#1#2#3{\setlength{\unitlength}{0.0105in}
  \rule[#2\unitlength]{0pt}{40\unitlength}
  \begin{picture}(40,0)(-#1,-#2)
   \put(0,0){\line(0,1){40}}
   \put(40,0){\line(0,1){40}}
   \put(0,0){\line(1,0){40}}
   \put(0,20){\line(1,0){40}}
   \put(0,40){\line(1,0){40}}
   \put(10,27){$\cdots$}
   \put(10,7){$\cdots$}
   \put(17,-20){\sc $#3$}
\end{picture}}

\def\genbox#1#2#3{\setlength{\unitlength}{0.0105in}
   \rule[-37\unitlength]{0pt}{60\unitlength}
   \underbrace
   {\sqqbox{0}{-17}{}{}{}
    \sqqboxd{0}{-17}{#2}
    \sqqbox{0}{-17}{}{}{}}\!
   \underbrace
   {\sqbox{0}{3}{}{}
    \sqboxd{0}{3}{#1}
    \sqbox{0}{3}{}{#3}}}

\def\sqbbox#1#2#3#4{\setlength{\unitlength}{0.02in}
  \rule[#2\unitlength]{0pt}{25\unitlength}
  \begin{picture}(20,0)(-#1,-#2)
   \put(0,0){\line(1,0){20}}
   \put(0,0){\line(0,1){20}}
   \put(20,20){\line(-1,0){20}}
   \put(20,20){\line(0,-1){20}}
   \put(8,8){\tiny $#3$}
   \put(21,21){\sc $#4$}
\end{picture}}

\def\sqqbbox#1#2#3#4#5{\setlength{\unitlength}{0.02in}
  \rule[#2\unitlength]{0pt}{40\unitlength}
  \begin{picture}(20,0)(-#1,-#2)
   \put(0,0){\line(0,1){40}}
   \put(20,0){\line(0,1){40}}
   \put(0,0){\line(1,0){20}}
   \put(0,20){\line(1,0){20}}
   \put(0,40){\line(1,0){20}}
   \put(8,28){\tiny $#3$}
   \put(8,8){\tiny $#4$}
   \put(21,41){\sc $#5$}
\end{picture}}

\def\YBR#1#2#3#4#5#6#7#8#9{
\setlength{\unitlength}{0.0125in}
\begin{picture}(300,100)(-10,-10)
 \thicklines
 \multiput(60,60)( 0,-60){2}{\line(1,0){40}}
 \multiput(60,60)(60,-30){2}{\line(-2,-3){20}}
 \multiput(40,30)(60, 30){2}{\line(2,-3){20}}
 \put( 40,30){\line(1,0){40}}
 \put(100,60){\line(-2,-3){20}}
 \put(100, 0){\line(-2,3){20}}
 \put( 80,30){\circle*{4}}
 \put(140,30){$=$}
 \multiput(190,60)( 0,-60){2}{\line(1,0){40}}
 \multiput(190,60)(60,-30){2}{\line(-2,-3){20}}
 \multiput(170,30)(60, 30){2}{\line(2,-3){20}}
 \put(190,60){\line(2,-3){20}}
 \put(190, 0){\line(2,3){20}}
 \put(250,30){\line(-1,0){40}}
 \put(210,30){\circle*{4}}
 \multiput( 57,-7)(130,0){2}{\ma{\ra{\twlrm \sc $#1$}}}
 \multiput(100,-7)(130,0){2}{\ma{\ra{\twlrm \sc $#2$}}}
 \multiput(122,28)(130,0){2}{\ma{\ra{\twlrm \sc $#3$}}}
 \multiput(101,63)(130,0){2}{\ma{\ra{\twlrm \sc $#4$}}}
 \multiput( 57,63)(130,0){2}{\ma{\ra{\twlrm \sc $#5$}}}
 \multiput( 34,28)(130,0){2}{\ma{\ra{\twlrm \sc $#6$}}}
 \put( 68,43){\ma{\ra{\twlrm \sc $#7$}}} 
 \put( 90,28){\ma{\ra{\twlrm \sc $#8$}}} 
 \put( 68,13){\ma{\ra{\twlrm \sc $#9$}}} 
 \put(218,43){\ma{\ra{\twlrm \sc $#9$}}} 
 \put(180,28){\ma{\ra{\twlrm \sc $#8$}}} 
 \put(218,13){\ma{\ra{\twlrm \sc $#7$}}} 
\end{picture}}

\def\inversion{
\setlength{\unitlength}{0.009in}
\rule[-9mm]{0mm}{20mm}
\begin{picture}(140,30)(-10,27)
 \thicklines
 \put(  0, 30){\line( 1, 1){ 30}}
 \put(  0, 30){\line( 1,-1){ 30}}
 \put( 30,  0){\line( 1, 1){ 60}}
 \put( 30, 60){\line( 1,-1){ 60}}
 \put( 90,  0){\line( 1, 1){ 30}}
 \put( 90, 60){\line( 1,-1){ 30}}
 \put( 60, 30){\circle*{6}}
 \put( -3, 27){\makebox(0,0)[rb]{\sc $a$}}
 \put( 27,-10){\ma{\sc $b$}}
 \put( 87,-10){\ma{\sc $b$}}
 \put(122, 27){\ma{\sc $c$}}
 \put( 27, 62){\ma{\sc $d$}}
 \put( 87, 62){\ma{\sc $d$}}
 \put( 34, 27){\makebox(0,0)[rb]{\sc $u$}}
 \put( 98, 27){\makebox(0,0)[rb]{\sc $-u$}}
 \put( 57, 20){\ma{\sc $g$}}
\end{picture}}

\def\invtwola{
\setlength{\unitlength}{0.009in}
\rule[-9mm]{0mm}{20mm}
\begin{picture}(140,30)(-10,27)
 \thicklines
 \put(  0, 30){\line( 1, 1){ 30}}
 \put(  0, 30){\line( 1,-1){ 30}}
 \put( 30,  0){\line( 1, 1){ 60}}
 \put( 30, 60){\line( 1,-1){ 60}}
 \put( 90,  0){\line( 1, 1){ 30}}
 \put( 90, 60){\line( 1,-1){ 30}}
 \put( 60, 30){\circle*{6}}
 \put( -3, 27){\makebox(0,0)[rb]{\sc $a$}}
 \put( 27,-10){\ma{\sc $c$}}
 \put( 87,-10){\ma{\sc $c$}}
 \put(122, 27){\ma{\sc $e$}}
 \put( 27, 62){\ma{\sc $d$}}
 \put( 87, 62){\ma{\sc $d$}}
 \put( 40, 27){\makebox(0,0)[rb]{\sc $-2\la$}}
 \put( 98, 27){\makebox(0,0)[rb]{\sc $2\la$}}
 \put( 55, 14){\ma{\sc $c^{\scs\prime}$}}
\end{picture}}

\def\face#1#2#3#4#5{
\setlength{\unitlength}{0.0155in}
\begin{picture}(40,30)(-5,13)
 \thicklines
 \put(15, 5){\makebox(0,0)[b]{\line(1,0){20}}}
 \put( 5, 5){\makebox(0,0)[b]{\line(0,1){20}}}
 \put(15,25){\makebox(0,0)[b]{\line(-1,0){20}}}
 \put(25,25){\makebox(0,0)[b]{\line(0,-1){20}}}
 \put( 4, 2){\makebox(0,0)[rb]{\sc $#1$}}
 \put(26, 2){\sc $#2$}
 \put(26,25){\sc $#3$}
 \put( 4,25){\makebox(0,0)[rb]{\sc $#4$}}
 \put(13,13){\sc $#5$}
\end{picture}}

\def\dface#1#2#3#4#5{
\setlength{\unitlength}{0.009in}
\rule[-9mm]{0mm}{20mm}
\begin{picture}(90,48)(-16,27)
 \thicklines
 \put( 0, 30){\line( 1,-1){ 30}}
 \put(30,  0){\line( 1, 1){ 30}}
 \put(30, 60){\line(-1,-1){ 30}}
 \put(60, 30){\line(-1, 1){ 30}}
 \put(-3, 27){\makebox(0,0)[rb]{\ra{\twlrm \sc $#1$}}}
 \put(27,-10){\ma{\ra{\twlrm \sc $#2$}}}
 \put(62, 27){\ma{\ra{\twlrm \sc $#3$}}}
 \put(27, 62){\ma{\ra{\twlrm \sc $#4$}}}
 \put(38, 27){\makebox(0,0)[rb]{\ra{\twlrm \sc $#5$}}}
\end{picture}}

\def\dfacesym#1#2#3#4#5{
\setlength{\unitlength}{0.009in}
\rule[-9mm]{0mm}{20mm}
\begin{picture}(90,48)(-16,27)
 \thicklines
 \put( 0, 30){\line( 1,-1){ 30}}
 \put(30,  0){\line( 1, 1){ 30}}
 \put(30, 60){\line(-1,-1){ 30}}
 \put(60, 30){\line(-1, 1){ 30}}
 \multiput(0,30)( 0.55, 0){9}{\makebox(0.4444,0.6667){\sevrm .}}
 \multiput(0,30)(-0.55, 0){9}{\makebox(0.4444,0.6667){\sevrm .}}
 \multiput(0,30)( 0, 0.55){9}{\makebox(0.4444,0.6667){\sevrm .}}
 \multiput(0,30)( 0,-0.55){9}{\makebox(0.4444,0.6667){\sevrm .}}
 \put(-7, 27){\makebox(0,0)[rb]{\ra{\twlrm \sc $#1$}}}
 \put(27,-10){\ma{\ra{\twlrm \sc $#2$}}}
 \put(62, 27){\ma{\ra{\twlrm \sc $#3$}}}
 \put(27, 62){\ma{\ra{\twlrm \sc $#4$}}}
 \put(38, 27){\makebox(0,0)[rb]{\ra{\twlrm \sc $#5$}}}
\end{picture}}

\def\dfaceasym#1#2#3#4#5{
\setlength{\unitlength}{0.009in}
\rule[-9mm]{0mm}{20mm}
\begin{picture}(90,48)(-16,27)
 \thicklines
 \put( 0, 30){\line( 1,-1){ 30}}
 \put(30,  0){\line( 1, 1){ 30}}
 \put(30, 60){\line(-1,-1){ 30}}
 \put(60, 30){\line(-1, 1){ 30}}
 \put(60, 30){\circle{6}}
 \put(-3, 27){\makebox(0,0)[rb]{\ra{\twlrm \sc $#1$}}}
 \put(27,-10){\ma{\ra{\twlrm \sc $#2$}}}
 \put(67, 27){\ma{\ra{\twlrm \sc $#3$}}}
 \put(27, 62){\ma{\ra{\twlrm \sc $#4$}}}
 \put(38, 27){\makebox(0,0)[rb]{\ra{\twlrm \sc $#5$}}}
\end{picture}}

\def\onebytwoface#1#2#3#4#5#6#7#8{
\setlength{\unitlength}{0.015in}
\rule[-6mm]{0mm}{13mm}
\begin{picture}(70,20)(-10,13)
 \thicklines
 \multiput(5,5)(0,20){2}{\line(1,0){40}}
 \multiput(5,5)(20,0){3}{\line(0,1){20}}
 \put( 3, 2){\makebox(0,0)[rb]{\sc $#1$}}
 \put(46, 2){\sc $#2$}
 \put(46,26){\sc $#3$}
 \put( 3,26){\makebox(0,0)[rb]{\sc $#4$}}
 \put(20,-2){\makebox(10,0)[b]{\sc $#5$}}
 \put(20,28){\makebox(10,0)[b]{\sc $#6$}}
 \put(10,13){\makebox(10,0)[b]{\sc $#7$}}
 \put(30,13){\makebox(10,0)[b]{\sc $#8$}}
\end{picture}}

\def\onebytwofacesym#1#2#3#4#5#6#7#8{
\setlength{\unitlength}{0.015in}
\rule[-6mm]{0mm}{13mm}
\begin{picture}(70,20)(-10,13)
 \thicklines
 \multiput(5,5)(0,20){2}{\line(1,0){40}}
 \multiput(5,5)(20,0){3}{\line(0,1){20}}
 \put( 3, 2){\makebox(0,0)[rb]{\sc $#1$}}
 \put(46, 2){\sc $#2$}
 \put(46,26){\sc $#3$}
 \put( 3,26){\makebox(0,0)[rb]{\sc $#4$}}
 \put(20,-2){\makebox(10,0)[b]{\sc $#5$}}
 \put(20,28){\makebox(10,0)[b]{\sc $#6$}}
 \put(10,13){\makebox(10,0)[b]{\sc $#7$}}
 \put(30,13){\makebox(10,0)[b]{\sc $#8$}}
 \multiput(24.8,5)( 0.4,-0.4){9}{\makebox(0.4444,0.6667){\sevrm .}}
 \multiput(24.8,5)(-0.4, 0.4){9}{\makebox(0.4444,0.6667){\sevrm .}}
 \multiput(24.8,5)( 0.4, 0.4){9}{\makebox(0.4444,0.6667){\sevrm .}}
 \multiput(24.8,5)(-0.4,-0.4){9}{\makebox(0.4444,0.6667){\sevrm .}}
\end{picture}}

\def\onebytwofaceasym#1#2#3#4#5#6#7#8{
\setlength{\unitlength}{0.015in}
\rule[-6mm]{0mm}{13mm}
\begin{picture}(70,20)(-10,13)
 \thicklines
 \multiput(5,5)(0,20){2}{\line(1,0){40}}
 \multiput(5,5)(20,0){3}{\line(0,1){20}}
 \put(25,25){\circle{5}}
 \put( 3, 2){\makebox(0,0)[rb]{\sc $#1$}}
 \put(46, 2){\sc $#2$}
 \put(46,26){\sc $#3$}
 \put(3, 26){\makebox(0,0)[rb]{\sc $#4$}}
 \put(21,-2){\makebox(10,0)[b]{\sc $#5$}}
 \put(21,28){\makebox(10,0)[b]{\sc $#6$}}
 \put(10,13){\makebox(10,0)[b]{\sc $#7$}}
 \put(30,13){\makebox(10,0)[b]{\sc $#8$}}
\end{picture}}

\def\onebytwofacesymasym{
\setlength{\unitlength}{0.015in}
\rule[-6mm]{0mm}{13mm}
\begin{picture}(70,20)(-10,13)
 \thicklines
 \multiput(5,5)(0,20){2}{\line(1,0){40}}
 \multiput(5,5)(20,0){3}{\line(0,1){20}}
 \put( 3, 2){\makebox(0,0)[rb]{\sc $a$}}
 \put(46, 2){\sc $b$}
 \put(46,26){\sc $c$}
 \put( 3,26){\makebox(0,0)[rb]{\sc $d$}}
 \put(20,-2){\makebox(10,0)[b]{\sc $\alpha$}}
 \put(10,13){\makebox(10,0)[b]{\sc $u$}}
 \put(30,13){\makebox(10,0)[b]{\sc $u\!\+\!2\!\la$}}
 \multiput(24.8,5)( 0.4,-0.4){9}{\makebox(0.4444,0.6667){\sevrm .}}
 \multiput(24.8,5)(-0.4, 0.4){9}{\makebox(0.4444,0.6667){\sevrm .}}
 \multiput(24.8,5)( 0.4, 0.4){9}{\makebox(0.4444,0.6667){\sevrm .}}
 \multiput(24.8,5)(-0.4,-0.4){9}{\makebox(0.4444,0.6667){\sevrm .}}
 \put(25,25){\circle{5}}
\end{picture}}

\def\twobytwofacesym{
\setlength{\unitlength}{0.015in}
\rule[-9mm]{0mm}{20mm}
\begin{picture}(70,30)(-10,23)
 \thicklines
 \multiput(5,5)(0,20){3}{\line(1,0){40}}
 \multiput(5,5)(20,0){3}{\line(0,1){40}}
 \put(25,25){\circle*{5}}
 \put( 3, 2){\makebox(0,0)[rb]{\sc $a$}}
 \put(47, 2){\sc $b$}
 \put(47,23){\sc $c$}
 \put( 3,23){\makebox(0,0)[rb]{\sc $d$}}
 \put(47,46){\sc $e$}
 \put(20,47){\makebox(10,0)[b]{\sc $e^{\scs\prime}$}}
 \put( 3,46){\makebox(0,0)[rb]{\sc $f$}}
 \put(23,27){\makebox(0,0)[rb]{\sc $c^{\scs\prime}$}}
 \put(20,-2){\makebox(10,0)[b]{\sc $\alpha$}}
 \put(10,13){\makebox(10,0)[b]{\sc $u$}}
 \put(30,13){\makebox(10,0)[b]{\sc $u\!\+\! 2\!\la$}}
 \put(10,33){\makebox(10,0)[b]{\sc $v$}}
 \put(30,33){\makebox(10,0)[b]{\sc $v\!\+\! 2\!\la$}}
 \multiput(24.8,5)( 0.4,-0.4){9}{\makebox(0.4444,0.6667){\sevrm .}}
 \multiput(24.8,5)(-0.4, 0.4){9}{\makebox(0.4444,0.6667){\sevrm .}}
 \multiput(24.8,5)( 0.4, 0.4){9}{\makebox(0.4444,0.6667){\sevrm .}}
 \multiput(24.8,5)(-0.4,-0.4){9}{\makebox(0.4444,0.6667){\sevrm .}}
\end{picture}}

\def\onebythreefaceasym#1#2#3#4#5#6#7#8#9{
\setlength{\unitlength}{0.015in}
\rule[-6mm]{0mm}{13mm}
\begin{picture}(90,20)(-10,13)
 \thicklines
 \multiput(5,5)(0,20){2}{\line(1,0){60}}
 \multiput(5,5)(20,0){4}{\line(0,1){20}}
 \put(35,25){\oval(30,5)}
 \put( 3, 2){\makebox(0,0)[rb]{\sc $#1$}}
 \put(21,-1){\makebox(10,0)[b]{\sc $#2$}}
 \put(41,-1){\makebox(10,0)[b]{\sc $#3$}}
 \put(66, 2){\sc $#4$}
 \put(66,26){\sc $#5$}
 \put( 3,26){\makebox(0,0)[rb]{\sc $#6$}}
 \put(10,13){\makebox(10,0)[b]{\sc $#7$}}
 \put(30,13){\makebox(10,0)[b]{\sc $#8$}}
 \put(50,13){\makebox(10,0)[b]{\sc $#9$}}
\end{picture}}

\def\onebynnface{
\setlength{\unitlength}{0.02in}
\rule[-6mm]{0mm}{14mm}
\begin{picture}(120,20)(-10,13)
 \thicklines
 \multiput( 5,5)(0,20){2}{\line(1,0){90}}
 \multiput( 5,5)(20,0){3}{\line(0,1){20}}
 \multiput(75,5)(20,0){2}{\line(0,1){20}}
 \put(  1, 1){\sc $b_{\scs 1}$}
 \put( 96, 1){\sc $b$}
 \put( 96,27){\sc $a_{\scs n+1}$}
 \put(  1,27){\sc $a$}
 \put( 20,-1){\makebox(10,0)[b]{\sc $c_{\scs 1}^{\scs j}$}}
 \put( 40,-1){\makebox(10,0)[b]{\sc $c_{\scs 2}^{\scs j}$}}
 \put( 70,-1){\makebox(10,0)[b]{\sc $c_{\scs n}^{\scs j}$}}
 \put( 20,27){\makebox(10,0)[b]{\sc $a_{\scs 1}$}}
 \put( 40,27){\makebox(10,0)[b]{\sc $a_{\scs 2}$}}
 \put( 70,27){\makebox(10,0)[b]{\sc $a_{\scs n}$}}
 \put( 10,14){\makebox(10,0)[b]{\sc $u$}}
 \put( 30,14){\makebox(10,0)[b]{\sc $u\+2\la$}}
 \put( 55,13.5){\makebox(10,0)[b]{$\cdots$}}
 \put( 80,14){\makebox(10,0)[b]{\sc $u\+2n\la$}}
\end{picture}}

\def\onebynnfacesym{
\setlength{\unitlength}{0.02in}
\rule[-6mm]{0mm}{14mm}
\begin{picture}(120,20)(-10,13)
 \thicklines
 \multiput( 5,5)(0,20){2}{\line(1,0){90}}
 \multiput( 5,5)(20,0){3}{\line(0,1){20}}
 \multiput(75,5)(20,0){2}{\line(0,1){20}}
 \put(  1, 1){\sc $b_{\scs 1}$}
 \put( 96, 1){\sc $b$}
 \put( 96,27){\sc $a_{\scs n+1}$}
 \put(  1,27){\sc $a$}
 \put( 49,-3){$\alpha$}
 \put( 20,27){\makebox(10,0)[b]{\sc $a_{\scs 1}$}}
 \put( 40,27){\makebox(10,0)[b]{\sc $a_{\scs 2}$}}
 \put( 70,27){\makebox(10,0)[b]{\sc $a_{\scs n}$}}
 \put( 10,14){\makebox(10,0)[b]{\sc $u$}}
 \put( 30,14){\makebox(10,0)[b]{\sc $u\+2\la$}}
 \put( 55,13.5){\makebox(10,0)[b]{$\cdots$}}
 \put( 80,14){\makebox(10,0)[b]{\sc $u\+2n\la$}}
 \multiput(24.8,4.8)( 0.3,-0.3){9}{\makebox(0.4444,0.6667){\sevrm .}}
 \multiput(24.8,4.8)(-0.3, 0.3){9}{\makebox(0.4444,0.6667){\sevrm .}}
 \multiput(24.8,4.8)( 0.3, 0.3){9}{\makebox(0.4444,0.6667){\sevrm .}}
 \multiput(24.8,4.8)(-0.3,-0.3){9}{\makebox(0.4444,0.6667){\sevrm .}}
 \multiput(74.8,4.8)( 0.3,-0.3){9}{\makebox(0.4444,0.6667){\sevrm .}}
 \multiput(74.8,4.8)(-0.3, 0.3){9}{\makebox(0.4444,0.6667){\sevrm .}}
 \multiput(74.8,4.8)( 0.3, 0.3){9}{\makebox(0.4444,0.6667){\sevrm .}}
 \multiput(74.8,4.8)(-0.3,-0.3){9}{\makebox(0.4444,0.6667){\sevrm .}}
 \thinlines
 \put( 27.6,2.4){\line(1, 0){44.8}}
 \put( 27.6,7.6){\line(1, 0){44.8}}
\end{picture}}

\def\onebynfacesymsym{
\setlength{\unitlength}{0.02in}
\rule[-6mm]{0mm}{14mm}
\begin{picture}(130,20)(-5,13)
 \thicklines
 \multiput( 5,5)(0,20){2}{\line(1,0){110}}
 \multiput( 5,5)(20,0){3}{\line(0,1){20}}
 \multiput(75,5)(20,0){3}{\line(0,1){20}}
 \put(  1,27){\sc $a$}
 \put( 20,27){\makebox(10,0)[b]{\sc $a_{\scs 1}$}}
 \put( 40,27){\makebox(10,0)[b]{\sc $a_{\scs 2}$}}
 \put( 70,27){\makebox(10,0)[b]{\sc $a_{\scs n\-1}$}}
 \put( 90,27){\makebox(10,0)[b]{\sc $a_{\scs n}$}}
 \put(116,27){\sc $a_{\scs n\+1}$}
 \put(  1, 1){\sc $b_{\scs 1}$}
 \put(116, 1){\sc $b$}
 \put( 93,-3){$\delta$}
 \put( 49,-3){$\gamma$}
 \put( 10,14){\makebox(10,0)[b]{\sc $u$}}
 \put( 30,14){\makebox(10,0)[b]{\sc $u\+2\la$}}
 \put( 55,13.5){\makebox(10,0)[b]{$\cdots$}}
 \put( 80,17){\makebox(10,0)[b]{\tiny $u+$}}
 \put( 80,12){\makebox(10,0)[b]{\tiny $2(\!n\!\-\!1\!)\!\la$}}
 \put(100,14){\makebox(10,0)[b]{\sc $u\+2n\la$}}
 \multiput(24.8,4.8)( 0.3,-0.3){9}{\makebox(0.4444,0.6667){\sevrm .}}
 \multiput(24.8,4.8)(-0.3, 0.3){9}{\makebox(0.4444,0.6667){\sevrm .}}
 \multiput(24.8,4.8)( 0.3, 0.3){9}{\makebox(0.4444,0.6667){\sevrm .}}
 \multiput(24.8,4.8)(-0.3,-0.3){9}{\makebox(0.4444,0.6667){\sevrm .}}
 \multiput(74.8,4.8)( 0.3,-0.3){9}{\makebox(0.4444,0.6667){\sevrm .}}
 \multiput(74.8,4.8)(-0.3, 0.3){9}{\makebox(0.4444,0.6667){\sevrm .}}
 \multiput(74.8,4.8)( 0.3, 0.3){9}{\makebox(0.4444,0.6667){\sevrm .}}
 \multiput(74.8,4.8)(-0.3,-0.3){9}{\makebox(0.4444,0.6667){\sevrm .}}
 \multiput(94.8,4.8)( 0.3,-0.3){9}{\makebox(0.4444,0.6667){\sevrm .}}
 \multiput(94.8,4.8)(-0.3, 0.3){9}{\makebox(0.4444,0.6667){\sevrm .}}
 \multiput(94.8,4.8)( 0.3, 0.3){9}{\makebox(0.4444,0.6667){\sevrm .}}
 \multiput(94.8,4.8)(-0.3,-0.3){9}{\makebox(0.4444,0.6667){\sevrm .}}
 \thinlines
 \put( 27.6,2.4){\line(1, 0){44.8}}
 \put( 27.6,7.6){\line(1, 0){44.8}}
\end{picture}}

\def\onebynfacesym{
\setlength{\unitlength}{0.02in}
\rule[-6mm]{0mm}{14mm}
\begin{picture}(130,20)(-5,13)
 \thicklines
 \multiput( 5,5)(0,20){2}{\line(1,0){110}}
 \multiput( 5,5)(20,0){3}{\line(0,1){20}}
 \multiput(75,5)(20,0){3}{\line(0,1){20}}
 \put(  1,27){\sc $a$}
 \put( 20,27){\makebox(10,0)[b]{\sc $a_{\scs 1}$}}
 \put( 40,27){\makebox(10,0)[b]{\sc $a_{\scs 2}$}}
 \put( 70,27){\makebox(10,0)[b]{\sc $a_{\scs n\-1}$}}
 \put( 90,27){\makebox(10,0)[b]{\sc $a_{\scs n}$}}
 \put(116,27){\sc $a_{\scs n\+1}$}
 \put(  1, 1){\sc $b_{\scs 1}$}
 \put(116, 1){\sc $b$}
 \put( 58,-3){$\alpha$}
 \put( 10,14){\makebox(10,0)[b]{\sc $u$}}
 \put( 30,14){\makebox(10,0)[b]{\sc $u\+2\la$}}
 \put( 55,13.5){\makebox(10,0)[b]{$\cdots$}}
 \put( 80,17){\makebox(10,0)[b]{\tiny $u+$}}
 \put( 80,12){\makebox(10,0)[b]{\tiny $2(\!n\!\-\!1\!)\!\la$}}
 \put(100,14){\makebox(10,0)[b]{\sc $u\+2n\la$}}
 \multiput(24.8,4.8)( 0.3,-0.3){9}{\makebox(0.4444,0.6667){\sevrm .}}
 \multiput(24.8,4.8)(-0.3, 0.3){9}{\makebox(0.4444,0.6667){\sevrm .}}
 \multiput(24.8,4.8)( 0.3, 0.3){9}{\makebox(0.4444,0.6667){\sevrm .}}
 \multiput(24.8,4.8)(-0.3,-0.3){9}{\makebox(0.4444,0.6667){\sevrm .}}
 \multiput(94.8,4.8)( 0.3,-0.3){9}{\makebox(0.4444,0.6667){\sevrm .}}
 \multiput(94.8,4.8)(-0.3, 0.3){9}{\makebox(0.4444,0.6667){\sevrm .}}
 \multiput(94.8,4.8)( 0.3, 0.3){9}{\makebox(0.4444,0.6667){\sevrm .}}
 \multiput(94.8,4.8)(-0.3,-0.3){9}{\makebox(0.4444,0.6667){\sevrm .}}
 \thinlines
 \put( 27.6,2.4){\line(1, 0){64.8}}
 \put( 27.6,7.6){\line(1, 0){64.8}}
\end{picture}}

\def\twobynoneface{
\setlength{\unitlength}{0.02in}
\rule[-12mm]{0mm}{26mm}
\begin{picture}(140,30)(-10,23)
 \thicklines
 \put( 5,5){\line(1,0){90}}
 \multiput( 5,25)(0,20){2}{\line(1,0){110}}
 \multiput( 5, 5)(20,0){3}{\line(0,1){ 40}}
 \multiput(75, 5)(20,0){2}{\line(0,1){ 40}}
 \put(115,25){\line(0,1){ 20}}
 \multiput(25,25)(20,0){2}{\circle*{3}}
 \multiput(75,25)(20,0){2}{\circle*{3}}
 \put(  1, 1){\sc $b_{\scs 2}$}
 \put( -1,23){\sc $b_{\scs 1}$}
 \put( 96, 1){\sc $b$}
 \put( 96,20){\sc $d$}
 \put(116,23){\sc $b$}
 \put(116,47){\sc $a_{\scs n\+1}$}
 \put(  1,47){\sc $a$}
 \put( 20,-1){\makebox(10,0)[b]{\sc $c_{\scs 1}^{\scs j}$}}
 \put( 40,-1){\makebox(10,0)[b]{\sc $c_{\scs 2}^{\scs j}$}}
 \put( 70,-1){\makebox(10,0)[b]{\sc $c_{\scs n\-1}^{\scs j}$}}
 \put( 20,47){\makebox(10,0)[b]{\sc $a_{\scs 1}$}}
 \put( 40,47){\makebox(10,0)[b]{\sc $a_{\scs 2}$}}
 \put( 70,47){\makebox(10,0)[b]{\sc $a_{\scs n\-1}$}}
 \put( 90,47){\makebox(10,0)[b]{\sc $a_{\scs n}$}}
 \put( 10,14){\makebox(10,0)[b]{\sc $-2n\la$}}
 \put( 30,14){\makebox(10,0)[b]{\tiny $-2(\!n\!\-\!1\!)\!\la$}}
 \put( 55,13.5){\makebox(10,0)[b]{$\cdots$}}
 \put( 80,14){\makebox(10,0)[b]{\sc $-2\la$}}
 \put( 10,34){\makebox(10,0)[b]{\sc $u$}}
 \put( 30,34){\makebox(10,0)[b]{\sc $u\+2\la$}}
 \put( 55,33.5){\makebox(10,0)[b]{$\cdots$}}
 \put( 80,37){\makebox(10,0)[b]{\tiny $u+$}}
 \put( 80,32){\makebox(10,0)[b]{\tiny $2(\!n\!\-\!1\!)\!\la$}}
 \put(100,34){\makebox(10,0)[b]{\sc $u\+2n\la$}}
\end{picture}}

\def\twobynonefacesym{
\setlength{\unitlength}{0.02in}
\rule[-12mm]{0mm}{26mm}
\begin{picture}(140,30)(-10,23)
 \thicklines
 \put( 5,5){\line(1,0){90}}
 \multiput( 5,25)(0,20){2}{\line(1,0){110}}
 \multiput( 5, 5)(20,0){3}{\line(0,1){ 40}}
 \multiput(75, 5)(20,0){2}{\line(0,1){ 40}}
 \put(115,25){\line(0,1){ 20}}
 \multiput(25,25)(20,0){2}{\circle*{3}}
 \multiput(75,25)(20,0){2}{\circle*{3}}
 \put(  1,47){\sc $a$}
 \put( 20,47){\makebox(10,0)[b]{\sc $a_{\scs 1}$}}
 \put( 40,47){\makebox(10,0)[b]{\sc $a_{\scs 2}$}}
 \put( 70,47){\makebox(10,0)[b]{\sc $a_{\scs n\-1}$}}
 \put( 90,47){\makebox(10,0)[b]{\sc $a_{\scs n}$}}
 \put(116,47){\sc $a_{\scs n\+1}$}
 \put( 96, 1){\sc $b$}
 \put( -1,23){\sc $b_{\scs 1}$}
 \put(  1, 1){\sc $b_{\scs 2}$}
 \put( 96,20){\sc $d$}
 \put(116,23){\sc $b$}
 \put( 49,-5){$\beta$}
 \put( 10,14){\makebox(10,0)[b]{\sc $-2n\la$}}
 \put( 30,14){\makebox(10,0)[b]{\tiny $-2(\!n\!\-\!1\!)\!\la$}}
 \put( 55,13.5){\makebox(10,0)[b]{$\cdots$}}
 \put( 80,14){\makebox(10,0)[b]{\sc $-2\la$}}
 \put( 10,34){\makebox(10,0)[b]{\sc $u$}}
 \put( 30,34){\makebox(10,0)[b]{\sc $u\+2\la$}}
 \put( 55,33.5){\makebox(10,0)[b]{$\cdots$}}
 \put( 80,37){\makebox(10,0)[b]{\tiny $u+$}}
 \put( 80,32){\makebox(10,0)[b]{\tiny $2(\!n\!\-\!1\!)\!\la$}}
 \put(100,34){\makebox(10,0)[b]{\sc $u\+2n\la$}}
 \multiput(24.8,4.8)( 0.3,-0.3){9}{\makebox(0.4444,0.6667){\sevrm .}}
 \multiput(24.8,4.8)(-0.3, 0.3){9}{\makebox(0.4444,0.6667){\sevrm .}}
 \multiput(24.8,4.8)( 0.3, 0.3){9}{\makebox(0.4444,0.6667){\sevrm .}}
 \multiput(24.8,4.8)(-0.3,-0.3){9}{\makebox(0.4444,0.6667){\sevrm .}}
 \multiput(74.8,4.8)( 0.3,-0.3){9}{\makebox(0.4444,0.6667){\sevrm .}}
 \multiput(74.8,4.8)(-0.3, 0.3){9}{\makebox(0.4444,0.6667){\sevrm .}}
 \multiput(74.8,4.8)( 0.3, 0.3){9}{\makebox(0.4444,0.6667){\sevrm .}}
 \multiput(74.8,4.8)(-0.3,-0.3){9}{\makebox(0.4444,0.6667){\sevrm .}}
 \thinlines
 \put( 27.6,2.4){\line(1, 0){44.8}}
 \put( 27.6,7.6){\line(1, 0){44.8}}
\end{picture}}

\def\twobynonefacesymsym{
\setlength{\unitlength}{0.02in}
\rule[-26mm]{0mm}{48mm}
\begin{picture}(130,40)(-5,10)
 \thicklines
 \multiput( 5,25)(0,20){2}{\line(1,0){110}}
 \multiput( 5,25)(20,0){3}{\line(0,1){ 20}}
 \multiput(75,25)(20,0){3}{\line(0,1){ 20}}
 \put( 95,25){\circle*{3}}
 \multiput(24.8,24.8)( 0.3,-0.3){9}{\makebox(0.4444,0.6667){\sevrm .}}
 \multiput(24.8,24.8)(-0.3, 0.3){9}{\makebox(0.4444,0.6667){\sevrm .}}
 \multiput(24.8,24.8)( 0.3, 0.3){9}{\makebox(0.4444,0.6667){\sevrm .}}
 \multiput(24.8,24.8)(-0.3,-0.3){9}{\makebox(0.4444,0.6667){\sevrm .}}
 \multiput(74.8,24.8)( 0.3,-0.3){9}{\makebox(0.4444,0.6667){\sevrm .}}
 \multiput(74.8,24.8)(-0.3, 0.3){9}{\makebox(0.4444,0.6667){\sevrm .}}
 \multiput(74.8,24.8)( 0.3, 0.3){9}{\makebox(0.4444,0.6667){\sevrm .}}
 \multiput(74.8,24.8)(-0.3,-0.3){9}{\makebox(0.4444,0.6667){\sevrm .}}
 \thinlines
 \put( 27.6,22.4){\line(1, 0){44.8}}
 \put( 27.6,27.6){\line(1, 0){44.8}}
 \put(  1,47){\sc $a$}
 \put( 20,47){\makebox(10,0)[b]{\sc $a_{\scs 1}$}}
 \put( 40,47){\makebox(10,0)[b]{\sc $a_{\scs 2}$}}
 \put( 70,47){\makebox(10,0)[b]{\sc $a_{\scs n\-1}$}}
 \put( 90,47){\makebox(10,0)[b]{\sc $a_{\scs n}$}}
 \put(116,47){\sc $a_{\scs n\+1}$}
 \put( -1,23){\sc $b_{\scs 1}$}
 \put(116,23){\sc $b$}
 \put( 98,20){\sc $d$}
 \put( 49,17){$\gamma$}
 \put( 10,34){\makebox(10,0)[b]{\sc $u$}}
 \put( 30,34){\makebox(10,0)[b]{\sc $u\+2\la$}}
 \put( 55,33.5){\makebox(10,0)[b]{$\cdots$}}
 \put( 80,37){\makebox(10,0)[b]{\tiny $u+$}}
 \put( 80,32){\makebox(10,0)[b]{\tiny $2(\!n\!\-\!1\!)\!\la$}}
 \put(100,34){\makebox(10,0)[b]{\sc $u\+2n\la$}}
 \thicklines
 \multiput(95,25)(9,-18){2}{\line(-2,-1){81}}
 \multiput(95,25)(-18,-9){2}{\line(1,-2){9}}
 \multiput(50, 2.5)(-18,-9){3}{\line(1,-2){9}}
 \multiput(41,-24.7)( 0.36,-0.12){9}{\makebox(0.4444,0.6667){\sevrm .}}
 \multiput(41,-24.7)(-0.36, 0.12){9}{\makebox(0.4444,0.6667){\sevrm .}}
 \multiput(41,-24.7)( 0.12, 0.36){9}{\makebox(0.4444,0.6667){\sevrm .}}
 \multiput(41,-24.7)(-0.12,-0.36){9}{\makebox(0.4444,0.6667){\sevrm .}}
 \multiput(86,-2.2)( 0.36,-0.12){9}{\makebox(0.4444,0.6667){\sevrm .}}
 \multiput(86,-2.2)(-0.36, 0.12){9}{\makebox(0.4444,0.6667){\sevrm .}}
 \multiput(86,-2.2)( 0.12, 0.36){9}{\makebox(0.4444,0.6667){\sevrm .}}
 \multiput(86,-2.2)(-0.12,-0.36){9}{\makebox(0.4444,0.6667){\sevrm .}}
 \thinlines
 \put(82.9,-1.1){\line(-2,-1){40.5}}
 \put(84.7,-5.2){\line(-2,-1){40.5}}
 \multiput(64,-3.5)(3,1.5){3}{$\cdot$}
 \put(105,5){\sc $b$}
 \put(  9,-13){\sc $b_{\scs 1}$}
 \put( 20,-38){\sc $b_{\scs 2}$}
 \put( 27, -4){\sc $d_{\scs 1}^{\scs\gamma}$}
 \put( 45,  5){\sc $d_{\scs 2}^{\scs\gamma}$}
 \put( 66, 17){\sc $d_{\scs n\-1}^{\scs\gamma}$}
 \put( 65,-21){$\beta$}
 \put( 22,-21.5){\makebox(10,0)[b]{\sc $-2n\la$}}
 \put( 40,-12.5){\makebox(10,0)[b]{\tiny $-2(\!n\!\-\!1\!)\!\la$}}
 \put( 85, 10){\makebox(10,0)[b]{\sc $-2\la$}}
\end{picture}}

\def\onebynfusion#1{
\setlength{\unitlength}{0.02in}
\rule[-6mm]{0mm}{14mm}
\begin{picture}(130,20)(-10,13)
 \thicklines
 \multiput( 5,5)(0,20){2}{\line(1,0){100}}
 \multiput( 5,5)(20,0){3}{\line(0,1){ 20}}
 \multiput(85,5)(20,0){2}{\line(0,1){ 20}}
 \put(  1, 1){\sc $a$}
 \put(106, 1){\sc $b$}
 \put(106,27){\sc $c$}
 \put(  1,27){\sc $d$}
 \put( 54,-3){$\alpha$}
 \put( 20,27){\makebox(10,0)[b]{\sc $e_{\scs 1}^{\scs #1}$}}
 \put( 40,27){\makebox(10,0)[b]{\sc $e_{\scs 2}^{\scs #1}$}}
 \put( 80,27){\makebox(10,0)[b]{\sc $e_{\scs n\-1}^{\scs #1}$}}
 \put( 10,14){\makebox(10,0)[b]{\sc $u$}}
 \put( 30,14){\makebox(10,0)[b]{\sc $u\+2\la$}}
 \put( 60,13.5){\makebox(10,0)[b]{$\cdots$}}
 \put( 90,17){\makebox(10,0)[b]{\tiny $u+$}}
 \put( 90,12){\makebox(10,0)[b]{\tiny $2(\!n\!\-\!1\!)\!\la$}}
 \multiput(24.8,4.8)( 0.3,-0.3){9}{\makebox(0.4444,0.6667){\sevrm .}}
 \multiput(24.8,4.8)(-0.3, 0.3){9}{\makebox(0.4444,0.6667){\sevrm .}}
 \multiput(24.8,4.8)( 0.3, 0.3){9}{\makebox(0.4444,0.6667){\sevrm .}}
 \multiput(24.8,4.8)(-0.3,-0.3){9}{\makebox(0.4444,0.6667){\sevrm .}}
 \multiput(84.8,4.8)( 0.3,-0.3){9}{\makebox(0.4444,0.6667){\sevrm .}}
 \multiput(84.8,4.8)(-0.3, 0.3){9}{\makebox(0.4444,0.6667){\sevrm .}}
 \multiput(84.8,4.8)( 0.3, 0.3){9}{\makebox(0.4444,0.6667){\sevrm .}}
 \multiput(84.8,4.8)(-0.3,-0.3){9}{\makebox(0.4444,0.6667){\sevrm .}}
 \thinlines
 \put( 27.6,2.4){\line(1, 0){54.8}}
 \put( 27.6,7.6){\line(1, 0){54.8}}
\end{picture}}

\def\onebynnfusion#1{
\setlength{\unitlength}{0.02in}
\rule[-6mm]{0mm}{14mm}
\begin{picture}(130,20)(-10,13)
 \thicklines
 \multiput( 5,5)(0,20){2}{\line(1,0){100}}
 \multiput( 5,5)(20,0){3}{\line(0,1){ 20}}
 \multiput(85,5)(20,0){2}{\line(0,1){ 20}}
 \put(  1, 1){\sc $a$}
 \put(106, 1){\sc $b$}
 \put(106,27){\sc $c$}
 \put(  1,27){\sc $d$}
 \put( 54,-3){$\alpha$}
 \put( 20,27){\makebox(10,0)[b]{\sc $e_{\scs 1}^{\scs #1}$}}
 \put( 40,27){\makebox(10,0)[b]{\sc $e_{\scs 2}^{\scs #1}$}}
 \put( 80,27){\makebox(10,0)[b]{\sc $e_{\scs n}^{\scs #1}$}}
 \put( 10,14){\makebox(10,0)[b]{\sc $u$}}
 \put( 30,14){\makebox(10,0)[b]{\sc $u\+2\la$}}
 \put( 60,13.5){\makebox(10,0)[b]{$\cdots$}}
 \put( 90,14){\makebox(10,0)[b]{\sc $u\+2n\la$}}
 \multiput(24.8,4.8)( 0.3,-0.3){9}{\makebox(0.4444,0.6667){\sevrm .}}
 \multiput(24.8,4.8)(-0.3, 0.3){9}{\makebox(0.4444,0.6667){\sevrm .}}
 \multiput(24.8,4.8)( 0.3, 0.3){9}{\makebox(0.4444,0.6667){\sevrm .}}
 \multiput(24.8,4.8)(-0.3,-0.3){9}{\makebox(0.4444,0.6667){\sevrm .}}
 \multiput(84.8,4.8)( 0.3,-0.3){9}{\makebox(0.4444,0.6667){\sevrm .}}
 \multiput(84.8,4.8)(-0.3, 0.3){9}{\makebox(0.4444,0.6667){\sevrm .}}
 \multiput(84.8,4.8)( 0.3, 0.3){9}{\makebox(0.4444,0.6667){\sevrm .}}
 \multiput(84.8,4.8)(-0.3,-0.3){9}{\makebox(0.4444,0.6667){\sevrm .}}
 \thinlines
 \put( 27.6,2.4){\line(1, 0){54.8}}
 \put( 27.6,7.6){\line(1, 0){54.8}}
\end{picture}}

\def\onebynnfaced{
\setlength{\unitlength}{0.02in}
\rule[-6mm]{0mm}{14mm}
\begin{picture}(130,20)(-10,13)
 \thicklines
 \multiput( 5,5)(0,20){2}{\line(1,0){100}}
 \multiput( 5,5)(20,0){3}{\line(0,1){ 20}}
 \multiput(85,5)(20,0){2}{\line(0,1){ 20}}
 \put(  1, 1){\sc $a$}
 \put(106, 1){\sc $b$}
 \put(106,27){\sc $c$}
 \put(  1,27){\sc $d$}
 \put( 20,-1){\makebox(10,0)[b]{\sc $e_{\scs 1}^{\scs j}$}}
 \put( 40,-1){\makebox(10,0)[b]{\sc $e_{\scs 2}^{\scs j}$}}
 \put( 80,-1){\makebox(10,0)[b]{\sc $e_{\scs n}^{\scs j}$}}
 \put( 20,27){\makebox(10,0)[b]{\sc $e_{\scs 1}^{\scs\beta}$}}
 \put( 40,27){\makebox(10,0)[b]{\sc $e_{\scs 2}^{\scs\beta}$}}
 \put( 80,27){\makebox(10,0)[b]{\sc $e_{\scs n}^{\scs\beta}$}}
 \put( 10,14){\makebox(10,0)[b]{\sc $u$}}
 \put( 30,14){\makebox(10,0)[b]{\sc $u\+2\la$}}
 \put( 60,13.5){\makebox(10,0)[b]{$\cdots$}}
 \put( 90,14){\makebox(10,0)[b]{\sc $u\+2n\la$}}
\end{picture}}

\def\zeronnfu{
\setlength{\unitlength}{0.025in}
\rule[-6mm]{0mm}{14mm}
\begin{picture}(180,20)(-10,13)
 \thicklines
 \multiput(  5, 5)(0,20){2}{\line(1,0){160}}
 \multiput(  5, 5)(20,0){5}{\line(0,1){ 20}}
 \multiput(125, 5)(20,0){3}{\line(0,1){ 20}}
 \multiput( 25,25)(40,0){2}{\circle{5}}
 \put(145,25){\circle{5}}
 \put( 10,14){\makebox(10,0)[b]{\sc $u$}}
 \put( 30,14){\makebox(10,0)[b]{\sc $u\+2\la$}}
 \put( 50,14){\makebox(10,0)[b]{\sc $u\+2\la$}}
 \put( 70,14){\makebox(10,0)[b]{\sc $u\+4\la$}}
 \put(100,13.5){\makebox(10,0)[b]{$\cdots$}}
 \put(130,14){\makebox(10,0)[b]{$\scs u\+2(\!n\-1\!)\la$}}
 \put(150,14){\makebox(10,0)[b]{\sc $u\+2n\la$}}
\end{picture}}

\def\splittz{
\setlength{\unitlength}{0.009in}
\rule[-12mm]{0mm}{26mm}
\begin{picture}(150,60)(-20,40)
 \thicklines
 \put(  0, 30){\line(1, 1){60}}
 \put(  0, 30){\line(1,-1){30}}
 \put( 30,  0){\line(1, 1){60}}
 \put( 30, 60){\line(1,-1){60}}
 \put( 60, 90){\line(1,-1){60}}
 \put( 90,  0){\line(1, 1){30}}
 \put( 60, 30){\circle*{6}}
 \put( 17, 27){\sc $-2\la$}
 \put( 73, 27){\sc $u\+2\la$}
 \put( 57, 57){\sc $u$}
 \put( 23, 63){\sc $a$}
 \put( 55, 14){\sc $a^{\scs\prime}$}
 \put( 27,-11){\sc $b$}
 \put( 87,-11){\sc $b$}
 \put(122, 27){\sc $c$}
 \put( 57, 93){\sc $d$}
 \put( 92, 63){\sc $c^{\scs\prime}$}
 \put(-10, 27){\sc $e$}
\end{picture}}

\def\pushtz{
\setlength{\unitlength}{0.009in}
\rule[-12mm]{0mm}{26mm}
\begin{picture}(150,70)(-20,40)
 \thicklines
 \put(  0, 60){\line(1, 1){30}}
 \put(  0, 60){\line(1,-1){60}}
 \put( 30, 90){\line(1,-1){60}}
 \put( 30, 30){\line(1, 1){60}}
 \put( 60,  0){\line(1, 1){60}}
 \put( 90, 90){\line(1,-1){30}}
 \put( 60, 60){\circle*{6}}
 \put( 83, 57){\sc $2\la$}
 \put( 24, 57){\sc $u$}
 \put( 44, 27){\sc $u\+2\la$}
 \put(-10, 57){\sc $a$}
 \put( 57,-11){\sc $b$}
 \put( 90, 21){\sc $c$}
 \put( 56, 66){\sc $c^{\scs\prime}$}
 \put( 27, 93){\sc $d$}
 \put( 87, 93){\sc $d$}
 \put(122, 57){\sc $e$}
 \put( 20, 21){\sc $\alpha$}
 \multiput(30,30)( 0.55, 0){9}{\makebox(0.4444,0.6667){\sevrm .}}
 \multiput(30,30)(-0.55, 0){9}{\makebox(0.4444,0.6667){\sevrm .}}
 \multiput(30,30)( 0, 0.55){9}{\makebox(0.4444,0.6667){\sevrm .}}
 \multiput(30,30)( 0,-0.55){9}{\makebox(0.4444,0.6667){\sevrm .}}
\end{picture}
\;\; = \;\;
\begin{picture}(150,70)(-20,40)
 \thicklines
 \put(  0, 30){\line(1, 1){60}}
 \put(  0, 30){\line(1,-1){30}}
 \put( 30,  0){\line(1, 1){60}}
 \put( 30, 60){\line(1,-1){60}}
 \put( 60, 90){\line(1,-1){60}}
 \put( 90,  0){\line(1, 1){30}}
 \put( 60, 30){\circle*{6}}
 \put( 23, 27){\sc $2\la$}
 \put( 84, 27){\sc $u$}
 \put( 44, 57){\sc $u\+2\la$}
 \put( 23, 63){\sc $a$}
 \put( 55, 14){\sc $a^{\scs\prime}$}
 \put( 27,-11){\sc $b$}
 \put( 87,-11){\sc $b$}
 \put(122, 27){\sc $c$}
 \put( 57, 93){\sc $d$}
 \put( 92, 63){\sc $e$}
 \put(-15, 27){\sc $\alpha$}
 \multiput(0,30)( 0.55, 0){9}{\makebox(0.4444,0.6667){\sevrm .}}
 \multiput(0,30)(-0.55, 0){9}{\makebox(0.4444,0.6667){\sevrm .}}
 \multiput(0,30)( 0, 0.55){9}{\makebox(0.4444,0.6667){\sevrm .}}
 \multiput(0,30)( 0,-0.55){9}{\makebox(0.4444,0.6667){\sevrm .}}
\end{picture}}

\def\zeroonefu{
\setlength{\unitlength}{0.009in}
\rule[-12mm]{0mm}{26mm}
\begin{picture}(150,75)(-20,40)
 \thicklines
 \put(  0, 60){\line(1, 1){30}}
 \put(  0, 60){\line(1,-1){60}}
 \put( 30, 90){\line(1,-1){60}}
 \put( 30, 30){\line(1, 1){60}}
 \put( 60,  0){\line(1, 1){60}}
 \put( 90, 90){\line(1,-1){30}}
 \put( 60, 60){\circle*{6}}
 \put( 83, 57){\sc $2\la$}
 \put( 24, 57){\sc $u$}
 \put( 44, 27){\sc $u\+2\la$}
 \put(-10, 57){\sc $a$}
 \put( 57,-11){\sc $b$}
 \put( 90, 21){\sc $c$}
 \put( 56, 66){\sc $c^{\scs\prime}$}
 \put( 27, 93){\sc $d$}
 \put( 87, 93){\sc $d$}
 \put(122, 57){\sc $e$}
 \put( 18, 21){\sc $a^{\scs\prime}$}
\end{picture}}

\def\splitzo{
\rule[-12mm]{0mm}{26mm}
\setlength{\unitlength}{0.009in}
\begin{picture}(150,70)(-20,40)
 \thicklines
 \put(  0, 60){\line(1, 1){30}}
 \put(  0, 60){\line(1,-1){60}}
 \put( 30, 90){\line(1,-1){60}}
 \put( 30, 30){\line(1, 1){60}}
 \put( 60,  0){\line(1, 1){60}}
 \put( 90, 90){\line(1,-1){30}}
 \put( 60, 60){\circle*{6}}
 \put( 83, 57){\sc $2\la$}
 \put( 24, 57){\sc $u$}
 \put( 44, 27){\sc $u\+2\la$}
 \put(-10, 57){\sc $a$}
 \put( 57,-11){\sc $b$}
 \put( 90, 21){\sc $c$}
 \put( 56, 66){\sc $c^{\scs\prime}$}
 \put( 27, 93){\sc $d$}
 \put( 87, 93){\sc $d$}
 \put(122, 57){\sc $e$}
 \put( 18, 21){\sc $a^{\scs\prime}$}
\end{picture}
\;\;\;\; = \;\;\;\;
\dface {c^{\scs\prime}_{\scs 1}}ced{2\la} \;\;\times\;\;
\onebytwofaceasym abcd{a'}{}u{u\!\+\!2\!\la}
}

\def\projzzo{
\rule[-12mm]{0mm}{26mm}
\setlength{\unitlength}{0.009in}
\begin{picture}(150,70)(-20,40)
 \thicklines
 \put(  0, 60){\line(1, 1){30}}
 \put(  0, 60){\line(1,-1){60}}
 \put( 30, 90){\line(1,-1){60}}
 \put( 30, 30){\line(1, 1){60}}
 \put( 60,  0){\line(1, 1){60}}
 \put( 90, 90){\line(1,-1){30}}
 \put( 60, 60){\circle*{6}}
 \put( 83, 57){\sc $2\la$}
 \put( 23, 57){\sc $2\la$}
 \put( 53, 27){\sc $4\la$}
 \put(-10, 57){\sc $d^{\scs\prime}$}
 \put( 57,-11){\sc $c$}
 \put( 90, 21){\sc $c^{\scs\prime\prime}$}
 \put( 57, 66){\sc $e$}
 \put( 27, 93){\sc $d$}
 \put( 87, 93){\sc $d$}
 \put(122, 57){\sc $d^{\scs\prime\prime}$}
 \put( 18, 21){\sc $c^{\scs\prime}$}
\end{picture}
\;\;\; =\;\;\;
\dface {e_{\scs 1}}{c^{\scs\prime\prime}}{d^{\scs\prime\prime}}{d}{2\la}
\;\;\times\;\; \onebytwofaceasym
{d^{\scs\prime}}c{c^{\scs\prime\prime}}d{c^{\scs\prime}}{}{2\la}{4\la}
}

\def\fuzzo{
\lefteqn{\hspace*{-8mm}
\rule[-16mm]{0pt}{32mm}
\setlength{\unitlength}{0.009in}
\begin{picture}(210,85)(-20,55)
 \thicklines
 \put(  0, 90){\line(1, 1){30}}
 \put(  0, 90){\line(1,-1){90}}
 \put( 30, 60){\line(1, 1){60}}
 \put( 30,120){\line(1,-1){90}}
 \put( 60, 30){\line(1, 1){90}}
 \put( 90,  0){\line(1, 1){90}}
 \put( 90,120){\line(1,-1){60}}
 \put(150,120){\line(1,-1){30}}
 \put( 60, 90){\circle*{6}}
 \put( 90, 60){\circle*{6}}
 \put(120, 90){\circle*{6}}
 \put(143, 87){\sc $2\la$}
 \put( 83, 87){\sc $2\la$}
 \put(113, 57){\sc $4\la$}
 \put( 24, 87){\sc $u$}
 \put( 44, 57){\sc $u\+2\la$}
 \put( 74, 27){\sc $u\+4\la$}
 \put(-10, 87){\sc $a$}
 \put( 18, 51){\sc $a^{\scs\prime}$}
 \put( 46, 21){\sc $a^{\scs\prime\prime}$}
 \put( 87,-11){\sc $b$}
 \put(120, 21){\sc $c$}
 \put(150, 51){\sc $c^{\scs\prime}$}
 \put(182, 87){\sc $c^{\scs\prime\prime}$}
 \multiput(27,123)(60,0){3}{\sc $d$}
 \put( 57, 96){\sc $e$}
 \put(117, 96){\sc $f$}
\end{picture}  \;\;\; = \;\;\;\;
\dface{f_{\scs 1}}{c^{\scs\prime}}{c^{\scs\prime\prime}}{d}{2\la}
\;\;\;\times
\begin{picture}(210,85)(-20,55)
 \thicklines
 \put(  0, 90){\line(1, 1){30}}
 \put(  0, 90){\line(1,-1){90}}
 \put( 30, 60){\line(1, 1){60}}
 \put( 30,120){\line(1,-1){90}}
 \put( 60, 30){\line(1, 1){60}}
 \put( 90,  0){\line(1, 1){60}}
 \put( 90,120){\line(1,-1){60}}
 \put( 60, 90){\circle*{6}}
 \put( 90, 60){\circle*{6}}
 \put(120, 90){\circle{6}}
 \put( 83, 87){\sc $2\la$}
 \put(113, 57){\sc $4\la$}
 \put( 24, 87){\sc $u$}
 \put( 44, 57){\sc $u\+2\la$}
 \put( 74, 27){\sc $u\+4\la$}
 \put(-10, 87){\sc $a$}
 \put( 18, 51){\sc $a^{\scs\prime}$}
 \put( 46, 21){\sc $a^{\scs\prime\prime}$}
 \put( 87,-11){\sc $b$}
 \put(120, 21){\sc $c$}
 \put(150, 51){\sc $c^{\scs\prime}$}
 \multiput(27,123)(60,0){2}{\sc $d$}
 \put( 57, 96){\sc $e$}
\end{picture}} \no\\*
& & = \; h(u)\:
\phi_{(0,1)}(a',a'',b)\:\frac{g(d,c')}{g(a',b)}\;
\dface{f_{\scs 1}}{c^{\scs\prime}}{c^{\scs\prime\prime}}{d}{2\la}
\;\times\!\!\!\!
\rule[-12mm]{0mm}{26mm}
\setlength{\unitlength}{0.009in}
\begin{picture}(150,70)(-20,40)
 \thicklines
 \put(  0, 60){\line(1, 1){30}}
 \put(  0, 60){\line(1,-1){60}}
 \put( 30, 90){\line(1,-1){60}}
 \put( 30, 30){\line(1, 1){60}}
 \put( 60,  0){\line(1, 1){60}}
 \put( 90, 90){\line(1,-1){30}}
 \put( 60, 60){\circle*{6}}
 \put( 24, 57){\sc $u$}
 \put( 44, 27){\sc $u\+3\la$}
 \put( 83, 57){\sc $3\la$}
 \put(-10, 57){\sc $a$}
 \put( 18, 21){\sc $a^{\scs\prime}$}
 \put( 57,-11){\sc $b$}
 \put( 90, 21){\sc $c$}
 \put(122, 57){\sc $c^{\scs\prime}$}
 \multiput(27,93)(60,0){2}{\sc $d$}
 \put( 57, 66){\sc $e$}
\end{picture}}

\def\proj{
\setlength{\unitlength}{0.0105in}
\rule[-25mm]{0mm}{55mm}
\begin{picture}(316,120)(-13,80)
 \thicklines
 \put(  0, 30){\line( 1, 1){150}}
 \put(  0, 30){\line( 1,-1){ 30}}
 \put( 30,  0){\line( 1, 1){150}}
 \put( 30, 60){\line( 1,-1){ 60}}
 \put( 60, 90){\line( 1,-1){ 90}}
 \put( 90,  0){\line( 1, 1){120}}
 \put( 90,120){\line( 1,-1){120}}
 \put(120,150){\line( 1,-1){150}}
 \put(150,  0){\line( 1, 1){ 90}}
 \put(150,180){\line( 1,-1){150}}
 \put(210,  0){\line( 1, 1){ 60}}
 \put(270,  0){\line( 1, 1){ 30}}
 \multiput( 60, 30)(60,0){4}{\circle*{5}}
 \multiput( 90, 60)(60,0){3}{\circle*{5}}
 \multiput(120, 90)(60,0){2}{\circle*{5}}
 \multiput(150,120)(60,0){1}{\circle*{5}}
 \put(147,183){\sc $a$}
 \put(180,152){\sc $a_{\scs 1}$}
 \put(210,122){\sc $a_{\scs 2}$}
 \put(270, 62){\sc $a_{\scs n\-1}$}
 \put(303, 27){\sc $a_{\scs n}$}
 \multiput( 27,-10)(60,0){5}{\sc $b$}
 \put(109,152){\sc $b_{\scs 1}$}
 \put( 79,122){\sc $b_{\scs 2}$}
 \put( 10, 62){\sc $b_{\scs n\-1}$}
 \put(-13, 27){\sc $b_{\scs n}$}
 \put(148,147){\sc $u$}
 \put(166,117){\sc $u\+2\la$}
 \put(220, 57){\tiny $u\+2(\!n\!\-\!2\!)\la$}
 \put(250, 27){\tiny $u\+2(\!n\!\-\!1\!)\la$}
 \multiput( 17, 27)(60,0){4}{\sc $-2\la$}
 \multiput( 47, 57)(60,0){3}{\sc $-4\la$}
 \put( 98,117){\sc $-\!2(\!n\!\-\!1\!)\la$}
\end{picture}}

\def\rowprojn{
\setlength{\unitlength}{0.0105in}
\rule[-25mm]{0mm}{55mm}
\begin{picture}(316,120)(-13,80)
 \thicklines
 \put(  0, 30){\line( 1, 1){150}}
 \put(  0, 30){\line( 1,-1){ 30}}
 \put( 30,  0){\line( 1, 1){150}}
 \put( 30, 60){\line( 1,-1){ 60}}
 \put( 60, 90){\line( 1,-1){ 90}}
 \put( 90,  0){\line( 1, 1){120}}
 \put( 90,120){\line( 1,-1){120}}
 \put(120,150){\line( 1,-1){150}}
 \put(150,  0){\line( 1, 1){ 90}}
 \put(150,180){\line( 1,-1){150}}
 \put(210,  0){\line( 1, 1){ 60}}
 \put(270,  0){\line( 1, 1){ 30}}
 \multiput( 60, 30)(60,0){4}{\circle*{5}}
 \multiput( 90, 60)(60,0){3}{\circle*{5}}
 \multiput(120, 90)(60,0){2}{\circle*{5}}
 \multiput(150,120)(60,0){1}{\circle*{5}}
 \put(147,183){\sc $a$}
 \put(180,152){\sc $a_{\scs 1}$}
 \put(210,122){\sc $a_{\scs 2}$}
 \put(270, 62){\sc $a_{\scs n\-1}$}
 \put(303, 27){\sc $a_{\scs n}$}
 \multiput( 27,-10)(60,0){5}{\sc $b$}
 \put(109,152){\sc $b_{\scs 1}$}
 \put( 79,122){\sc $b_{\scs 2}$}
 \put( 10, 62){\sc $b_{\scs n\-1}$}
 \put(-13, 27){\sc $b_{\scs n}$}
 \multiput( 17, 27)(60,0){5}{\sc $-2\la$}
 \multiput( 47, 57)(60,0){4}{\sc $-4\la$}
 \multiput( 98,117)(60,0){2}{\sc $-\!2(\!n\!\-\!1\!)\la$}
 \put(136,147){\sc $-2n\la$}
\end{picture}}

\def\columnproj{
\setlength{\unitlength}{0.0105in}
\rule[-25mm]{0mm}{55mm}
\begin{picture}(316,120)(-13,80)
 \thicklines
 \put(  0,150){\line( 1,-1){150}}
 \put(  0,150){\line( 1, 1){ 30}}
 \put( 30,180){\line( 1,-1){150}}
 \put( 30,120){\line( 1, 1){ 60}}
 \put( 60, 90){\line( 1, 1){ 90}}
 \put( 90,180){\line( 1,-1){120}}
 \put( 90, 60){\line( 1, 1){120}}
 \put(120, 30){\line( 1, 1){150}}
 \put(150,180){\line( 1,-1){ 90}}
 \put(150,  0){\line( 1, 1){150}}
 \put(210,180){\line( 1,-1){ 60}}
 \put(270,180){\line( 1,-1){ 30}}
 \multiput( 60,150)(60,0){4}{\circle*{5}}
 \multiput( 90,120)(60,0){3}{\circle*{5}}
 \multiput(120, 90)(60,0){2}{\circle*{5}}
 \multiput(150, 60)(60,0){1}{\circle*{5}}
 \multiput( 27,183)(60,0){5}{\sc $a$}
 \put(303,146){\sc $a_{\scs 1}$}
 \put(270,111){\sc $a_{\scs 2}$}
 \put(210, 51){\sc $a_{\scs n\-1}$}
 \put(180, 21){\sc $a_{\scs n}$}
 \put(147,-10){\sc $b$}
 \put(-13,147){\sc $b_{\scs 1}$}
 \put( 19,111){\sc $b_{\scs 2}$}
 \put( 70, 51){\sc $b_{\scs n\-1}$}
 \put(109, 21){\sc $b_{\scs n}$}
 \put(147, 27){\sc $u$}
 \put(166, 57){\sc $u\+2\la$}
 \put(219,117){\tiny $u\+2(\!n\!\-\!2\!)\la$}
 \put(249,147){\tiny $u\+2(\!n\!\-\!1\!)\la$}
 \multiput( 17,147)(60,0){4}{\sc $-2\la$}
 \multiput( 47,117)(60,0){3}{\sc $-4\la$}
 \put( 98, 57){\sc $-\!2(\!n\!\-\!1\!)\la$}
\end{picture}}

\def\columnprojn{
\setlength{\unitlength}{0.0105in}
\rule[-25mm]{0mm}{55mm}
\begin{picture}(316,120)(-13,80)
 \thicklines
 \put(  0,150){\line( 1,-1){150}}
 \put(  0,150){\line( 1, 1){ 30}}
 \put( 30,180){\line( 1,-1){150}}
 \put( 30,120){\line( 1, 1){ 60}}
 \put( 60, 90){\line( 1, 1){ 90}}
 \put( 90,180){\line( 1,-1){120}}
 \put( 90, 60){\line( 1, 1){120}}
 \put(120, 30){\line( 1, 1){150}}
 \put(150,180){\line( 1,-1){ 90}}
 \put(150,  0){\line( 1, 1){150}}
 \put(210,180){\line( 1,-1){ 60}}
 \put(270,180){\line( 1,-1){ 30}}
 \multiput( 60,150)(60,0){4}{\circle*{5}}
 \multiput( 90,120)(60,0){3}{\circle*{5}}
 \multiput(120, 90)(60,0){2}{\circle*{5}}
 \multiput(150, 60)(60,0){1}{\circle*{5}}
 \multiput( 27,183)(60,0){5}{\sc $a$}
 \put(303,146){\sc $a_{\scs 1}$}
 \put(270,111){\sc $a_{\scs 2}$}
 \put(210, 51){\sc $a_{\scs n\-1}$}
 \put(180, 21){\sc $a_{\scs n}$}
 \put(147,-10){\sc $b$}
 \put(-13,147){\sc $b_{\scs 1}$}
 \put( 19,111){\sc $b_{\scs 2}$}
 \put( 70, 51){\sc $b_{\scs n\-1}$}
 \put(109, 21){\sc $b_{\scs n}$}
 \multiput( 17,147)(60,0){5}{\sc $-2\la$}
 \multiput( 47,117)(60,0){4}{\sc $-4\la$}
 \multiput( 98, 57)(60,0){2}{\sc $-\!2(\!n\!\-\!1\!)\la$}
 \put(136, 27){\sc $-2n\la$}
\end{picture}}

\def\mnface{
\setlength{\unitlength}{0.0125in}
\rule[-8mm]{0pt}{18mm}
\begin{picture}(70,30)(-10,22)
 \thicklines
 \multiput(5,5)(0,40){2}{\line(1,0){40}}
 \multiput(5,5)(40,0){2}{\line(0,1){40}}
 \put( 0, 1){\sc $a$}
 \put(24,-1){\sc $\alpha$}
 \put(46, 1){\sc $b$}
 \put(47,24){\sc $\nu$}
 \put(46,46){\sc $c$}
 \put(24,48){\sc $\beta$}
 \put( 0,46){\sc $d$}
 \put(-2,24){\sc $\mu$}
 \put(23,23){\sc $u$}
\end{picture}}

\def\mnfusion{
\setlength{\unitlength}{0.0125in}
\begin{picture}(345,150)(0,140)
\thicklines
\multiput(10,10)(0,60){3}{\line(1,0){60}}
\multiput(10,220)(0,60){2}{\line(1,0){60}}
\multiput(220,10)(0,60){3}{\line(1,0){120}}
\multiput(220,220)(0,60){2}{\line(1,0){120}}
\multiput(10,10)(60,0){2}{\line(0,1){120}}
\multiput(10,220)(60,0){2}{\line(0,1){60}}
\multiput(220,10)(60,0){3}{\line(0,1){120}}
\multiput(220,220)(60,0){3}{\line(0,1){60}}
\multiput(68.125, 10)(0,60){3}{\multiput(0,0)(7.5,0){21}{\line(1,0){3.75}}}
\multiput(68.125,220)(0,60){2}{\multiput(0,0)(7.5,0){21}{\line(1,0){3.75}}}
\multiput( 10,128.125)(60,0){2}{\multiput(0,0)(0,7.5){16}{\line(0,1){3.75}}}
\multiput(220,128.125)(60,0){3}{\multiput(0,0)(0,7.5){16}{\line(0,1){3.75}}}
\multiput(70,70)(0,60){2}{\circle*{6}}
\put(70,220){\circle*{6}}
\multiput(220,70)(60,0){2}{\multiput(0,0)(0,60){2}{\circle*{6}}}
\multiput(280,2200)(0,60){2}{\circle*{6}}
\put(  0,  0){$a$}
\put( -5, 67){$e_2^j$}
\put( -5,217){$e_m^j$}
\put(  0,283){$d$}
\put( 65, -2){$f_2^i$}
\put(215, -2){$f_{n-1}^i$}
\put(275, -2){$f_n^i$}
\put(341, -2){$b$}
\put( 65,285){$f_2^\mu$}
\put(215,285){$f_{n-1}^\mu$}
\put(275,285){$f_n^\mu$}
\put(341,283){$c$}
\put(343, 67){$e_2^\nu$}
\put(343,127){$e_3^\nu$}
\put(343,217){$e_m^\nu$}
\put( 36,247){$u$}
\put(225,247){$\scriptstyle u+2(n-2)\lambda$}
\put(285,247){$\scriptstyle u+2(n-1)\lambda$}
\put( 15, 97){$\scriptstyle u+2(2-m)\lambda$}
\put(225, 97){$\scriptstyle u+2(n-m)\lambda$}
\put(282, 97){$\scriptscriptstyle u+2(n-m+1)\lambda$}
\put( 15, 37){$\scriptstyle u+2(1-m)\lambda$}
\put(222, 37){$\scriptscriptstyle u+2(n-m-1)\lambda$}
\put(285, 37){$\scriptstyle u+2(n-m)\lambda$}
\end{picture}
}

\def\Young{
\setlength{\unitlength}{0.0125in}
\begin{picture}(510,120)(0,0)
\put(0,0){\line(1,0){300}}
\multiput(0,60)(0,60){2}{\line(1,0){510}}
\multiput(0,0)(60,0){3}{\line(0,1){120}}
\multiput(240,0)(60,0){2}{\line(0,1){120}}
\put(360,60){\line(0,1){60}}
\multiput(450,60)(60,0){2}{\line(0,1){60}}
\multiput(170,24)(0,60){2}{\multiput(0,0)(8,0){5}{.}}
\multiput(393,74)(8,0){3}{.}
\put(4,27){$\scriptstyle u+2(m+n)\lambda$}
\put(62,27){$\scriptscriptstyle u+2(m+n-1)\lambda$}
\put(245,27){$\scriptstyle u+2(n+1)\lambda$}
\put(2,87){$\scriptscriptstyle u+2(m+n-1)\lambda$}
\put(62,87){$\scriptscriptstyle u+2(m+n-2)\lambda$}
\put(252,87){$\scriptstyle u+2n\lambda$}
\put(305,87){$\scriptstyle u+2(n-1)\lambda$}
\put(472,87){$u$}
\end{picture}
}

\def\mndim#1#2#3#4#5{\setlength{\unitlength}{0.0115in}
\rule[-11mm]{0mm}{24mm}
\begin{picture}(42,45)(21,46)
\thicklines
\put(35,80){\line(-1,-2){15}}
\put(20,50){\line( 1,-2){15}}
\put(35,20){\line( 1, 2){15}}
\put(50,50){\line(-1, 2){15}}
\put(45,65){\sc $#1$}
\put(23,47){\sc $(\!n,\!m\!)$}
\put(45,28){\sc $#2$}
\put(18,28){\sc $\rho$}
\put(18,65){\sc $\gamma$}
\put(53,47){\sc $#5$}
\put(11,47){\sc $d$}
\put(32,83){\sc $#3$}
\put(32,10){\sc $#4$}
\end{picture}}

\def\nmfusion{
\setlength{\unitlength}{0.0125in}
\begin{picture}(210,111)(0,0)
\thicklines
\put( 20, 60){\line( 0, 1){45}}
\put( 20, 60){\line( 1, 0){210}}
\put(110, 60){\line( 0, 1){45}}
\put( 20,105){\line( 1, 0){210}}
\put(230, 60){\line( 0, 1){45}}
\put(110, 60){\line(-3,-1){90}}
\put(110, 60){\line( 4,-1){120}}
\put( 20, 30){\line( 4,-1){120}}
\put(230, 30){\line(-3,-1){90}}
\put(110, 60){\circle*{6}}
\put( 47, 80){\sc $W_{(n,0)}(u)$}
\put(123, 80){\sc $W_{(m,0)}(u+(2n+1)\la)$}
\put( 80, 26){\sc $W^{(m,0)}_{(n,0)}(-(2n+1)\la)$}
\put( 10, 54){$a$}
\put( 10,105){$d$}
\put(233,105){$c$}
\put(233, 54){$b$}
\put( 11, 27){$a$}
\put(233, 27){$b$}
\put( 65, 39){\sc $\alpha$}
\put( 60, 53){\sc $\alpha$}
\put(153, 40){\sc $\beta$}
\put(170, 53){\sc $\beta$}
\put(107, 48){\sc $c'$}
\end{picture}}

\begin{document}
\vspace{0.5cm}
\begin{center}
\title{Fusion of Dilute $A_L$ Lattice Models}

\author{Yu-kui Zhou\footnote{Email: zhouy@maths.anu.edu.au}$^,$\footnote{
       On leave of absence from {\sl Institute of Modern Physics,
                      Northwest University, Xian 710069, China}}}

\address{Mathematics Department, The Australian National University,\\
          Canberra, ACT 0200, Australia   }

\author{Paul A.\ Pearce\footnote{Email: pap@maths.mu.oz.au}}

\address{Mathematics Department, University of Melbourne,\\
          Parkville, Victoria 3052, Australia  }

\author{Uwe Grimm\footnote{Email: grimm@phys.uva.nl}}

\address{Instituut voor Theoretische Fysica, Universiteit van Amsterdam,\\
         Valckenierstraat 65, 1018 XE Amsterdam, The Netherlands}
\end{center}

\begin{abstract}
The fusion procedure is implemented for the dilute $A_L$ lattice
models and a fusion hierarchy of functional equations with
an $su(3)$ structure is derived for the fused transfer matrices.
We also present the Bethe ansatz equations for the dilute $A_L$ lattice
models and discuss their connection with the fusion hierarchy.
The solution of the fusion hierarchy for the eigenvalue spectra
of the dilute $A_L$ lattice models will be presented in a subsequent
paper.
\end{abstract}

\bigskip
hepth/9506108
\bigskip

\section{Introduction}
\setcounter{equation}{0}

The dilute \ade lattice models \cite{WNS:92,Roche:92} are exactly solvable
\cite{Baxter:82} restricted solid-on-solid (RSOS) models on the square
lattice. These
models resemble the \ade lattice models of Pasquier \cite{Pasquier:87} in
that the spins or heights take their values on a Dynkin diagram of a
classical \ade Lie algebra. The properties of these two families of \ade
models, however, are quite distinct and recent studies have shown that the
dilute \ade models exhibit some new and very interesting aspects.

First and foremost, in contrast to the $A_L$ model of Andrews, Baxter and
Forrester \cite{ABF:84}, the dilute $A_L$ models, with $L$ odd, can be
solved \cite{WPSN:94} off-criticality in the presence of a
symmetry-breaking field. In particular, in an appropriate regime, the
dilute $A_3$ model lies in the universality class of the Ising model in a
magnetic field and gives the magnetic exponent $\delta=15$
\cite{WPSN:94}. In addition, Zamolodchikov \cite{Zam:89} has argued that
the magnetic Ising model in the scaling region is described by an
$E_8$ scattering
theory. Accordingly, related $E_8$ structures have recently been uncovered
\cite{BNW:94, WaPe:95} in the dilute $A_3$ model. Lastly, the dilute \ade
lattice models give \cite{OBPe:95} lattice realizations of the complete
unitary minimal series of conformal field theories \cite{CIZ:87}. This
again is not the case for the \ade models of Pasquier.

An important step in the study of the $A_L$ models of Andrews, Baxter and
Forrester (ABF) was the fusion \cite{DJMO:86} of the elementary weights to
form new solutions of the Yang-Baxter equations. Subsequently, it was shown
\cite{BaRe:89} that the transfer matrices of the fused ABF models satisfy
special functional relations which can be solved for the eigenvalue
spectra of these models. Moreover, at criticality, these equations can be
solved for the central charges \cite{BaRe:89} and conformal weights
\cite{KlPe:92}.

Following the developments for the $A_L$ models of Andrews, Baxter and
Forrester, we carry out in this paper the fusion procedure for the dilute
$A_L$ lattice models and derive a fusion hierarchy of functional
relations satisfied by the fused transfer matrices. The solution of this
fusion hierarchy for the eigenvalue spectra, central charges and conformal
weights will be given in a subsequent paper \cite{ZhPe:95}.

Historically, the fusion procedure was first introduced in \cite{KRS:81}.
It has since been successfully applied
\cite{Cherednik:82,DJMO:86,JKMO:88b,ZhHo:89,ZhPe:94,Zhou:94} to many solvable
models in two-dimensional statistical mechanics.

The layout of this paper is as follows. In section~1.1 we define
the dilute lattice models. In section~1.2 we present the Bethe ansatz
for the commuting transfer matrices. Then in section~1.3 we discuss
the fusion hierarchy and its connection to the Bethe ansatz.
In section~2 we construct the fused face weights for elementary fusion.
In section~3 we give in detail the procedure for constructing
the completely symmetrically, $(n,0)$ and antisymmetrically $(0,n)$
fused face weights. This is generalized in section~4 to construct the
fused face weights for arbitrary fusion of mixed type $(n,m)$.
Finally, in section~5, we present the fusion hierarchy of functional
relations and Bethe ansatz for the general fusion of mixed type $(n,m)$.
We summarize our results in section~6.

\subsection{Dilute \mbox{\protect\boldmath $A_L$} Lattice Models}

\begin{figure}[t]
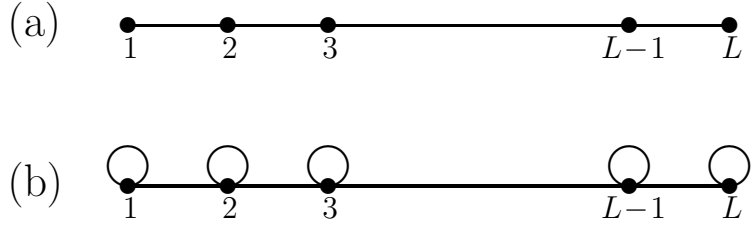

\begin{center}
\dynkin2
\caption{\label{fig1} The $A_L$ Dynkin diagram (a) and the effective
         adjacency diagram of the dilute $A_L$ models (b).}
\end{center}
\end{figure}

The dilute $A_L$ lattice models \cite{WNS:92}
are restricted solid--on--solid (RSOS) models with $L$ heights
built on the $A_L$ Dynkin diagram as shown in Fig~\ref{fig1}(a).
The elements $A_{a,b}$ of the adjacency matrix for this diagram are given by
\begin{equation}
A_{a,b} \; = \; A_{b,a} \; = \; \left\{ \ba{ll}
 1,\qquad & \mbox{$|a-b|=1$ } \\
 0,\qquad & \mbox{otherwise.} \ea  \right.
\label{ABFadj}
\end{equation}
The face weights of the  dilute $A_L$ models are
\addtolength{\jot}{2mm}
\be
\hs{-0.2}\wt Waaaau & = &
\frac{\tha{6\la-u}\tha{3\la+u}}{\tha{6\la}\tha{3\la}} \no \\
\hs{-0.2}& &\hs{-0.5}-\left(
 \frac{S_{a+1}}{S_a}\frac{\thd{2a\la-5\la}}{\thd{2a\la+\la}}
 +\frac{S_{a-1}}{S_a}\frac{\thd{2a\la+5\la}}{\thd{2a\la-\la}}\right)
\frac{\tha{u}\tha{3\la-u}}{\tha{6\la}\tha{3\la}} \no \\
\hs{-0.2}\wt Waaa{a\pm 1}u & = & \wt Wa{a\pm 1}aau \;\; = \;\;
\frac{\tha{3\la-u}\thd{\pm 2a\la+\la-u}}{\tha{3\la}\thd{\pm 2a\la+\la}}\no \\
\hs{-0.2}\wt W{a\pm 1}aaau & = & \wt Waa{a\pm 1}au \;\; =\;\;
\left(\frac{S_{a\pm 1}}{S_a}\right)^{1/2}
\frac{\tha{u}\thd{\pm 2a\la-2\la+u}}{\tha{3\la}\thd{\pm 2a\la+\la}} \no \\
\hs{-0.2}\wt Wa{a\pm 1}{a\pm 1}au & = & \wt Waa{a\pm 1}{a\pm 1}u \no \\
\hs{-0.2}& = & \left(\frac{\thd{\pm 2a\la+3\la}\thd{\pm 2a\la-\la}}
           {\vartheta_4^2(\pm 2a\la+\la)}\right)^{1/2}
\frac{\tha{u}\tha{3\la-u}}{\tha{2\la}\tha{3\la}} \label{weights} \\
\hs{-0.2}\wt Wa{a\mp 1}a{a\pm 1}u & = &
\frac{\tha{2\la-u}\tha{3\la-u}}{\tha{2\la}\tha{3\la}} \no \\
\hs{-0.2}\wt W{a\pm 1}a{a\mp 1}au & = &
-\left(\frac{S_{a-1}S_{a+1}}{S^2_a}\right)^{1/2}
\frac{\tha{u}\tha{\la-u}}{\tha{2\la}\tha{3\la}} \no \\
\hs{-0.2}\wt W{a\pm 1}a{a\pm 1}au & = &
\frac{\tha{3\la-u}\tha{\pm 4a\la+2\la+u}}{\tha{3\la}\tha{\pm 4a\la+2\la}}
+\frac{S_{a\pm 1}}{S_a}
\frac{\tha{u}\tha{\pm 4a\la-\la+u}}{\tha{3\la}\tha{\pm 4a\la+2\la}} \no \\
\hs{-0.2}& = & \frac{\tha{3\la+u}\tha{\pm 4a\la-4\la+u}}
{\tha{3\la}\tha{\pm 4a\la-4\la}} \no \\
& & \hs{-1}+\left(\frac{S_{a\mp 1}}{S_a}\frac{\tha{4\la}}{\tha{2\la}}
-\frac{\thd{\pm 2a\la-5\la}}{\thd{\pm 2a\la+\la}} \right)
\frac{\tha{u}\tha{\pm 4a\la-\la+u}}{\tha{3\la}\tha{\pm 4a\la-4\la}}\no
\ee
\addtolength{\jot}{-2mm}
Here the crossing factors are
\be
S_a & = & (-1)^{\displaystyle a} \;\frac{\tha{4a\la}}{\thd{2a\la}}
\ee
and $\tha{u}$, $\thd{u}$ are standard theta functions of nome $p$ with $|p|<1$.
Note that the effective adjacency matrix in the face weights
is $I+A$, see Fig~\ref{fig1}(b). We denote by $\val a$ the number
of allowed neighbours of height $a$,
\be
\val a & = & \sum_{b} \; \left(I+A\right)_{a,b} .
\label{valdef}
\ee
Obviously, we have \mbox{$\val a \leq 3$} for the dilute $A_L$ models.
The weights (\ref{weights}) are normalized such that
\be
\wt Wabcd0 & = & (I+A)_{a,b}\: (I+A)_{a,d}\: \delta_{a,c} \; ,
\label{norm}
\ee
are invariant under reflections along the diagonals
\be
\wt Wabcdu & = & \wt Wcbadu\;\; =\;\;\wt Wadcbu\;\; =\;\;\wt Wcdabu
\label{refl}
\ee
and satisfy the crossing symmetry
\be
\wt Wabcd{3\la-u} & = &
{\left(\frac{S_{a}S_{c}}{S_{b}S_{d}}\right)}^{1/2}\; \wt Wdabcu
\;\; .
\label{cross}
\ee

The dilute models admit four different physical branches.
The spectral parameter $u$ and the crossing parameter $\la$
in the four branches take values
\renewcommand{\arraystretch}{2}
\be
\ba{lr@{\;<u<\;}ll}
\mbox{branch 1:} & 0 & 3\la &
\disp\la=\frac{\pi}{4}\frac{L}{L+1} \\
\mbox{branch 2:} & 0 & 3\la &
\disp\la=\frac{\pi}{4}\frac{L+2}{L+1} \\
\mbox{branch 3:} & -\pi+3\la & 0 &
\disp\la=\frac{\pi}{4}\frac{L+2}{L+1} \\
\mbox{branch 4:} & \qquad -\pi+3\la & 0 \qquad  &
\disp\la=\frac{\pi}{4}\frac{L}{L+1} \ea
\ee
\renewcommand{\arraystretch}{1}

At criticality, the face weights simplify to \cite{WNS:92,Roche:92}
\be
& & \wt Wabcdu =\face abcd{u}
= \;\rho_1(u)
  \delta_{a,b,c,d} +\rho_2(u) \delta_{a,b,c} A_{a,d}+\rho_3(u)
\delta_{a,c,d} A_{a,b} \no\\*
& &\quad \mbox{} +\sqrt{S_a \over S_b}\rho_4 (u) \delta_{b,c,d} A_{a,b}
     +\sqrt{S_c \over S_a}\rho_5(u) \delta_{a,b,d} A_{a,c}
     +\rho_6(u) \delta_{a,b} \delta_{c,d} A_{a,c}  \label{cweights}\\*
& &\quad \mbox{}  +\rho_7(u) \delta_{a,d} \delta_{c,b} A_{a,b} +\rho_8(u)
\delta_{a,c} A_{a,b} A_{a,d} +
       \sqrt{{S_a S_c \over S_b S_d}} \rho_9(u) \delta_{b,d} A_{a,b}
A_{b,c} \no
\ee
with
\be
\rho_1 (u)&=&{\sin 2\la \sin 3\la +\sin u \sin (3\la-u)
                        \over \sin 2\la \sin 3\la}  \no\\
\rho_2 (u)&=&\rho_3 (u)\;\;=\;\;{\sin (3\la-u)\over \sin 3\la}\no \\
\rho_4 (u)&=&\rho_5 (u)\;\;=\;\;{\sin u \over \sin 3\la} \no \\
\rho_6 (u)&=&\rho_7 (u)\;\;=\;\;{\sin u \sin (3\la-u) \over \sin
                                     2\la \sin 3\la}  \\
\rho_8 (u)&=&{\sin (2\la-u) \sin (3\la-u)\over\sin
                                     2\la\sin 3\la} \no \\
\rho_9 (u)&=&-{\sin u \sin(\la-u) \over \sin 2\la \sin 3\la}. \no
\ee
Moreover, the crossing factors reduce to the
nonnegative elements of the Perron-Frobenius eigenvector of the
adjacency matrix given by
\be
\sum_b A_{a,b}S_b = 2\cos\left({\pi\over L\+1}\right)\; S_a \;.
\ee

\subsection{Commuting Transfer Matrices and Bethe Ansatz Equations}
\label{in}

The dilute $A_L$ models are exactly solvable because their face weights
satisfy the Yang-Baxter equations
\be
\lefteqn{\sum_{g=1}^{L} \wt Wabgfu \wt Wfgdev \wt Wgbcd{v-u}} \no\\
& = & \sum_{g=1}^{L} \wt Wfage{v-u} \wt Wabcgv \wt Wgcdeu
\ee
Diagrammatically, this equation is represented as follows
\be
 \YBR abcdefv{v-u}u \label{YBR}
\ee
where the solid circle denotes summation over heights.
This implies that the row transfer matrices $\T(u)$ commute.
Here the elements of $\T(u)$ are given by
\be
\langle \sigma | \T(u) | \sigma' \rangle =
\prod_{j=1}^{N} \wt W{\sigma_j}{\sigma_{j+1}}{\sigma_{j+1}'}{\sigma_j'}u
\ee
where the paths $\sigma=\{\sigma_1,\sigma_2,\ldots,\sigma_N\}$ and
$\sigma'=\{\sigma_1',\sigma_2',\ldots,\sigma_N'\}$
are allowed configurations of heights along a row with periodic
boundary conditions $\sigma_{N+1}=\sigma_1$ and $\sigma_{N+1}'=\sigma_1'$.

The eigenvalues $T(u)$ of the row transfer matrices $\T(u)$ can be
calculated using a Bethe ansatz. Explicitly, the eigenvalues are given
by \cite{BNW:94}
\be
T(u) & = & \omega\,\frac{s(u-2\la)s(u-3\la)}{s(2\la)s(3\la)}
           \frac{Q(u+\la)}{Q(u-\la)}
     +     \frac{s(u)s(3\la -u)}{s(2\la)s(3\la)}
           \frac{Q(u)Q(u-3\la)}{Q(u-\la)Q(u-2\la)} \no\\*
     &   & \qquad + \omega^{-1}\,
           \frac{s(u)s(u-\la)}{s(2\la)s(3\la)}
           \frac{Q(u-4\la)}{Q(u-2\la)} \label{BAE}
\ee
where
\be
s(u)=\tha{u}^{N}, \qquad
Q(u)=\prod_{j=1}^{N} \tha{u-u_j}
\ee
and the zeros $\{u_j\}$ satisfy the Bethe ansatz equations
\be
\omega^{-1} \left[\frac{\tha{u_j+\la}}{\tha{u_j-\la}}\right]^N & = &
- \prod_{k=1}^{N} \frac{\tha{u_j-u_k+2\la}\tha{u_j-u_k-\la}}
                       {\tha{u_j-u_k-2\la}\tha{u_j-u_k+\la}}
\ee
with $j=1,\ldots,N$ and $\omega=\exp(i\pi\ell/(L+1))$, $\ell=1,\ldots,L$.
At criticality, these equations reduce, apart from the phase factors,
to the Bethe ansatz equations of the Izergin-Korepin
model \cite{IzKor81,VichResh83}. The Bethe ansatz equations ensure that
the eigenvalues $T(u)$ are entire functions of $u$.

\subsection{Fusion Hierarchy}
\label{FH}

\begin{figure}[t]
\vspace*{-10mm}
\begin{center}
\leavevmode\epsfxsize=120mm\epsffile{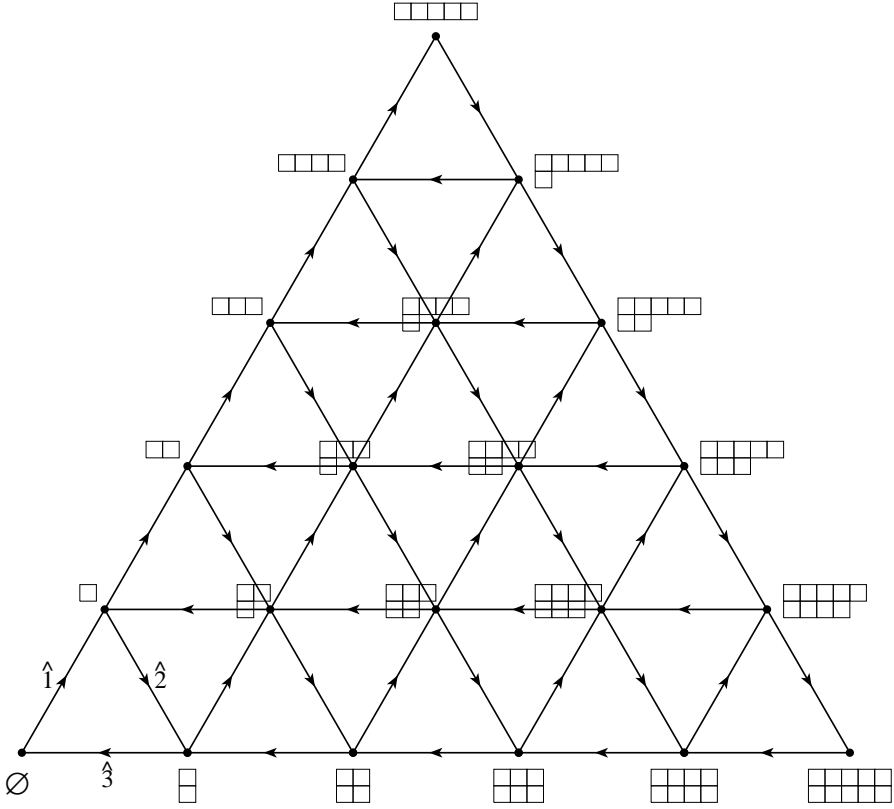}
\caption{\label{fig2}
         The $su(3)$ weight lattice at level $l=5$. Each dot
         corresponds to a member of the fusion hierarchy of the
         dilute $A_3$ model labeled by the Young diagram of the
         respective representation of $su(3)$.}
\end{center}
\end{figure}

Before discussing the fusion hierarchy, we recall some basic facts
concerning $su(3)$. Let $(n,m)$, where $n$ and $m$ are nonnegative integers,
denote the highest weight irreducible representations of $su(3)$.
Then the decomposition of the basic tensor product representations into
irreducible representations is given by
\be
(n,m)\otimes (1,0) & = & (n+1,m) \oplus (n-1,m+1) \oplus (n,m-1) \no\\*
(n,m)\otimes (0,1) & = & (n,m+1) \oplus (n+1,m-1) \oplus (n-1,m) .
\ee
The irreducible representations can be represented by Young tableaux
\be
(n,m) & = & \genbox{n}{m}{}
\ee
so that, for example,
\be
\lefteqn{\genbox{n}{m}{} \;\;\otimes\;\;\sqbox{0}{-7}{}{} \;\; = \;\;
\genbox{\!\!\!\!n\+1}{m}{} } \no\\*
& & \oplus \;\; \genbox{\!\!\!\!n\-1}{\!\!\!\!\!m\+1}{} \;\;\oplus
\sqbox{20}{3}{}{}\sqqbox{0}{-37}{}{}{}\!\genbox{n}{\!\!\!\!\!m\-1}{}
\ee
where the triple box in a column corresponds to the trivial representation
and can be omitted. The relation between these representations
is encapsulated in the $su(3)$ weight lattice, which is shown
in Fig~\ref{fig2} for the case of level $l=m+n=5$.

The Bethe ansatz equations (\ref{BAE}) can be thought of as matrix equations
in $\T(u)$ and an auxiliary matrix family $\Q(u)$ which commutes with
$\T(u)$. These matrix equations imply the fusion hierarchy
\be
\T^{(n,0)}_0\T^{(1,0)}_n  & = &\T^{(n-1,1)}_0+\T^{(n+1,0)}_0  \label{fh1}\\
\T^{(0,m)}_1\T^{(0,1)}_0  & = &\T^{(0,m+1)}_0+f_0 \T^{(1,m-1)}_1
\label{fh2}\\
\T^{(n,0)}_0\T^{(0,m)}_{n}& = &\T^{(n,m)}_{0}
  +f_{n-1}\T^{(n-1,0)}_0 \T^{(0,m-1)}_{n+1} \label{fh3}
\ee
where $\T^{(n,m)}_k=\T^{(n,m)}(u+2k\la)$ is the row transfer matrix of the
fused model of fusion type $(1,0)\times(n,m)$ in the horizontal and
vertical directions. Here we have suppressed the horizontal fusion level
and in this case $\T^{(0,0)}_k=\one$,
$\T^{(1,0)}_k=\T(u+2k\la)$ and $f_k=f(u+2k\la)$ with
\be
f(u) = (-1)^N
\frac{s(u-3\la)s(u-2\la)s(u-\la)s(u+2\la)s(u+3\la)s(u+4\la)}
     {s(2\la)^3 s(3\la)^3}.
\ee
We will later show that the fusion equations (\ref{fh1})--(\ref{fh3}) hold
for arbitrary fusion of type $(n',m')\times(n,m)$, again with the
horizontal fusion type suppressed. In this general case, the fused face
weights involve a rectangular block of $(n'+2m')\times(n+2m)$ elementary
faces. For the moment, however, we only consider $(n',m')=(1,0)$ for
simplicity.

To derive the fusion hierarchy, we use semi-standard Young tableaux
\cite{KiRe:87,BaRe:89,KuSu:94} and set
\be
\sqbox{0}{-7}{1}{k}
          & = &  \omega\,\frac{s(u+2k\la-2\la)s(u+2k\la-3\la)}{s(2\la)s(3\la)}
           \frac{Q(u+2k\la+\la)}{Q(u+2k\la-\la)} \no\\*
\sqbox{0}{-7}{2}{k}
          & = &  \frac{s(u+2k\la)s(u+2k\la-3\la)}{s(2\la)s(3\la)}
           \frac{Q(u+2k\la)Q(u+2k\la-3\la)}{Q(u+2k\la-\la)Q(u+2k\la-2\la)}
          \label{boxes}\\*
\sqbox{0}{-7}{3}{k}
          & = & \omega^{-1}\,
           \frac{s(u+2k\la)s(u+2k\la-\la)}{s(2\la)s(3\la)}
           \frac{Q(u+2k\la-4\la)}{Q(u+2k\la-2\la)} \no
\ee
so that
\be
T^{(1,0)}_0 & = & \sqbox{0}{-7}{1}{0} \; +\; \sqbox{0}{-7}{2}{0} \; +\;
    \sqbox{0}{-7}{3}{0} \; =\;
\sum \sqbox{0}{-7}{}{0}
\ee
where such summations are performed over all allowed numberings of the
boxes using the numbers $1$, $2$, and $3$.
For a general Young tableau, the numbers must not decrease moving to
the right along a row and must strictly increase moving down a column:
\be
\sqqbox{0}{-17}{1}{2}{}
\sqqbox{0}{-17}{1}{3}{}
\sqqbox{0}{-17}{2}{3}{}
\sqbox{0}{3}{2}{}
\sqbox{0}{3}{3}{}
\ee
Such a Young tableau denotes the product of the eight labeled boxes as
given by (\ref{boxes}) where it is understood that the relative shifts in the
arguments are given by
\be
\sqqbbox{0}{-17}{\!\!\!\!u\+8\la}{\!\!\!\!\!\!u\+10\la}{}
\sqqbbox{0}{-17}{\!\!\!\!u\+6\la}{\!\!\!\!u\+8\la}{}
\sqqbbox{0}{-17}{\!\!\!\!u\+4\la}{\!\!\!\!u\+6\la}{}
\sqbbox{0}{3}{\!\!\!\!u\+2\la}{}
\sqbbox{0}{3}{u}{0}
\ee
and the zero superscript gives the shift in the top right box.

Using this notation, the eigenvalues of the fused row transfer
matrix at level $(n,m)$ can be written as
\be
T^{(n,m)}_{0} & = & \sum \; \genbox{n}{m}{0}
\ee
where the number of terms in the sum is given by the dimension
of the irreducible representations of $su(3)$
\be
\mbox{dim}(n,m) = (n+1)(m+1)(n+m+2)/2.
\ee
It is straightforward to show that these satisfy the fusion
equations (\ref{fh1})--(\ref{fh3}) with
\be
f_{0}  =  \sqbox{0}{13}{1}{0}\sqqbox{-20}{-27}{2}{3}{}\hs{-0.25}, \hs{0.15}
T^{(0,1)}_0  =\;
\sqqbox{0}{-17}{1}{2}{0} \; +\;
\sqqbox{0}{-17}{1}{3}{0} \; +\;
\sqqbox{0}{-17}{2}{3}{0} \;= \;(-1)^N
\frac{s(u\+2\la)s(u\-3\la)}{s(2\la)s(3\la)} T^{(1,0)}_{1/2} \label{T01}
\ee
The product of factors on the left side of the fusion equations
can be precisely partitioned into the two terms on the right side.
Although it is not evident from these equations, the $T^{(n,m)}_0$
inductively defined are in fact entire functions of $u$. Furthermore,
these equations close at level $n+m=2L$ with
\be
T^{(n,m)}_0 = 0 \qquad \mbox{if $n+m\geq 2L$}.
\ee
These facts will be established for the dilute $A_L$ lattice models
by carrying out the fusion procedure
directly at the level of the face weights in the sequel.

The fusion equations completely determine the fused transfer matrices.
Indeed, the solution of (\ref{fh1})--(\ref{fh3}) for the fused transfer
matrices $T^{(n,m)}_0$ can be written in the determinantal form
\be \hs{-0.5}
T^{(n,m)}_0  =
\hbox{\small $ \rule[-115pt]{0.15mm}{230pt}
\matrix{
T^{(0,1)}_{n\!+m\!-1} & f_{n\!+m\!-2} & & & & \vl & & & & &\cr
T^{(1,0)}_{n\!+m\!-1} & \ddots & \ddots & & & \vl & & & & &\cr
1 & \ddots & \ddots & \ddots & & \vl & & & & & \cr
& \ddots & \ddots &  \ddots & f_{n} & \vl & & & & &\cr
& & 1 & T^{(1,0)}_{n\!+1} & T^{(0,1)}_{n} & \vl & f_{n\!-1} & & & &\cr
\multispan{11} \hl \cr
& & & & 1 & \vl & T^{(1,0)}_{n\!-1} & T^{(0,1)}_{n\!-2} & f_{n\!-3} & &\cr
& & & & & \vl & 1 & \ddots & \ddots & \ddots &\cr
& & & & & \vl & & \ddots & \ddots & \ddots & f_{0} \cr
& & & & & \vl & & & 1 & T^{(1,0)}_{1} & T^{(0,1)}_{0} \cr
& & & & & \vl & & & & 1 & T^{(1,0)}_{0} } \rule[-115pt]{0.15mm}{230pt} $}
\label{determ}
\ee
This can be directly verified using standard properties of determinants.
In particular, it immediately follows from (\ref{BAE}) and (\ref{T01})
that $T^{(n,m)}_0$ are entire functions of $u$.
\section{Elementary Fusion}
\setcounter{equation}{0}

Fusion is a process \cite{KRS:81} to build up new solutions of the
Yang-Baxter equation. The essential idea is to form
$Z$-invariant \cite{Baxter78,Baxter:82} $p\times q$ blocks of elementary
face weights. Then new solutions of the Yang-Baxter equation with distinct
critical behavior are obtained by applying suitable projectors.
To fuse the dilute $A_L$ models, we will follow the detailed methods of
Zhou and Pearce \cite{ZhPe:94}. However, we present the basic example of
$1\times 2$ fusion in some detail to allow us to introduce our notation
properly and to keep the paper self-contained.

In this section we consider the elementary fusion of a row of two and
three faces corresponding to fusion levels $(2,0)$, $(0,1)$ and $(0,0)$,
the latter corresponding to a Young diagram with three vertically arranged
boxes, which in the $su(3)$ case reduces to the trivial representation.
Thereafter, in the following section, we treat the more general case of
level $(n,0)$ and $(0,n)$ fusion, in both horizontal and vertical directions.
Subsequently, in section~4, we present the general case of fusion level
$(n,m)$.

\subsection{Projectors}
\label{subpro}
Let us define local face transfer operators $X_j(u)$ with elements
\be
\langle \sigma | X_j(u) | \sigma' \rangle =
\wt W{\sigma_j}{\sigma_{j+1}}{\sigma_j'}{\sigma_{j-1}}u \;
\prod_{k\neq j}\delta_{\sigma_j,\sigma_j'}
\label{localfaceop}
\ee
where $\sigma$ and $\sigma'$ are allowed paths.
The matrix $X_j(u)$ is block diagonal and we denote
the blocks for fixed $j$ by
\be
X^{(b,d)}(u)\;\; = \dface {}b{}d{u\hs{0.1}} \label{blocks}
\ee
The dimension of this block is given by the number of allowed two-step
paths from $d$ to $b$ which is \mbox{$[(I+A)^{2}]_{b,d}\leq 3$}.

In addition to the Yang-Baxter equation, the face weights of
the dilute $A_L$ models also satisfy the local inversion relation
\be
\sum_{g=1}^L \wt Wabgdu \wt Wgbcd{-u} \;\; = \;\;
\inversion \;\; = \;\; \rho(u)\rho(-u)\delta_{a,c}
\label{invrel}
\ee
where
\be
\rho(u) = \frac{\tha{2\la-u}\tha{3\la-u}}{\tha{2\la}\tha{3\la}}.
\ee
In terms of matrix multiplication of the face transfer operators,
this relation takes the form
\be
X^{(b,d)}(u) X^{(b,d)}(-u) = \rho(u)\rho(-u) I.
\label{blockinvrel}
\ee
It follows that the four block matrices
\be
\dface {}b{}d{3\la}\; ,\;  \dface {}b{}d{-3\la} \qquad\qquad
\dface {}b{}d{2\la}\; ,\;  \dface {}b{}d{-2\la}
\ee
are singular. In fact, after appropriate normalization,
$X^{(b,d)}(3\la)$ and $X^{(b,d)}(2\la)$ are projection operators.
Although the other two are not strictly projectors, we will subsequently
refer to these four singular operators as projectors. Within each pair,
these operators are orthogonal as a consequence of the inversion relation.

Either of the above pairs of projectors can be used to construct
new solvable models by the fusion procedure \cite{Zhou:94}. These two
types of fusion are very different in nature. The first pair of projectors
leads to $su(2)$ type fusion with adjacency matrices $A^{(n)}$
of the fused lattice models at level $n$
given by the $su(2)$ fusion rules
\be
& A^{(0)} = I , \qquad A^{(1)} = I+A & \no \\*
& A^{(n)} A^{(1)} = A^{(n-1)} + A^{(n+1)} \quad n=1,2,3,\ldots &
\ee
without any closure. The second pair of projectors leads to $su(3)$ type
fusion with adjacency matrices $A^{(n,m)}$ of the fused lattice models
at level $(n,m)$ given by $su(3)$ fusion rules
\be
&  A^{(n,m)} = 0 \quad \mbox{if $n<0$ or $m<0$}, \qquad
   A^{(n,m)}=A^{(m,n)}  & \no \\*
&  A^{(0,0)} = I, \qquad A^{(1,0)} = I+A &  \label{adjfusion}\\*
&  A^{(n,m)} A^{(1,0)} = A^{(n+1,m)} + A^{(n-1,m+1)} + A^{(n,m-1)},
\quad n,m=0,1,2,\ldots & \no
\ee
These equations close with
\be
A^{(n,m)} = 0 \quad \mbox{for $n+m\geq 2L$}.
\ee

Note that the fusion hierarchy equations (\ref{fh1})--(\ref{fh3})
yield valid adjacency equations if the fused transfer matrices are replaced
by the fused adjacency matrices, the shifts are discarded and
the functions $f_k$ are set to one.
The elements of the fused adjacency matrices can in general be
nonnegative integers greater than one. In this case we distinguish
the edges of the adjacency diagram
joining two given sites by bond variables $\alpha =1,2,\ldots$ If there
is just one edge then the corresponding bond variable is $\alpha=1$.
More generally, the entries $A^{(n,m)}_{a,b}$
of the fused adjacency matrices give the number of admissible bonds
joining states $a$ and $b$ in the fused models at level $(n,m)$.
Explicitly, the fused adjacency matrices $A^{(n,m)}=A^{(m,n)}$
for $A_3$ are given by
\be
\mbox{\small $\pmatrix{1 & 0 & 0 \cr 0 & 1 & 0 \cr 0 & 0 & 1}$} & = &
A^{(0,0)} \; = \; A^{(5,0)} \; = \; A^{(0,5)}\no \\
\mbox{\small $\pmatrix{1 & 1 & 0 \cr 1 & 1 & 1 \cr 0 & 1 & 1}$} & = &
A^{(1,0)} \; = \; A^{(4,1)} \; = \; A^{(0,4)}\; = \;
A^{(4,0)} \; = \; A^{(1,4)}\; = \; A^{(0,1)} \no \\
\mbox{\small $\pmatrix{1 & 1 & 1 \cr 1 & 2 & 1 \cr 1 & 1 & 1}$} & = &
A^{(2,0)} \; = \; A^{(3,2)}\; = \; A^{(0,3)} \; = \;
A^{(3,0)}\; = \; A^{(2,3)}\; = \; A^{(0,2)}   \no \\
\mbox{\small $\pmatrix{1 & 2 & 1 \cr 2 & 2 & 2 \cr 1 & 2 & 1}$} & = &
A^{(1,1)} \; = \; A^{(3,1)}\; = \; A^{(1,3)} \no \\
\mbox{\small $\pmatrix{1 & 2 & 2 \cr 2 & 3 & 2 \cr 2 & 2 & 1}$} & = &
A^{(1,2)} \; = \; A^{(2,1)} \; = \; A^{(2,2)} \label{adj3}
\ee
where the $\Z_3$ symmetry of the weight lattice
about the fusion level $(m,n)$ is apparent, see Fig~\ref{fig2}.

Let us list some properties of the projectors useful for the
implementation of $su(3)$ fusion. These either follow from the explicit form
of the face weights or from the inversion relation (\ref{invrel})
with $u=-2\la$.  The first group is obtained by inserting
the explicit face weights at $u=-2\la$ for an given value of $a$.
This gives \addtolength{\jot}{4mm}
\be
\dface {b\!\pm\!1}b{b\!\pm\!1}{\hs{-0.2}b\!\pm\! 2}{2\la}
  \;\; & = & 0 \no \\
\dface bbc{\hs{-0.2}b\!\pm\! 1}{2\la} \;\; & = &
  \sqrt{\frac{\thd{2b\la\mp\la}}{\thd{2b\la\pm 3\la}}}
  \;\;\;\;\;\dface {b\!\pm\! 1}bc{\hs{-0.2}b\!\pm\! 1}{2\la}
\label{PP} \\
\dface {b\!\pm\!1}bcb{2\la} \;\; & = &
  A_{b\!\pm\! 1,b}\:\frac{\tha{\la}}{\tha{2\la}}\:
  \frac{S(b\!\pm\! 1,b)}{S(b,b)}\; \dface bbcb{2\la} \no
\ee
for any value of $c$.
Analogously, we obtain a second group of relations by inserting
the explicit face weights at $u=2\la$ for a given value of $c$.
This gives
\be
& & \dface{b\!\pm\!1}b{b\!\pm\!1}{\hs{-0.2}b\!\pm\! 2}{-2\la}
  \;\;\;\; = \;\;\frac{\tha{4\la}\tha{5\la}}{\tha{2\la}\tha{3\la}} \no\\
& & \dface abb{\hs{-0.2}b\!\pm\! 1}{-2\la} \;\;\;\; = \;\;
  -\:\sqrt{\frac{\thd{2b\la\pm 3\la}}{\thd{2b\la\mp\la}}}
   \;\dface ab{b\!\pm\! 1}{\hs{-0.2}b\!\pm\! 1}{-2\la} \label{PM}\\
\lefteqn{\hspace*{-31mm}
   A_{b,b\-1}\,\frac{S(b\-1,b)}{S(b\+ 1,b)}
   \dface ab{b\-1}b{-2\la}\;\;\,+\,
   \frac{\tha{2\la}}{\tha{\la}}\,\frac{S(b,b)}{S(b\+1,b)}
   \dface abbb{-2\la}\!+\,
   A_{b,b\+1}\dface ab{b\+1}b{-2\la}\;\;=\: 0} \no
\ee
where $a$ is arbitrary. \addtolength{\jot}{-4mm}
Here we introduced the compact notation
\be
S(a,b) & = & \sqrt{S_a} \thd{2b\la+(b-a)\la} .
\ee

\subsection{Level (2,0) Fusion}
\label{20fusion}

The projector of symmetric $1\times 2$ fusion is the local face transfer
operator $X_j(-2\la)$, or more explicitly, the blocks $X^{(b,d)}(-2\la)$
in (\ref{blocks}). The action of these projectors is to project out certain
two-step paths from $d$ to $b$. The number of remaining paths are then
given by the entries of the fused adjacency matrix $A^{(2,0)}$.

To be more precise, let us denote the set of two-step paths $(a,a',b)$
from $a$ to $b$ by $\mbox{path}(a,b;2)$. We refer to a path $(a,a',b)$ as
{\em dependent} on a set of paths $\{(a,a'_i,b)\}$ (with respect to
$X^{(b,a)}(-2\la)$) if there exist $\phi(a,a'_i,b)$ such that
\be
\dface cb{a^{\scs\prime}}a{-2\la} & = &
\sum_{i} \phi(a,a'_i,b) \dface cb{a^{\scs\prime}_{\scs i}}a{-2\la}
\label{deppath}
\ee
holds for any $c$. In other words, a set of paths $\{(a,a'_i,b)\}$ is called
{\em independent} if
\be
\sum_{i} \phi(a,a'_i,b)
\dface cb{a^{\scs\prime}_{\scs i}}a{-2\la} & = & 0
\label{indeppath}
\ee
implies that all coefficients vanish, i.e., $\phi(a,a'_i,b)=0$. Note that this
definition of dependent and independent paths obviously depends on the
projector under consideration.

Given a dependent set of paths, the choice of a maximal independent subset
is by no means unique. However, this does not matter for our purpose as
different choices result in equivalent fused models related to each other by a
local gauge transformation. This allows us to use arbitrary sets of independent
paths in the construction of the fused face weights. We denote such a set of
independent two-step paths from $a$ to $b$ (w.r.t.\ $X^{(b,a)}(-2\la)$)
by $\mbox{indpath}_{(2,0)}[a,b]$. The actual number of these paths is of course
determined by the number of non-zero eigenvalues\footnote{Remember that
$X^{(b,a)}(-2\la)$ is not strictly a projector, hence its eigenvalues do not
have to be 0 or 1.} of the  block $X^{(b,a)}(-2\la)$. The eigenvectors for
the zero eigenvalue of each block  $X^{(b,a)}(-2\la)$ are given in the second
and third equations in (\ref{PM}). As shown before, these immediately follow
from the inversion relation. Note, however, that we do {\em not} have to use
the explicit form of the eigenvectors for non-zero eigenvalues in what follows.

For our present discussion, this means that there is only one case where
we have more than one independent path to consider, namely if $a=b$ and
$\val a=3$, i.e., if \mbox{$2\leq a\leq L-1$}. In this case there are three
allowed paths $(a,a-1,a)$, $(a,a,a)$ and $(a,a+1,a)$ which are dependent by
(\ref{PM}), so one is left with two independent paths in this case. For the
fused face weights, this implies that one has to consider two different kinds
of bonds labeled by a {\em bond variable} $\alpha$ which takes two values
$\alpha=1,2$ corresponding to the two independent paths.

The allowed two-step paths on the adjacency diagram of the dilute $A_L$ models
shown in Fig~\ref{fig1}(b) are
\be
\mbox{path}(a,b;2) & = & \left\{
\ba{l@{\hspace{5mm}}l}
\{(a,a\!\pm\! 1,a\!\pm\! 2)\} & \mbox{if $b=a\pm 2$} \\
\{(a,a,a\!\pm\! 1),(a,a\!\pm\! 1,a\!\pm\! 1)\} & \mbox{if $b=a\pm 1$} \\
\{(1,2,1),(1,1,1)\} & \mbox{if $a=b=1$} \\
\{(L,L,L),(L,L\- 1,L)\} & \mbox{if $a=b=L$} \\
\{(a,a\+ 1,a),(a,a,a),(a,a\- 1,a)\} & \mbox{if $a=b$, $\val a=3$}
\ea \right. \label{paths}
\ee
{}From these we choose the independent paths for the construction of
the $(2,0)$ fused face weights as follows
\be
\mbox{indpath}_{(2,0)}[a,b] & = & \left\{\ba{l@{\hspace{5mm}}l}
\{(a,a\!\pm\! 1,a\!\pm\! 2)\} & \mbox{if $b=a\pm 2$} \\
\{(a,a\+1,a\+1)\} & \mbox{if $b=a+1$} \\
\{(a,a,a\-1)\} & \mbox{if $b=a-1$} \\
\{(1,2,1)\} & \mbox{if $a=b=1$} \\
\{(L,L\-1,L)\} & \mbox{if $a=b=L$} \\
\{(a,a\+1,a),(a,a\-1,a)\} & \mbox{if $a=b$, $\val a=3$}
\ea \right. \label{indpaths20}
\ee
or, in other words,
\be
\mbox{indpath}_{(2,0)}[a,b] & = &
\left\{(a,a',b)\in\mbox{path}(a,b;2)\, |\, a'\neq\min(a,b)\right\}
\, . \mbox{\hspace*{10mm}}
\label{ip20}
\ee
Obviously, the number of independent paths is
\be
\left|\mbox{indpath}_{(2,0)}[a,b]\right| & = & A^{(2,0)}_{a,b}
\ee
where $A^{(2,0)}$ is the fused adjacency matrix of (\ref{adjfusion}).
We define $\phi_{(2,0)}(a,a',b|\alpha)$ as the coefficients of
the path $(a,a',b)$ in terms of the ``basis'' of independent paths
$\mbox{indpath}_{(2,0)}[a,b]$, where \mbox{$1\leq\alpha\leq A^{(2,0)}_{a,b}$}
labels the elements of $\mbox{indpath}_{(2,0)}[a,b]$, respectively. This means
\be
\dface cb{a^{\scs\prime}}a{-2\la} & = &
\sum_{\alpha=1}^{A^{(2,0)}_{a,b}} \phi_{(2,0)}(a,a',b|\alpha)
\dface cb{a^{\scs\prime}_{\scs\alpha}}a{-2\la}
\label{phidef}
\ee
where $(a,a'_{\alpha},b)$ is the corresponding independent path. The
coefficients can be read off from (\ref{PM}), explicitly they are given by
\be
\renewcommand{\arraystretch}{2}
\phi_{(2,0)}(a,a',b|\alpha) = \left\{\ba{l@{\hspace{5mm}}l}
  \delta_{\alpha,1} & \mbox{if $|a-b|=2$} \\
  \(-\sqrt{\disp\frac{\thd{2a'_{\alpha}\la+\la}}
                     {\thd{2a'_{\alpha}\la-3\la}}}\)^{a'_{\alpha}-a'}
  \delta_{\alpha,1} & \mbox{if $|a-b|=1$} \\
  \(-\disp\frac{\tha{\la}}{\tha{2\la}}
    \disp\frac{S(a'_{\alpha},a)}{S(a,a)}\)^{\delta_{a,a'}}\delta_{\alpha,1}
    & \mbox{if $a\=b$, $\val a\=2$} \\
  \(-\disp\frac{\tha{\la}}{\tha{2\la}}
    \disp\frac{S(a'_{\alpha},a)}{S(a,a)}
    \)^{\delta_{a,a'}}
    & \mbox{if $a\=b$, $\val a\=3$, $|a'_{\alpha}\-a'|\!<\!2$} \\
  0  & \mbox{otherwise} \ea \right. \label{coeff20}
\renewcommand{\arraystretch}{1}
\ee
where it is understood that $\phi_{(2,0)}(a,a',b|\alpha)=0$
if $(a,a',b)\not\in\mbox{path}(a,b;2)$ and $a'_{\alpha}$ is defined as above.

To phrase it differently, and maybe more clearly, the blocks
$X^{(b,a)}(-2\la)$ are square matrices in the basis given by
$\mbox{path}(a,b;2)$. From the inversion relation, one obtains eigenvectors
of $X^{(b,a)}(-2\la)$ with eigenvalue $0$ (if $|b-a|<2$), (\ref{PM}) give
their components in the basis of the paths up to an arbitrary normalization.
The $\phi_{(2,0)}(a,a',b|\alpha)$ defined above are nothing else than the
components of $\alpha$ linearly independent vectors which span the orthogonal
complement of the zero eigenvalue eigenspace of $X^{(b,a)}(-2\la)$. Note that
these are in general not eigenvectors of $X^{(b,a)}(-2\la)$ and therefore the
vectors for different $\alpha$ are not necessarily mutually orthogonal; for
instance, for $a=b$, $\val a=3$, the three vectors $|\alpha\rangle$
($\alpha=0,1,2$, $\alpha=0$ denoting the eigenvector with zero eigenvalue)
have the following components in the basis $\mbox{path}(a,a;2)$ given
in (\ref{paths})
\be
|0\rangle & = &
\left(S(a+1,a),\frac{\tha{2\la}}{\tha{\la}}S(a,a),S(a-1,a)\right) \no\\
|1\rangle & = &
\left(1,-\frac{\tha{\la}}{\tha{2\la}}\frac{S(a+1,a)}{S(a,a)},0\right) \\
|2\rangle & = &
\left(0,-\frac{\tha{\la}}{\tha{2\la}}\frac{S(a-1,a)}{S(a,a)},1\right) \no
\ee
where $|1\rangle$ and $|2\rangle$ are both orthogonal to $|0\rangle$
but not orthogonal to each other. We can split any summation over
paths $(a,a',b)$ into a summation over the zero eigenvalue eigenvectors
of $X^{(b,a)}(-2\la)$ and its orthogonal complement which yields a sum
over $\alpha$ with coefficients $\phi_{(2,0)}(a,a',b|\alpha)$ by choosing
a suitable basis. Clearly, (\ref{PP}) are just
\be
\sum_{a'} \phi_{(2,0)}(a,a',b|\alpha)\;\;
\dface{a^{\scs\prime}}{b}{c}{a}{2\la} \; = \;\;
\dfacesym{\alpha}{b}{c}{a}{2\la}\; =\;\; 0 \label{PPPP}
\ee
for $\alpha=1,2$ and any value of $c$. We refer to the sum
with coefficients $\phi_{(2,0)}(a,a',b|\alpha)$
as the {\em symmetric sum} and denote it by a cross with label $\alpha$.
In particular, we find the following decomposition (``split property'')
\be
\splittz & \;\; = \;\; & \sum_{a^{\prime}}
\dface eb{a^{\scs\prime}}a{-2\la}
\onebytwoface abcd{a^{\scs\prime}}{c^{\scs\prime}}u{u\!\+\!2\!\la}
\no\\*
& \;\; = \;\; &
\sum_{a^{\prime}}\; \sum_{\alpha}\;\phi_{(2,0)}(a,a',b|\alpha)\;
\dface eb{a^{\scs\prime}_{\scs\alpha}}a{-2\la}
\onebytwoface abcd{a^{\scs\prime}}{c^{\scs\prime}}u{u\!\+\!2\!\la}
\no\\*
& \;\; = \;\; &
\sum_{\alpha} \dface eb{a^{\scs\prime}_{\scs\alpha}}a{-2\la}
\onebytwofacesym abcd{\alpha}{c^{\scs\prime}}u{u\!\+\!2\!\la}
\label{splittwozero}
\ee
where in the last step we performed the summation over $a'$. Here, the sum
on the RHS includes only the independent paths
$(a,a'_{\alpha},b)$, $\alpha\in\{1,2\}$.

After these preliminary remarks, we are finally in a position to define
the fused face weights for the elementary symmetric fusion. These are
basically the symmetric sums on the RHS of (\ref{splittwozero}), but we
still have to make a choice how to relate the bond variable $\beta$ to the
value of $c'$. In the case $c=d$ and $\val c =3$ this choice is not completely
free. Due to our selection of independent paths we have to exclude
$c'=c$, because the path $(c,c,c)$ has non-zero coefficients in terms of
the two independent paths. For definiteness and simplicity, we choose $c'$
such that $(d,c',c)$ is path labeled by $\beta$ in the set
$\mbox{indpath}_{(2,0)}[d,c]$.

\begin{lemma}[Elementary Symmetric Fusion] \label{lemma1}
If\ $(a,b)$ and\ $(d,c)$\ are\ admissible \newline
edges at fusion level $(2,0)$,
we define the $1\times 2$ fused weights by
\be
\hs{-1} \wf {W_{(2,0)}}ubcd{\alpha}{}{\beta}{\vspace*{-3mm}} & = & \!\!
\onebytwofacesym abcd{\alpha}{\beta}u{u\!\+\!2\!\la}
\;\; =\;
\onebytwofacesym abcd{\alpha}{c^{\scs\prime}_{\scs\beta}}u{u\!\+\!2\!\la}
\no\\* & = &
{\disp \sum_{a'} }\;\phi_{(2,0)}(a,a',b|\alpha)\;
\wt Wa{a'}{c'_{\beta}}d{u} \wt W{a'}bc{c'_{\beta}}{u\+2\la}
\label{twozero}
\ee
where the sum is over all allowed spins $a'$, the bond variables take
values $1\leq\alpha\leq A^{(2,0)}_{a,b}$, $1\leq\beta\leq A^{(2,0)}_{c,d}$,
and where the coefficients $\phi_{(2,0)}(a,a',b|\alpha)$ are those of
(\ref{coeff20}). Furthermore, the value of $c'$ on the RHS is chosen such that
$(d,c',c)\in\mbox{\rm indpath}_{(2,0)}[d,c]$, with $\beta$ being the label of
this particular  element of\/ $\mbox{\rm indpath}_{(2,0)}[d,c]$. The weights
so defined satisfy the Yang-Baxter equation. In particular, we note that
\be
 \wf {W_{(2,0)}}{0}bcd{\alpha}{\vspace*{-3mm}}{\beta}{} \;\; =\;\;
 \wf {W_{(2,0)}}{\la}bcd{\alpha}{\vspace*{-3mm}}{\beta}{} \;\; =\;\; 0
\label{twozerozero}
\ee
for all $a,b,c,d,\alpha,\beta$.
\end{lemma}

To show that the Yang-Baxter equation is indeed satisfied, we proceed as
follows. {}From the Yang-Baxter equation of the elementary weights
and (\ref{PPPP}), one obtains
\be
\pushtz \;\; =\;\;  0  \;\; = \;\; \invtwola
\label{push}
\ee
This establishes the ``push-through'' property, i.e., the dependence of
the symmetric sum in (\ref{push}) on the path $(d,c',c)$ is obviously the
same as that of the projector $X^{(c,d)}(-2\la)$, see (\ref{PM}). Explicitly,
this yields
\be
& & \hs{1.7}
\onebytwofacesym abc{c\!\pm\! 1}{\alpha}{c}u{u\!\+\!2\!\la}\; = \;\;
-\;\sqrt{\frac{\thd{2c\la\pm 3\la}}{\thd{2c\la\mp \la}}}\;\;
\onebytwofacesym abc{c\!\pm\! 1}{\alpha}{c\!\pm\! 1}u{u\!\+\!2\!\la}\no\\*
& & \label{twozerofu} \\*
& & \hspace*{-7mm}
A_{c,c\-1}\frac{S(c\-1,c)}{S(c\+1,c)}
\!\!\onebytwofacesym abcc{\alpha}{c\-1}u{u\!\+\!2\!\la}\!\!\!
+\:\frac{\tha{2\la}}{\tha{\la}}\frac{S(c,c)}{S(c\+1,c)}
\!\!\onebytwofacesym abcc{\alpha}cu{u\!\+\!2\!\la}\!\!\!
+\: A_{c,c\+1}\!\!
\!\!\onebytwofacesym abcc{\alpha}{c\+1}u{u\!\+\!2\!\la}\!\!\! =\: 0 \no
\ee
or, for short,
\be
\onebytwofacesym abcd{\alpha}{c^{\scs\prime}}u{u\!\+\!2\!\la} \; = \;\;
\sum_{\beta}\; \phi_{(2,0)}(d,c',c|\beta)
\onebytwofacesym abcd{\alpha}{c^{\scs\prime}_{\scs\beta}}u{u\!\+\!2\!\la}
\label{pushshort}
\ee
In complete analogy to the decomposition in (\ref{splittwozero}), one finds
\be
\twobytwofacesym & = & \sum_{\beta}
\onebytwofacesym abcd{\alpha}{c^{\scs\prime}_{\scs\beta}}u{u\!\+\! 2\!\la}
\times
\onebytwofacesym dcef{\beta}{e^{\scs\prime}}v{v\!\+\! 2\!\la}
\ee
and therefore the Yang-Baxter equation for the fused weights
follows immediately from that for the elementary faces (\ref{YBR}).
This also shows why it is most convenient to choose the $c'$ in the
definition of the fused weights (\ref{twozero}) to correspond to an
independent path and that one has to be careful if there is more
than one independent path.

{}From the initial condition (\ref{norm}), the fused weights
(\ref{twozero}) at $u=0$ reduce to the symmetric sum of the
elementary face weight at $u=2\la$ which vanishes due to (\ref{PPPP}).
Similarly, one obtains the second part of (\ref{twozerozero})
using the crossing symmetry (\ref{cross}). This implies that the
fused weights contain an overall factor of $\tha{u}\tha{u-\la}$.

\subsection{Level (0,1) Fusion}
\label{01fusion}

The antisymmetric $1\times 2$ fusion is constructed by using
the blocks $X^{(c,d)}(2\la)$ as projectors.
This means that the fused weights are basically given by the products
\be
\zeroonefu
\ee
Here, things are somewhat simpler then for the $(2,0)$ fusion, since the
blocks $X^{(c,d)}(2\la)$, up to normalization, really are projectors onto
at most one-dimensional spaces. In particular, this implies that we do not
need to introduce a bond variable.

Note that we define the $(0,1)$ fusion by summing over the variable $c'$
instead $a'$. This convention allows us to use the same product of
elementary face weights in both cases (otherwise we would have to
interchange $u$ and $u+2\la$). This also means that the definition
of independent paths $\mbox{indpath}_{(0,1)}[d,c]$ apparently involves
``the other side'' of the projector which however makes no difference
since the weights (\ref{weights}) are symmetric under reflection,
see (\ref{refl}). Let us choose the following sets of independent paths
\be
\mbox{indpath}_{(0,1)}[d,c] & = &
\left\{(d,c',c)\in\mbox{path}(d,c;2) | c'=\min(c,d)\right\}
\label{indpaths01}
\ee
which is just the complement of the set of independent paths for level
$(2,0)$ fusion, see (\ref{indpaths20})--(\ref{ip20}). Clearly,
the corresponding adjacency matrix $A^{(0,1)}$ for the fused weights
coincides with that of the elementary face weights, i.e.,
$A^{(0,1)}=A^{(1,0)}$ (\ref{adjfusion}).
{}From (\ref{PP}), we obtain\renewcommand{\arraystretch}{2}
\be
\phi_{(0,1)}(d,c',c) = \left\{\ba{l@{\hspace{5mm}}l}
  \disp \(\frac{\thd{2c'_1\la+3\la}}{\thd{2c'_1\la-\la}}\)^{(c'-c'_1)/2}
   & \mbox{if $|c-d|=1$} \\
  \disp \(\frac{\tha{\la}}{\tha{2\la}}\frac{S(c',d)}{S(d,d)}\)^{|c'-d|}
   & \mbox{if $c=d$} \\
  0 & \mbox{otherwise}\ea \right.
\label{phi01}
\ee
\renewcommand{\arraystretch}{1}
for all $(d,c',c)\in\mbox{path}(d,c;2)$ and zero otherwise,
where $c'_1=\min(c,d)$.

We refer to the sum over $c'$ with coefficients $\phi_{(0,1)}(d,c',c)$ as
an {\em antisymmetric sum} and denote it by a circle (without any further
label) in our diagrammatical notation. Obviously, (\ref{PM}) becomes
\be
\sum_{c} \phi_{(0,1)}(d,c',c)
\;\dface ac{c^{\scs\prime}}d{-2\la}\;\; = \;\;
\dfaceasym ac{}d{-2\la} \;\; = \;\; 0 \label{PMPM}
\ee
and the split relation analogous to (\ref{splittwozero}) is simply
\be
\splitzo
\label{splitzeroone}
\ee
where $(d,c'_{1},c)$ is the corresponding independent path, i.e.,
\mbox{$c'_{1}=\min(c,d)$}. As before, we define the fused weights to be
the object on the RHS of the above equation, up to a choice on the values
of $a'$. By the same arguments as above, the dependence on $a'$ can be read
off directly from (\ref{PP}), yielding
\be
\onebytwofaceasym a{a\!\pm\! 1}cda{}u{u\!\+\! 2\la} & \; = \;\; &
\sqrt{\frac{\thd{2a\la\mp\la}}{\thd{2a\la\pm 3\la}}}\;
\onebytwofaceasym a{a\!\pm\! 1}cd{a\!\pm\! 1}{}u{u\!\+\! 2\la} \no\\
\onebytwofaceasym aacd{a\!\pm\! 1}{}u{u\!\+\! 2\la} & \; = \;\; &
A_{a\!\pm\! 1,a}\frac{\tha{\la}}{\tha{2\la}}\,
\frac{S(a\!\pm\! 1,a)}{S(a,a)}\;
\onebytwofaceasym aacda{}u{u\!\+\! 2\la}
\label{pushzeroone}
\ee
which is just
\be
\onebytwofaceasym abcd{a^{\scs\prime}}{}u{u\!\+\! 2\la} & \; = \;\; &
\phi_{(0,1)}(a,a',b)\;
\onebytwofaceasym abcd{a^{\scs\prime}_{\scs 1}}{}u{u\!\+\! 2\la}
\ee
with \mbox{$a'_1=\min(a,b)$}. Again, we choose $a'$
in (\ref{splitzeroone}) such that
$(a,a',b)\in\mbox{indpath}_{(0,1)}[a,b]$ to define the fused weights.

\begin{lemma}[Elementary Antisymmetric Fusion]\label{lemma2}
Define
\be
\wt {W_{(0,1)}}abcdu & = &
\!\!\onebytwofaceasym abcd{}{}u{u\!\+\! 2\la} \;\; =\;
\onebytwofaceasym abcd{a^{\scs\prime}}{}u{u\!\+\! 2\la} \no\\*
& = & \sum_{c'}\;\phi_{(0,1)}(d,c',c)\;\wt Wa{a'}{c'}d{u}
      \wt W{a'}bc{c'}{u\+ 2\la}
\label{zeroone}
\ee
where we sum over all allowed values of $c'$ and where
$a'=\min(a,b)$. Then the fused weights satisfy
\be
 \wt {W_{(0,1)}}abcdu & = & -\, s^1_1\, s^1_{-3/2}\:
     \frac{g(d,c)}{g(a,b)}\;\wt Wabcd{u\+\la} \no \\*
 &  =  &  -\, r^{1}_{1}\:\frac{g(d,c)}{g(a,b)}\;\wt Wabcd{u\+\la}
\label{zeroonered}
\ee
where the gauge factors $g(a,b)$ are given by
\renewcommand{\arraystretch}{2}
\be
g(a,b)& = &
\left(\frac{\tha{2\la}}{\tha{\la}}\right)^{|a-b|}\;\times\;
\left\{ \ba{l@{\hspace{5mm}}l}
\disp\sqrt{\frac{\thd{2b\la+\la}}{\thd{2b\la-\la}}} & \mbox{for $a>b$} \\
\disp\sqrt{\frac{\thd{2b\la-\la}}{\thd{2b\la-3\la}}} & \mbox{for $a<b$} \\
\disp\sqrt{\frac{\thd{2b\la+\la}\thd{2b\la-\la}}{\vartheta_{4}^{2}(2b\la)}} &
\mbox{for $a=b$} \ea \right.
\label{zeroonegauge}
\ee
\renewcommand{\arraystretch}{1}
and $s^m_k$ and $r^m_k$ are functions of $u$ defined as
\be
s^m_k & = & \prod_{j=0}^{m-1}
 \frac{\tha{u+2(k-j)\la}}{\sqrt{\tha{2\la}\tha{3\la}}} \\
r^m_k & = & s^m_k\, s^m_{k-5/2}
 \;\; = \;\; \prod_{j=0}^{m-1}
 \frac{\tha{u+2(k\-j)\la}\tha{u+(2(k\-j)\-5)\la}}{\tha{2\la}\tha{3\la}} \; .
\label{s}\ee
In particular, this implies that
\be
\wt {W_{(0,1)}}abcd{\-2\la} \;\; = \;\;
\wt {W_{(0,1)}}abcd{3\la} \;\; = \;\; 0
\label{zeroonezero}
\ee
for all $a,b,c,d$.
\end{lemma}

In other words, up to a gauge and some overall factors the $(0,1)$
fused weights are nothing but the elementary face weights (\ref{weights})
shifted by $\lambda$. This of course already implies that the Yang-Baxter
equation is fulfilled which alternatively follows from the push-through
properties (\ref{pushzeroone}) as in the symmetric case. We omit the proof
of the lemma since it reduces to the explicit computation of the fused weights.
However, the appearance of the overall factor \mbox{$r^1_1=s^1_1s^1_{-3/2}$}
in (\ref{zeroonered}) can be understood directly from (\ref{zeroonezero})
which follows from (\ref{PMPM}) using the initial condition (\ref{norm}) and
crossing symmetry (\ref{cross}).

The symmetric and antisymmetric fusion are ``orthogonal'' in the sense that
\be
\onebytwofacesymasym = \;\;\; 0
\ee
for all $a,b,c,d$ and $\alpha$.
This follows,  for example from (\ref{push}), which basically
is just the inversion relation (\ref{blockinvrel}) for $u=\pm2\la$.
But this is not all, the action of the two projectors is also complementary
in the sense that there is no non-trivial subspace that is annihilated by
both the blocks $X^{(b,d)}(2\la)$ and $X^{(b,d)}(-2\la)$,
provided that \mbox{$|b-d|\leq2$}. This follows from the inversion relation
(\ref{blockinvrel}) since the RHS has only simple zeros in the spectral
parameter $u$. If both blocks really were projectors,
this would mean that they add up to the identity matrix. Although this
is not the case here, we can choose the
independent paths such that the two sets of independent paths
$\mbox{indpath}_{(2,0)}[a,b]$ and $\mbox{indpath}_{(0,1)}[a,b]$ are
disjoint and that their union consists of all allowed paths, i.e.,
\be
\mbox{indpath}_{(2,0)}[a,b]\;\cap\;\mbox{indpath}_{(0,1)}[a,b]
& = & \emptyset\no\\*
\mbox{indpath}_{(2,0)}[a,b]\;\cup\;\mbox{indpath}_{(0,1)}[a,b]
& = & \mbox{path}(a,b;2)
\ee
which is exactly what we did, see (\ref{indpaths20})--(\ref{ip20})
and (\ref{indpaths01}). In particular, this implies the relation
\be
{\left(A^{(1,0)}\right)}^2 & = & A^{(2,0)} \;\; +\;\; A^{(0,1)}
\ee
for the fused adjacency matrices.

\subsection{Antisymmetric 1$\times$3 Fusion: Level (0,0)}
\label{00fusion}

Before we move on to the higher level symmetric fusion, we first have a
look at the completely antisymmetric $1\times 3$ fused weights. The
corresponding projector is the following product of elementary face weights
\be
\projzzo
\label{prozerozero}
\ee
where the above equality is just (\ref{splitzeroone}) with $u=2\la$
and $e_1$ is defined accordingly.

Of course, (\ref{prozerozero}) means that we can equally well regard
the antisymmetric sum on the RHS as the projector defining the fused
weights. By Lemma~(\ref{lemma2}), it is proportional to the elementary
face weights at $u=3\la$ which in turn is related by crossing symmetry
to $u=0$ and hence proportional to $\delta_{c,d}$. Furthermore, the local
face operators (\ref{localfaceop}) at $u=3\la$ represent projectors onto
at most one dimension. This implies that the same is true for our
projector (\ref{prozerozero}), viewed as a matrix acting from path
$(d,d',c',c)$ to $(d,d'',c'',c)$, hence we have to deal with only one
independent path here. Let us choose this path to be $(c,c,c,c)$. Then
the only non-vanishing coefficients $\phi_{(0,0)}(d,d',c',c)$ are given by
$\phi_{(0,0)}(c,c,c,c)=1$ and
\be
\lefteqn{\hspace*{-10mm}
\phi_{(0,0)}(c,c\!\pm\!1,c,c) \;\; = \;\;
\phi_{(0,0)}(c,c,c\!\pm\!1,c)} \no \\*[2mm]
& & = \;\; \sqrt{\frac{\thd{2c\la\mp\la}}{\thd{2c\la\pm3\la}}}\;
    \phi_{(0,0)}(c,c\!\pm\!1,c\!\pm\!1,c) \;\; = \;\;
    \frac{\tha{\la}}{\tha{2\la}} \frac{S(c\!\pm\!1,c)}{S(c,c)}
\ee

The fused weights are deduced from the product of elementary
faces with the projector of (\ref{prozerozero}).
{}From the above remarks, we find
\be
\fuzzo
\ee
where
\be
h(u) & = & -\:\frac{\tha{\la}\tha{4\la}\tha{u\-\la}\tha{u\+4\la}}
        {\vartheta_{1}^{2}(2\la)\,\vartheta_{1}^{2}(3\la)}.
\ee
Hence the fused weights are basically the same as those of antisymmetric
$su(2)$ type fusion with the projector $X^{(c,d)}(3\la)$. This
shows that the fused weights are essentially trivial, in particular
they vanish unless $c=d$ and also $a=b$, which follows
using the Yang-Baxter equation (\ref{YBR}).
We obtain the following result for the fused weights.

\begin{lemma}[Antisymmetric 1$\times$3 Fusion]\label{lemma3}
The non-zero fused weights defined by
\be
\lefteqn {\hspace*{-9mm} \wt {W_{(0,0)}}aaccu \;\;\; = \;
\onebythreefaceasym a{}{}accu{u\!\+\!2\la}{u\!\+\!4\la} \;\;\; = \;
\onebythreefaceasym aaaaccu{u\!\+\!2\la}{u\!\+\!4\la}} \no\\*
& & \hspace*{-7mm} = \;\;\sum_{d',c'}\;\phi_{(0,0)}(c,d',c',c)\;
\wt Waa{d'}cu\wt Waa{c'}{d'}{u\+2\la}\wt Waac{c'}{u\+4\la}
\label{eq:zerozeroonefu}
\ee
have the following simple form
\be
\wt {W_{(0,0)}}aaccu & = & f^{1}_{0}\:\frac{g(c)}{g(a)}
\ee
where we define functions $f^{m}_{k}$ of $u$ by
\be
f^m_k & = & (-1)^m\, r^m_{k+1}\, r^m_{k+3/2}\, r^m_{k+2} \;\; = \;\;
(-1)^m\, s^m_{k-3/2}\, s^m_{k-1}\, s^m_{k-1/2}\, s^m_{k+1}\,
  s^m_{k+3/2}\, s^m_{k+2}  \label{ff}
\ee
and $g(a)$ is given by
\be
g(a)=\sqrt{\frac{\thd{2a\la+\la}\thd{2a\la-\la}}{\vartheta_{4}^{2}(2a\la)}}\;.
\ee
\end{lemma}

Clearly, one has again the split and push-through properties for the fused
weights as in the two previous cases. However, we do not list them here
as there is nothing to prove apart from an explicit calculation of the weights.

Note that the $g(a)$ act as a simple gauge. This means that the fused weights
are trivial, in the sense that the transfer matrix constructed from these
weights is a multiple of the identity matrix. This is in  agreement with the
expected $su(3)$ structure of the fusion hierarchy of the
dilute $A_L$ models (\ref{T01}).
\section{Fusion at Levels ($n$,0) and (0,$n$)}
\setcounter{equation}{0}

We now want to apply the fusion procedure to higher levels,
starting with symmetric fusion.
To do so we start by generalizing our notation for
paths and independent paths.

\subsection{Projectors and Paths}
\label{PN}

Let us define graphically
\be
P_{(n,0)}(u)^{(a,a_1,a_2,\ldots,a_n,b)}_{(a,b_1,b_2,\ldots,b_n,b)}
\;\;\;\; = \proj
\label{projdef}
\ee
where we regard $P_{(n,0)}(u)$ as an operator acting on the
$(n\+1)$-step path  $(a,a_1,a_2,\ldots,a_n,b)$ to produce the path
$(a,b_1,b_2,\ldots,b_n,b)$.
As we will see in what follows, $P_{(n,0)}(u)$ will give
the weights of level $(n,0)$ fusion,
and $P_{(n,0)}(-2n\la)$ corresponds to the ``projector'' of
symmetric level $(n\+1,0)$ fusion.

Clearly, $P_{(1,0)}(u)$ is just an elementary block, and $P_{(2,0)}(u)$
is related to the $1\times 2$ symmetric fusion presented in
section~\ref{20fusion}. In particular, (\ref{splittwozero})
becomes
\be
P_{(2,0)}(u)^{(d,c',c,b)}_{(d,a,e,b)} & = &
\sum_{\alpha}\; P_{(1,0)}(-2\la)^{(a,a'_{\alpha},b)}_{(a,e,b)}\;
\wf {W_{(2,0)}}ubcd{\alpha}{}{\beta}{\vspace*{-3mm}}
\ee
where the sum labeled by the bond variable $\alpha\in\{1,2\}$
is over all independent paths
\mbox{$(a,a'_{\alpha},b)\in\mbox{indpath}_{(2,0)}[a,b]$}.

Now consider $P_{(n,0)}(u)$. By re-arranging elementary faces using
the Yang-Baxter equation (\ref{YBR}), it is easy to see that any
two adjacent faces with the spectral parameters $u+2j\la$ and
$u+2(j\+1)\la$ in (\ref{projdef}) can be considered as an instance of level
$(2,0)$ fusion. Therefore the properties (\ref{splittwozero}) and
(\ref{twozerofu}) imply the relations
\be
\lefteqn{\hspace*{-12mm}
P_{(n,0)}(u)_{(a,b_1,\ldots,b_j,b_{j+1},b_{j+2},\ldots,b_n,b)}^{
              (a,a_1,\ldots,a_j,a_{j+2},a_{j+2},\ldots,a_n,b)}} \no \\*
& & = \;\;
-\;\sqrt{\frac{\thd{2a_{j+2}\la\pm3\la}}{\thd{2a_{j+2}\la\mp\la}}}\;
P_{(n,0)}(u)_{(a,b_1,\ldots,b_j,b_{j+1},b_{j+2},\ldots,b_n,b)}^{
              (a,a_1,\ldots,a_j,a_j,a_{j+2},\ldots,a_n,b)}
\label{P1}
\ee
for $a_j-a_{j+2}=\pm 1$ and
\be
\lefteqn{\hspace*{-15mm}
A_{a_{j},a_{j}\-1}\;\frac{S(a_j\-1,a_j)}{S(a_j\+1,a_j)}\;
P_{(n,0)}(u)_{(a,b_1,\ldots,b_j,b_{j+1},b_{j+2},\ldots,b_n,b)}^{
              (a,a_1,\ldots,a_j,a_j-1,a_{j\+2},\ldots,a_n,b)}} \no\\*
& & + \;\;\frac{\tha{2\la}}{\tha{\la}}\;\frac{S(a_j,a_j)}{S(a_j\+1,a_j)}\;
P_{(n,0)}(u)_{(a,b_1,\ldots,b_j,b_{j\+1},b_{j\+2},\ldots,b_n,b)}^{
               (a,a_1,\ldots,a_j,a_j,a_{j\+2},\ldots,a_n,b)} \no\\*
& & + \;\; A_{a_{j},a_{j}\+1}\;
P_{(n,0)}(u)_{(a,b_1,\ldots,b_j,b_{j+1},b_{j+2},\ldots,b_n,b)}^{
               (a,a_1,\ldots,a_j,a_j+1,a_{j\+2},\ldots,a_n,b)}
\;\;\;= \;\;\; 0
\label{P2}
\ee
for $a_j=a_{j+2}$.
These two equations take over both the role of (\ref{PM})
in level $(n\+1,0)$ fusion (with $u=-2n\la$) and, via a generalized
split property (\ref{splittwozero}), of (\ref{twozerofu}) in
level $(n,0)$ fusion.
Of course, (\ref{P1}) and (\ref{P2}) only express that any
antisymmetric sum of
$P_{(n,0)}(u)^{(a,a_1,\ldots,a_n,b)}_{(a,b_1,\ldots,b_n,b)}$ over
any of the variables $a_j$ vanishes. Therefore we can summarize them by
\be
\sum_{a_j}\; \phi_{(0,1)}(a_{j-1},a_j,a_{j+1})\;
P_{(n,0)}(u)^{(a,a_1,\ldots,a_n,b)}_{(a,b_1,\ldots,b_n,b)} & = & 0
\ee
for $1\leq j\leq n$, where we set $a_0=a$ and $a_{n+1}=b$.

As before, we denote the set of $n$-step paths from $a$ to $b$
on the effective adjacency diagram of Fig~\ref{fig1}(b) by
$\mbox{path}(a,b;n)$, or $(a,b;n)$ for short.
The number of such paths is given by
\be
\left|\mbox{path}(a,b;n)\right| & = &  |(a,b;n)| \;\; = \;\;
{\left[{\left(I+A\right)}^{n}\right]}_{a,b}
\ee
i.e., by the corresponding element of the $n$-th power of the
effective adjacency matrix \mbox{$A^{(1,0)}=I+A$}. In this basis,
$P_{(n,0)}(u)$ becomes a square matrix which we can choose to be
block-diagonal by an appropriate ordering of paths.
Let us introduce $(a,b;n|j)$ as a short-hand notation for the
$j$-th path in the set $\mbox{path}(a,b;n)$.

The notion of independent paths also generalizes immediately
from the discussion of section~2. Here, a set of $(n\+1)$-step paths
$\{(a,a^i_1,\ldots,a^i_n,b)\}$ is {\em independent}
(w.r.t.\ $P_{(n,0)}(-2n\la)$) if, for any path $(a,b_1,\ldots,b_n,b)$,
\be
\sum_{i}\; \phi(a,a^i_1,\ldots,a^i_n,b)\;
P_{(n,0)}(-2n\la)_{(a,b_1,\ldots,b_n,b)}^{(a,a^i_1,\ldots,a^i_n,b)}
& = & 0
\label{nzasym}
\ee
implies that all coefficients $\phi(a,a^i_1,\ldots,a^i_n,b)$ vanish.
We denote by $\mbox{indpath}_{(n+1,0)}[a,b]=[a,b;(n\+1,0)]$
a maximal set of independent
paths from $a$ to $b$, by $|[a,b;(n\+1,0)]|$ the number of its elements,
and abbreviate its $\alpha$-th element by
$[a,b;(n\+1,0)|\alpha]$ (in order to avoid confusion, we will always
use Greek letters for indices referring to independent paths).

As in section~2, we define coefficients $\phi_{(n+1,0)}$ by
\be
P_{(n,0)}(-2n\la)^{(a,b;n+1|j)}_{(a,b;n+1|k)} & = &
\sum_{\alpha=1}^{|{\scs [a,b;(n\+1,0)]}|}
\phi_{(n+1,0)}[a,b](j|\alpha)\;
P_{(n,0)}(-2n\la)^{[a,b;(n+1,0)|\alpha]}_{(a,b;n+1|k)}
\ee
so the $\phi_{(n+1,0)}[a,b](j|\alpha)$ are the ``coordinates''
of the path $(a,b;n\+1|j)$
in the basis of independent paths $[a,b;(n\+1,0)|\alpha]$.

{}From these definitions, we immediately obtain the generalization of
the split property (\ref{splittwozero}) to the $(n,0)$ case. By definition,
we have
\addtolength{\jot}{4mm}
\be
\lefteqn{
P_{(n+1,0)}(u)_{(a,b_1,\ldots,b_n,b_{n+1},b)}^{(a,a_1,\ldots,a_n,a_{n+1},b)}
} \no\\*
& = & \!\!\!\!\!\sum_{j=1}^{|{\scs (b_1,b;n\+1)}|}
P_{(n,0)}(-2n\la)_{(b_1,\ldots,b_{n+1},b)}^{(b_1,b;n+1|j)}
\onebynnface \no\\*
& = & \!\!\!\!\!\sum_{j=1}^{|{\scs (b_1,b;n\+1)}|}\,
\sum_{\alpha=1}^{|{\scs [b_1,b;(n\+1,0)]}|}\!\!\!
P_{(n,0)}(-2n\la)_{(b_1,\ldots,b_{n+1},b)}^{[b_1,b;(n+1,0)|\alpha]}\,
\phi_{(n+1,0)}[b_1,b](j|\alpha) \hspace*{-25pt}
\makebox[160pt][l]{\onebynnface} \no\\*
& = & \!\!\!\!\!\sum_{\alpha=1}^{|{\scs [b_1,b;(n\+1,0)]}|}\;
P_{(n,0)}(-2n\la)_{(b_1,\ldots,b_{n+1},b)}^{[b_1,b;(n+1,0)|\alpha]}
\onebynnfacesym \label{splitnzero}
\ee
where the $c_k^j$ are given by \addtolength{\jot}{-4mm}
\be
(b_1,c_1^j\ldots,c_n^j,b) & = & (b_1,b;n\+1|j) \; .
\ee
In the last line of (\ref{splitnzero}), we have introduced a symbol
for the ``symmetric sum'' over $c_1^j,\ldots,c_n^j$ labeled by $\alpha$,
which we will use in what follows to define the $(n\+1,0)$ fused weights.

We also need to generalize the push-through
property (\ref{pushshort}).
We use  the recursive nature of the fusion procedure to
prove the push-through property by induction and at the same time
obtain an expression for the number of independent paths.
Suppose that the push-through property holds for the symmetric sum
of $n$ elementary faces
\be
\lefteqn{\hspace*{-10mm}\onebynfusion{j}} \no\\*
 & = & \sum_{\beta=1}^{|{\scs [d,c;(n,0)]}|}
\phi_{(n,0)}[d,c](j|\beta) \onebynfusion{\beta}
\label{pushn}
\ee
where the paths are labeled by
\mbox{$(d,c;n|j)=(d,e_1^j,\ldots,e_{n-1}^j,c)$} and
$[d,c;(n,0)|\beta]=(d,e_1^{\beta},\ldots,e_{n-1}^{\beta},c)$.
As in (\ref{splitnzero}), but splitting the product in a
different place, we have \addtolength{\jot}{4mm}
\be
\lefteqn{
P_{(n+1,0)}(u)_{(a,b_1,\ldots,b_n,b_{n+1},b)}^{(a,a_1,\ldots,a_n,a_{n+1},b)}
} \no\\
& = & \sum_{j=1}^{|{\scs (b_2,b;n)}|}
P_{(n-1,0)}(-2(n\-1)\la)_{(b_2,\ldots,b_{n+1},b)}^{(b_2,b;n|j)}
\twobynoneface \no\\*
& = & \sum_{\beta=1}^{|{\scs [b_2,b;(n,0)]}|}
P_{(n-1,0)}(-2(n\-1)\la)_{(b_2,\ldots,b_{n+1},b)}^{[b_2,b;(n,0)|\beta]}
\twobynonefacesym \no\\*
& = & \sum_{\beta=1}^{|{\scs [b_2,b;(n,0)]}|}
\sum_{\gamma=1}^{|{\scs [b_1,d;(n,0)]}|}
P_{(n-1,0)}(-2(n\-1)\la)_{(b_2,\ldots,b_{n+1},b)}^{[b_2,b;(n,0)|\beta]}
\!\!\twobynonefacesymsym \no\\*
& = &
\sum_{\gamma=1}^{|{\scs [b_1,d^{\delta};(n,0)]}|}\;
\sum_{d^{\delta}}\;\;
P_{(n,0)}(-2n\la)_{(b_1,\ldots,b_{n+1},b)}^{([b_1,d^{\delta};(n,0)|\gamma],b)}
\onebynfacesymsym \no\\*
& = & \sum_{\alpha=1}^{|{\scs [b_1,b;(n\+1,0)]}|}\;
P_{(n,0)}(-2n\la)_{(b_1,\ldots,b_{n+1},b)}^{[b_1,b;(n+1,0)|\alpha]}
\onebynfacesym \label{pushnn}
\ee
where \addtolength{\jot}{-4mm} we used the push-through property
(\ref{pushn}), the split property and the elementary symmetric fusion
of Lemma~\ref{lemma1}. The last line is just equation~(\ref{splitnzero}) again.
This clearly reflects the recursive nature of the construction
of fused weights:
the projector of fusion at level $(n\+1,0)$ is basically a face weight
of level $(n,0)$ fusion. We also see that in order to find the
independent paths at level $(n\+1,0)$, we only have to consider
paths which are independent at level $(n,0)$ and append one step to those.
This also establishes the push-trough property for the symmetric
sum of $n+1$ elementary faces by using the push-through properties
(\ref{pushn}) for $n$ faces and (\ref{pushshort}) for the
elementary fusion of two faces.

The number of independent paths is of course given by the difference
of the total number of paths and the number of independent equations
obtained from (\ref{P1}) and (\ref{P2}). This is just the matrix
element $A^{(n+1,0)}_{a,b}$ of the fused adjacency matrix defined
in (\ref{adjfusion}), i.e.,
\be
\left|\mbox{indpath}_{(n+1,0)}[a,b]\right| & = &
\left|[a,b;(n\+1,0)]\right| \;\; = \;\; A^{(n+1,0)}_{a,b} \; .
\ee
Of course, we in fact have to prove that the recursive
formula (\ref{adjfusion}) is  true and that it defines the
adjacency matrices of the fused weights which we construct.
This can be done by induction.
However, even for the special case we have considered so far
we need to be able to say something about fusion levels $(n,1)$,
because these enter in (\ref{adjfusion}).
Although these fusion levels
will not be discussed until section~4, we briefly sketch the argument
and it will become clear that the definitions of section~4 ensure that
it is correct.

Assume that we know the set $\mbox{indpath}_{(n,0)}[a,a']$,
$n\geq 1$, and that it contains $A^{(n,0)}_{a,a'}$ elements.
This is clearly fulfilled for $n=1$.
Then the number of all allowed
\mbox{$(n\+1)$}-step paths from $a$ to $b$ which are obtained by appending
one step to the paths in $\mbox{indpath}_{(n,0)}[a,a']$, is
obviously given by
\be
\sum_{a'}\; A^{(n,0)}_{a,a'}\; A^{(1,0)}_{a',b} & = &
{\left(A^{(n,0)}\cdot A^{(1,0)}\right)}_{a,b} \; .
\ee
However, not all of these are independent. How many relations are there?
To see this, we have to look at the last two steps of the paths, and count
how many of these are dependent as two-step paths,
keeping track of the number of paths in $\mbox{indpath}_{(n,0)}[a,a']$
which terminate in the corresponding manner.
To count these directly appears to be complicated, but
remember that for two-step paths one could choose the independent
paths for antisymmetric fusion as the complement of those for symmetric
fusion. This means that we arrive at a set of independent paths
$\mbox{indpath}_{(n+1,0)}[a,a']$ if we exclude all those paths
which are independent w.r.t.\ fusion containing $(n\-1)$ symmetric
and one antisymmetric sum at the end. We will see later that this
is exactly the definition of level $(n\-1,1)$ fused weights, and
hence there are $A^{(n-1,1)}_{a,b}$  such paths. Therefore
\be
\left|\mbox{indpath}_{(n+1,0)}[a,b]\right| & = &
{\left(A^{(n,0)}\cdot A^{(1,0)}\; - \; A^{(n-1,1)}\right)}_{a,b}
\;\; = \;\; A^{(n+1,0)}_{a,b}
\ee
in agreement with (\ref{adjfusion}).

\subsection{Level (\mbox{\protect\boldmath $n$},0)
            and (0,\mbox{\protect\boldmath $n$}) Fusion}\label{0nfusion}

We start by giving the $(n\+1,0)$ fused weights.
\begin{lemma}[Symmetric Fusion of a Single Row] \label{lemma4}
Define the fused weights at level $(n\+1,0)$ by
\addtolength{\jot}{4mm}
\be
\lefteqn{\wf{W_{(n+1,0)}}{u}{b}{c}{d}{\alpha}{\vspace*{-3mm}}{\beta}{}
\;\; =  \hs{-0.1} \onebynnfusion{\beta}} \no\\*
& & = \;\;\sum_{j=1}^{|{\scs (a,b;n+1)}|}\;
  \phi_{(n+1,0)}[a,b](j|\alpha) \onebynnfaced \no\\
&&= \;\;\sum_{j=1}^{|{\scs (a,b;n+1)}|}\;
  \phi_{(n+1,0)}[a,b](j|\alpha)\;\prod_{k=0}^{n}\;
  \wt W{e^j_{k}}{\hspace*{-3mm}e^j_{k+1}}{\hspace*{-3mm}
      e^\beta_{k+1}}{e^\beta_k}{\!u\+2k\lambda}
\label{nzerofusion}
\ee
\addtolength{\jot}{-4mm}
where
\begin{eqnarray*}
(a\equiv e_0^j,e_1^j\ldots,e_n^j,e_{n+1}^j\equiv b) & = &
(a,b;n\+1|j) \;, \\
(d\equiv e_0^{\beta},e_1^{\beta},\ldots,e_n^{\beta},
 e_{n+1}^{\beta}\equiv c) & = & [d,c;(n\+1,0)|\beta] \;,
\end{eqnarray*}
and the bond variables $\alpha$ and $\beta$ take values
\[
1\,\leq\,\alpha\,\leq\,\left|[a,b;(n\+1,0)]\right|\,=\,A^{(n+1,0)}_{a,b}
\;\;\;,\;\;\;
1\,\leq\,\beta\,\leq\,\left|[d,c;(n\+1,0)]\right|\,=\,A^{(n+1,0)}_{c,d}\; .
\]
These weights satisfy the Yang-Baxter equation (\ref{fuYBR}).
\end{lemma}
This lemma follows in the same way as Lemma~\ref{lemma1} for the
case of level $(2,0)$ fusion by using the
split (\ref{splitnzero}) and push-through (\ref{pushn}) properties.
For a formal proof one should use induction over $n$.

The fused weights for level $(0,n)$ are constructed as follows.
We consider a row of $2n$ elementary faces with spectral parameters
arranged according to the sequence
($u$, $u+2\la$, $u+2\la$, $u+4\la$, $u+4\la$, \ldots,
 $u+2(n-1)\la$, $u+2(n-1)\la$, $u+2n\la$).
We then perform an antisymmetric
fusion on each pair of adjacent faces with spectral parameters
differing by $2\la$, i.e.,
\be
\zeronnfu \label{zeronnfu}
\ee
By Lemma~\ref{lemma2} this yields, apart from  gauge and overall factors,
a row of $n$ elementary faces with spectral parameters
($u+\la$, $u+3\la$, \ldots, $u+(2n-3)\la$, $u+(2n-1)\la$).
These are then fused by the symmetric $(n,0)$ fusion process
described above. Altogether, this means that the $(0,n)$ fused weights
differ from the weights at fusion level $(n,0)$ only by
a shift in the spectral parameter, a gauge transformation and some
overall factors. Keeping track of the accumulated factors, one obtains
\be
\wf {W_{(0,n)}}{u}bcd{\alpha}{\vspace*{-3mm}}{\beta}{}
& = & \frac{g(d,c|\beta)}{g(a,b|\alpha)}\;
\left(\prod_{k=1}^{n}(-r^1_k)\right)\;
\wf {W_{(n,0)}}{u+\la}bcd{\alpha}{\vspace*{-3mm}}{\beta}{}
\label{zeron}
\ee
where $g(a,b|\alpha)$ denotes the product of the gauge factors
of (\ref{zeroonegauge}) along the independent path $[a,b;(n,0)|\alpha]$.
Strictly speaking, $\alpha$ and $\beta$ have different meaning on
the two sides of equation~(\ref{zeron}), labeling independent paths at
level $(0,n)$ on the left and independent paths at level $(n,0)$ on
the right. However, our construction of the fused weights gives a
one-to-one correspondence between these paths which is implied in
equation~(\ref{zeron}).

The fusion of a single column of elementary faces is basically
the same as that of a single row. Again we can make a choice on which side
of the column we perform the symmetric sum, here we choose the left side.
The operator in (\ref{projdef}) is then replaced by
\be
P^{(n,0)}(u)^{(a,a_1,a_2,\ldots,a_n,b)}_{(a,b_1,b_2,\ldots,b_n,b)}
\;\;\;\; = \columnproj
\label{columnprojdef}
\ee
where here and in what follows we use upper indices to denote the
fusion level in the vertical direction. Of course, $P^{(n,0)}(-2n\la)$ is
now the projector of level $(n\+1,0)$ fusion of $n\+1$ elementary faces in a
single column. But, from the Yang-Baxter equation, one finds
\be
\lefteqn{\hspace{-10mm}
P_{(n,0)}(-2n\la)^{(a,a_1,\ldots,a_n,b)}_{(a,b_1,\ldots,b_n,b)}
\;\;\;\; = \hspace{-3mm} \rowprojn } \no\\
& & \hspace{-10mm} = \hspace{-8mm} \columnprojn \hspace{-8mm} = \;\;\;\;
P^{(n,0)}(-2n\la)^{(a,a_1,\ldots,a_n,b)}_{(a,b_1,\ldots,b_n,b)}
\label{propro}\ee
so they are in fact identical.

This means we can use the same paths and independent paths to construct
the vertically fused weights at level $(n,0)$ as in the horizontal case.
Of course, this also holds true for the above discussion about
the level $(0,n)$ fused weights. Therefore, we will not go into more
detail here and instead move on to discuss directly the more general case
of symmetric fusion of rectangular blocks of elementary faces.

\subsection{Symmetric Fusion of Rectangular Blocks}

So far, we have considered fusion of a single row and
a single column of weights with fusion types labeled by the irreducible
representations $(n,0)$ and $(0,n)$ of $su(3)$.
In fact, we want to fuse the dilute models in both the
horizontal and vertical direction simultaneously.
The corresponding fused weights
are labeled by two irreducible representations of $su(3)$
(respectively their Young tableaux) where we
use the convention that the lower index of the weights corresponds
to the horizontal and the upper index to the vertical fusion level.
In general, these fused weights have bond variables not only
on the horizontal, but also on the vertical bonds.
Let us now consider the symmetric fusion of a rectangular
$m$ by $n$ block of elementary faces. Given two  positive integers
$m,n>1$ the fused face weights are defined by
\be
\lefteqn{\hspace*{-10mm}
\wf {W^{(m,0)}_{(n,0)}}ubcd{\!\!\!\alpha\!\!\!}{\nu}{\!\!\!\beta\!
  \!\!}{\mu} \;\;=\; \mnface }  \no\\
&&\hspace*{-10mm}=\sum_{j=1}^{|(a,d;m)|}\!\phi_{(m,0)}[a,d]{(j|\mu)}\!
                 \sum_{\alpha_2,\ldots,\alpha_{m}}\,\prod_{k=1}^{m}\,
   \row {W_{(n,0)}}{u\-2(m\-k)\lambda}{e^j_{k}}{e^\nu_k}{e^\nu_{k+1}}{
   e^j_{k\+1}}{\hspace*{-3mm}\alpha_k\hspace*{-3mm}}{\hspace*{-3mm}
  \alpha_{k+1}\hspace*{-3mm}}  ,\label{(m,0)(n,0)}
\ee
where
\begin{eqnarray*}
(a\equiv e_1^j,e_2^j\ldots,e_{m}^j,e_{m+1}^j\equiv d)
& = & (a,d;m|j) \;, \\
(b\equiv e_1^j,e_2^{\nu},\ldots,e_{m}^{\nu},e_{m+1}^{\nu}\equiv c)
& = & [b,c;(m,0)|\nu] \;,
\end{eqnarray*}
and the four bond variables $\alpha$, $\beta$, $\mu$ and $\nu$ take values
\begin{eqnarray*}
&&1\,\leq\,\alpha\,\leq\,\left|[a,b;(n,0)]\right|\;=\,A^{(n,0)}_{a,b}\h ,\h
1\,\leq\,\beta\,\leq\,\left|[d,c;(n,0)]\right|\;=\,A^{(n,0)}_{c,d} \\
&&1\,\leq\,\mu\,\leq\,\left|[a,d;(m,0)]\right|\,=\,A^{(m,0)}_{a,d}\h,\h
1\,\leq\,\nu\,\leq\,\left|[b,c;(m,0)]\right|\,=\,A^{(m,0)}_{b,c}\; .
\end{eqnarray*}
For the bond variables $\alpha_1,\alpha_2,\ldots,\alpha_{m+1}$ in
(\ref{(m,0)(n,0)}) we have that $\alpha=\alpha_1$, $\beta=\alpha_{m+1}$
and the other internal bond variables $\alpha_k$ are summed
over $1,\ldots,A_{e^j_k,e_k^\nu}^{(n,0)}$.
The $(n,0)$ fusion of $1\times {n}$ faces is given by (\ref{nzerofusion}).
Of course these rectangular fused face weights could be obtained by
fusing columns together instead of rows.

The fused face weights (\ref{(m,0)(n,0)}) are shown diagrammatically in
Fig~\ref{mnfusion}.
\begin{figure}[bt]
\[
\sum_{i,j}\:
\phi{\scriptstyle _{(n,0)}[a,b](i|\alpha)}\;
\phi{\scriptstyle _{(m,0)}[a,d](j|\mu)}
\rule[-44mm]{0mm}{86mm} \mnfusion
\]
\caption{\label{mnfusion}
  Diagrammatic representation of the face weights of
   the $n\times m$ fully symmetrically fused models. Sites indicated
       with a solid circle are summed over all possible spin states and
       the spins along the outer edges are given by
       $(a,e_2^j\ldots,e_{m}^j,d) \; = \; (a,d;m|j)$,
       $(b,e_2^{\nu},\ldots,e_{m}^{\nu},c) \; = \; [b,c;(m,0)|\nu]$,
       $(a,f_2^i\ldots,f_{n}^i,b) \; = \; (a,b;n|i)$ and
       $(d,f_2^{\mu},\ldots,f_{n}^{\mu},c) \; = \; [d,c;(n,0)|\mu]$.}
\end{figure}
These weights depend on both the spin variables $a,b,c,d$ and
the bond variables $\alpha,\beta,\mu,\nu$. Moreover, according to
section~\ref{0nfusion} the fused weights of symmetric $(m,0)\times(0,n)$
fusion and $(0,m)\times(0,n)$ fusion can be simply
derived from those of symmetric $(m,0)\times(n,0)$ fusion
\addtolength{\jot}{4mm}
\be
\wf {W^{(m,0)}_{(0,n)}}{u}bcd{\alpha}{\nu}{\beta}{\mu} & \sim &
\left(\prod_{k=1}^{n} (-1)^m r^m_k\right) \;
\wf {W^{(m,0)}_{(n,0)}}{u+\lambda}bcd{\alpha}{\nu}{\beta}{\mu} \\
\wf {W^{(0,m)}_{(0,n)}}{u}bcd{\alpha}{\nu}{\beta}{\mu} & \sim &
\left(\prod_{k=1}^{n} f^m_{k-2}\right)\;
\wf {W^{(m,0)}_{(n,0)}}{u}bcd{\alpha}{\nu}{\beta}{\mu}
\label{(0,m),(0,n)}
\ee
where \addtolength{\jot}{-4mm}
$\sim$ means equality up to gauge factors, compare equation~(\ref{zeron})
above, and the functions $r^m_k$ and $f^m_k$
are defined in (\ref{s}) and (\ref{ff}), respectively.

The fused blocks (\ref{(m,0)(n,0)}) are built up from rows of
fused faces. All properties held by single row of fused faces carry over
to the fused blocks (\ref{(m,0)(n,0)}). From the push-through
property (\ref{pushn}) we have the following lemma.
\begin{lemma}
Decomposing the path $(d,c;n|j)$ in terms of the independent paths
\newline $[d,c;(n,0)|\beta]$ gives
\be
\wf {W^{(m,0)}_{(n,0)}}{u}bcd{\alpha}{\nu}{j}{\mu}\;\; =\;\;
    \sum_{\beta=1}^{A^{(n,0)}_{(d,c)}}
\phi_{(n,0)}{[d,c](j|\beta)}\;\:
\wf {W^{(m,0)}_{(n,0)}}{u}bcd{\alpha}{\nu}{\beta}{\mu}
\label{eq:mbynfusionprop1}
\ee
Similarly, decomposing the path $(b,c;m|j)$ in terms of the independent
paths $[b,c;(m,0)|\nu]$ gives
\be
\wf {W^{(m,0)}_{(n,0)}}{u}bcd{\alpha}{j}{\beta}{\mu}\;\; =\;\;
  \sum_{\nu=1}^{A^{(m,0)}_{(b,c)}}
\phi_{(m,0)}[b,c](j|\nu)\;\:
\wf {W^{(m,0)}_{(n,0)}}{u}bcd{\alpha}{\nu}{\beta}{\mu}\; .
\label{eq:mbynfusionprop2}
\ee
\end{lemma}
This lemma implies the following theorem.
\begin{theorem}
For a triple of positive integers $m,n,l$, the fused face weights
\samepage (\ref{(m,0)(n,0)}) satisfy the Yang-Baxter equation
\be
\lefteqn{\hs{-0.1}
\sum_{(\eta_1,\eta_2,\eta_3)}\!\sum_g \mbox{\small $
   \Wb fugde{\mh\eta_1\mh}{\eta_3}{\mh\mu\mh}{\vs{-.1}\nu}
   \Wbml g{u\-v}bcd{\mh\eta_2\mh}{\beta}{\mh\gamma\mh}{\vs{-.1}\eta_3}
\Wbln a{v}bgf{\mh\alpha\mh}{\eta_2}{\mh\eta_1\mh}{\vs{-.1}\rho} $}} \no\\
\label{fuYBR} \\
\hs{-0.1} &=&
\hs{-0.1}\sum_{(\eta_1,\eta_2,\eta_3)}\!\sum_g\mbox{\small $
    \Wb aubcg{\mh\alpha\mh}{\beta}{\mh\eta_1\mh}{\vs{-.1}\eta_3}
    \Wbml f{u\-v}age{\mh\rho\mh}{\eta_3}{\mh\eta_2\mh}{\vs{-.1}\nu}
    \Wbln gvcde{\mh\eta_1\mh}{\gamma}{\mh\mu\mh}{\vs{-.1}\eta_2}$} \; . \no
\ee
\end{theorem}
\section{Fusion at Level ($n$,$m$)}
\setcounter{equation}{0}
The fused face weights at fusion level $(n,m)$ are more complicated to
construct. For clarity let us first consider the case  $n=m=1$.
Then we will generalize the construction to the general fusion level $(n,m)$.

\subsection{Level (1,1) Fusion}
Consider the following product of elementary face weights
\be
\setlength{\unitlength}{0.009in}
\begin{picture}(200,60)(0,60)
\thicklines
\put( 90,120){\line( 1,-1){90}}
\put(180, 30){\line(-1,-1){30}}
\put(150,  0){\line(-1, 1){90}}
\put( 60, 90){\line(-1,-1){30}}
\put( 30, 60){\line( 1,-1){60}}
\put( 90,  0){\line( 1, 1){60}}
\put(120, 90){\line(-1,-1){90}}
\put( 30,  0){\line(-1, 1){30}}
\put(  0, 30){\line( 1, 1){30}}
\put( 60, 90){\line( 1, 1){30}}
\put( 90, 60){\circle*{6}}
\put( 60, 30){\circle*{6}}
\put(120, 30){\circle*{6}}
\put( 87,123){\sc $d$}
\put( 53, 92){\sc $a$}
\put(121, 92){\sc $c^{\scs\prime}$}
\put( 22, 62){\sc $a^{\scs\prime\prime}$}
\put(151, 62){\sc $c^{\scs\prime\prime}$}
\put( -8, 28){\sc $e$}
\put(183, 28){\sc $c$}
\multiput(27,-11)(60,0){3}{\sc $b$}
\put( 87, 86){\sc $u$}
\put( 84, 26){\sc $2\lambda$}
\put(134, 26){\sc $u\!+\!2\lambda$}
\put(104, 56){\sc $u\!+\!4\lambda$}
\put( 16, 26){\sc $-4\lambda$}
\put( 46, 56){\sc $-2\lambda$}
\end{picture}
\rule[-16mm]{0mm}{33mm}
=\;\;\sum_f
\setlength{\unitlength}{0.009in}
\begin{picture}(230,60)(-15,60)
\thicklines
\put( 90,120){\line( 1,-1){90}}
\put(180, 30){\line(-1,-1){30}}
\put(150,  0){\line(-1, 1){90}}
\put( 60, 90){\line( 1, 1){30}}
\put(120, 90){\line(-1,-1){90}}
\put(180, 90){\line( 1,-1){30}}
\put(210, 60){\line(-1,-1){30}}
\put(180, 90){\line(-1,-1){60}}
\put( 90, 60){\line(-1,-1){60}}
\put( 30,  0){\line(-1, 1){30}}
\put(  0, 30){\line( 1, 1){60}}
\put( 30, 60){\line( 1,-1){30}}
\put( 90, 60){\circle*{6}}
\put(150, 60){\circle*{6}}
\put( 87,123){\sc $d$}
\put( 53, 92){\sc $a$}
\put(122, 92){\sc $c^{\scs\prime}$}
\put(177, 93){\sc $c^{\scs\prime}$}
\put( 22, 62){\sc $a^{\scs\prime\prime}$}
\put(213, 58){\sc $c^{\scs\prime\prime}$}
\put( -8, 28){\sc $e$}
\put(181, 22){\sc $c$}
\multiput(60, 22)( 50,0){2}{\sc $f$}
\multiput(27,-11)(120,0){2}{\sc $b$}
\put( 86, 44){\sc $a^{\scs\prime}$}
\put(147, 70){\sc $g$}
\put( 87, 86){\sc $u$}
\put(174, 56){\sc $2\lambda$}
\put( 16, 26){\sc $-4\lambda$}
\put( 46, 56){\sc $-2\lambda$}
\put(104, 56){\sc $u\!+\!2\lambda$}
\put(134, 26){\sc $u\!+\!4\lambda$}
\end{picture} \;\;\;\; .
\label{fusion(1,1)}
\ee
This product can be simplified using the elementary antisymmetric fusion
of section~\ref{01fusion}. Applying equations~(\ref{splitzeroone}) and
(\ref{pushzeroone}), and subsequently Lemma~\ref{lemma2}, the product
(\ref{fusion(1,1)}) reduces to
\be
\lefteqn{\hspace*{-10mm}
\dface {c_{\scs 1}}c{c^{\scs\prime\prime}}{c^{\scs\prime}}{2\la}\;\;
\sum_f
\setlength{\unitlength}{0.009in}
\begin{picture}(195,60)(-10,60)
\thicklines
\put( 90,120){\line( 1,-1){90}}
\put(180, 30){\line(-1,-1){30}}
\put(150,  0){\line(-1, 1){90}}
\put( 60, 90){\line( 1, 1){30}}
\put(120, 90){\line(-1,-1){90}}
\put(150, 60){\line(-1,-1){30}}
\put( 90, 60){\line(-1,-1){60}}
\put( 30,  0){\line(-1, 1){30}}
\put(  0, 30){\line( 1, 1){60}}
\put( 30, 60){\line( 1,-1){30}}
\put( 90, 60){\circle*{6}}
\put(150, 60){\circle{6}}
\put( 87,123){\sc $d$}
\put( 53, 92){\sc $a$}
\put(122, 92){\sc $c^{\scs\prime}$}
\put( 22, 62){\sc $a^{\scs\prime\prime}$}
\put( -8, 28){\sc $e$}
\put(181, 22){\sc $c$}
\multiput(60, 22)( 50,0){2}{\sc $f$}
\multiput(27,-11)(120,0){2}{\sc $b$}
\put( 86, 44){\sc $a^{\scs\prime}$}
\put( 87, 86){\sc $u$}
\put( 16, 26){\sc $-4\lambda$}
\put( 46, 56){\sc $-2\lambda$}
\put(104, 56){\sc $u\!+\!2\lambda$}
\put(134, 26){\sc $u\!+\!4\lambda$}
\end{picture}   =  \rule[-16mm]{0mm}{33mm} \;\;\;\;
\dface {c_{\scs 1}}c{c^{\scs\prime\prime}}{c^{\scs\prime}}{2\la}
\setlength{\unitlength}{0.009in}
\begin{picture}(200,60)(-15,60)
\thicklines
\put( 90,120){\line( 1,-1){90}}
\put(180, 30){\line(-1,-1){30}}
\put(150,  0){\line(-1, 1){90}}
\put( 60, 90){\line( 1, 1){30}}
\put(120, 90){\line(-1,-1){90}}
\put(150, 60){\line(-1,-1){30}}
\put( 90, 60){\line(-1,-1){60}}
\put( 30,  0){\line(-1, 1){30}}
\put(  0, 30){\line( 1, 1){60}}
\put( 30, 60){\line( 1,-1){30}}
\put( 90, 60){\circle*{6}}
\put(150, 60){\circle{6}}
\put( 60, 30){\circle{6}}
\put( 87,123){\sc $d$}
\put( 53, 92){\sc $a$}
\put(122, 92){\sc $c^{\scs\prime}$}
\put( 22, 62){\sc $a^{\scs\prime\prime}$}
\put( -8, 28){\sc $e$}
\put(181, 22){\sc $c$}
\put(110, 22){\sc $f_{\scs 1}$}
\multiput(27,-11)(120,0){2}{\sc $b$}
\put( 86, 44){\sc $a^{\scs\prime}$}
\put( 87, 86){\sc $u$}
\put( 16, 26){\sc $-4\lambda$}
\put( 46, 56){\sc $-2\lambda$}
\put(104, 56){\sc $u\!+\!2\lambda$}
\put(134, 26){\sc $u\!+\!4\lambda$}
\end{picture}} \no\\
& = & \; h'(u)\;\; \frac{g(c^{\prime},c)}{g(a,e)}\; \;\phi_{(0,1)}(a,a'',e)
\;\;\;\;\dface {c_{\scs 1}}c{c^{\scs\prime\prime}}{c^{\scs\prime}}{2\la}\;
\setlength{\unitlength}{0.009in}
\rule[-12mm]{0mm}{26mm}
\begin{picture}(150,60)(-20,40)
 \thicklines
 \put(  0, 30){\line(1, 1){60}}
 \put(  0, 30){\line(1,-1){30}}
 \put( 30,  0){\line(1, 1){60}}
 \put( 30, 60){\line(1,-1){60}}
 \put( 60, 90){\line(1,-1){60}}
 \put( 90,  0){\line(1, 1){30}}
 \put( 60, 30){\circle*{6}}
 \put( 17, 27){\sc $-3\la$}
 \put( 73, 27){\sc $u\+3\la$}
 \put( 57, 57){\sc $u$}
 \put( 23, 63){\sc $a$}
 \put( 55, 14){\sc $a^{\scs\prime}$}
 \put( 27,-11){\sc $b$}
 \put( 87,-11){\sc $b$}
 \put(122, 27){\sc $c$}
 \put( 57, 93){\sc $d$}
 \put( 92, 63){\sc $c^{\scs\prime}$}
 \put(-10, 27){\sc $e$}
\end{picture} \label{fusion11}
\ee
where $c_1=\min(c,c')$, the gauge factors $g(a,b)$ are given
by (\ref{zeroonegauge}), the coefficients of independent paths
$\phi_{(0,1)}(a,a',b)$ are listed in (\ref{phi01}) and
\[
h'(u)\;=\;\frac{\tha{2\la}\tha{7\la}\tha{u\-\la}\tha{u\+4\la}}
     {\vartheta_{1}^{2}(2\la)\,\vartheta_{1}^{2}(3\la)}\;.
\]
So we see that the fusion of three faces essentially reduces
to that of two faces, similar to what we observed in section~\ref{00fusion}.
Hence the fused weights at level $(1,1)$ are basically those of
elementary symmetric $su(2)$ type fusion \cite{Zhou:94} with the projector
$X^{(b,a)}(-3\lambda)$.

Since it is our primary interest to derive relations between fused row
transfer matrices, we do not care about the gauge factors in what follows.
That means we can use (\ref{fusion11}) to {\em define\/}
the fused weights using independent paths w.r.t.\ the
projector $X^{(b,a)}(-3\lambda)$. Clearly, these are two-step paths whereas
the fusion with the $\pm 2\lambda$ projectors is built on three-step paths,
but they are of course in one-to-one correspondence which can be seen
form (\ref{fusion11}). Although, to be precise, we should distinguish the
two sets by using another notation, we will not do so because we do
not want to introduce more symbols here.

As the complimentary projector $X^{(b,a)}(3\lambda)$ projects onto
at most one-dimensional subspaces, we only have to exclude at most one
path from all possible paths to obtain an independent set of paths
for the projector $X^{(b,a)}(-3\lambda)$. Let us choose
\be
\mbox{indpath}_{(1,1)}[a,b] & = &
\left\{ \ba{l@{\hspace*{10mm}}l}
\mbox{path}(a,b;2) & \mbox{if $a\neq b$} \\
\displaystyle \left\{(a,a',a)\in\mbox{path}(a,a;2)\,|\, a'\neq a\right\}
& \mbox{if $a=b$}\; . \ea \right.
\ee
The corresponding coefficients $\phi_{(1,1)}(a,a',b|\alpha)$ can be
obtained from the analogues of (\ref{PM}) which follow from the
inversion relations (\ref{blockinvrel}) and the face weights at
spectral parameter $u=3\lambda$. Explicitly, they read
\be
\phi_{(1,1)}(a,a',b|\alpha) & = & \left\{ \ba{l@{\hspace*{10mm}}l}
\delta_{a',a_{\alpha}} & \mbox{if $a\neq b$} \\
\displaystyle A^{(1,0)}_{a',a'_{\alpha}}\, (-1)^{\delta_{a,a'}}\,
{\left(\frac{S(a'_{\alpha})}{S(a')}\right)}^{1/2} &
\mbox{if $a=b$} \ea \right. \label{coeff11}
\ee
where $(a,a'_{\alpha},b)$ is the element of $\mbox{indpath}_{(1,1)}[a,b]$
labeled by $\alpha$. As before, the independent paths can again be used to
split the projector from the fused faces
\be
\setlength{\unitlength}{0.009in}
\rule[-12mm]{0mm}{26mm}
\begin{picture}(150,60)(-20,40)
 \thicklines
 \put(  0, 30){\line(1, 1){60}}
 \put(  0, 30){\line(1,-1){30}}
 \put( 30,  0){\line(1, 1){60}}
 \put( 30, 60){\line(1,-1){60}}
 \put( 60, 90){\line(1,-1){60}}
 \put( 90,  0){\line(1, 1){30}}
 \put( 60, 30){\circle*{6}}
 \put( 17, 27){\sc $-3\la$}
 \put( 73, 27){\sc $u\+3\la$}
 \put( 57, 57){\sc $u$}
 \put( 23, 63){\sc $a$}
 \put( 55, 14){\sc $a^{\scs\prime}$}
 \put( 27,-11){\sc $b$}
 \put( 87,-11){\sc $b$}
 \put(122, 27){\sc $c$}
 \put( 57, 93){\sc $d$}
 \put( 92, 63){\sc $c^{\scs\prime}$}
 \put(-10, 27){\sc $e$}
\end{picture}
= \;\; \sum_{a'}\;\phi_{(1,1)}(a,a',b|\alpha)\;
\dface eb{a^{\scs\prime}_{\scs\alpha}}a{-3\la}
\onebytwoface abcd{a^{\scs\prime}}{c^{\scs\prime}}u{u\!\+\!3\!\la} \; .
\label{split11}
\ee
The fused weights are defined in the following Lemma.
\begin{lemma}\label{lem7} If $(a,b)$ and $(d,c)$ are admissible edges
at fusion level $(1,1)$, then
\be
\lefteqn{
\wf {W_{(1,1)}}ubcd{\alpha}{}{\beta}{\vspace*{-0.3cm}} \;\; = \;\;
-r^1_2 \onebytwofacesym abcd{\alpha}{\beta}u{u\!\+\!3\!\la} = \;\; -r^1_2
\onebytwofacesym abcd{\alpha}{c^{\scs\prime}_{\scs\beta}}u{u\!\+\!3\!\la}}
\no\\*
& = & -r^1_2\;\;\sum_{a'}\;\,\phi_{(1,1)}(a,a',b|\alpha)\;\,
\wt Wa{a'}{c'_{\beta}}d{u} \;\wt W{a'}bc{c'_{\beta}}{u\+3\la}
\label{eq:oneonefu}
\ee
where the sum is over all allowed spins $a'$, the bond variables take
values $1\leq\alpha\leq A^{(1,1)}_{a,b}$, $1\leq\beta\leq A^{(1,1)}_{c,d}$,
and where the coefficients $\phi_{(1,1)}(a,a',b|\alpha)$ are those
of (\ref{coeff11}). Furthermore, the value of $c'_{\beta}$ on the RHS is
chosen such that $(d,c'_{\beta},c)\in\mbox{\rm indpath}_{(1,1)}[d,c]$, with
$\beta$ labeling the particular element of $\mbox{\rm indpath}_{(1,1)}[d,c]$.
The such defined weights satisfy the Yang-Baxter equation (\ref{fuYBR1}).
In particular, we note that for all fixed $a,b,c,d,\alpha,\beta$ we have
\be
\wf {W_{(1,1)}}{u}bcd{\alpha}{\vspace*{-0.3cm}}{\beta}{} \; =\; 0
\h\h \mbox{for \ $u=0,\lambda,-4\lambda$}\, . \label{oneonezeros}
\ee
\end{lemma}
Here we also used the same symbol for symmetric sum, which now
refers to the symmetric sum defined by the projector $X^{(b,a)}(-3\lambda)$.
Note that we kept the spectral parameter-dependent function $-r^1_2$ in the
definition of the fused weights, which directly explains two of the zeros
of (\ref{oneonezeros}). The zero at $u=0$ is obvious from the orthogonality
of the projectors $X^{(b,a)}(\pm 3\lambda)$.

\subsection{Fusion at Level
            (\mbox{\protect\boldmath $n$},
             \mbox{\protect\boldmath $m$})}

Fusion at level ($n,m$) can be constructed by generalizing the
fusion procedure of level ($1,1$) fusion. Firstly, let us consider
the one row fusion. Let $Y=(n,m)$ be a Young diagram with $n+2m$
nodes in Fig~\ref{fig:Young}. We use $2m$ faces with spectral
parameter shifts listed in the left  double rows of the Young diagram
in Fig~\ref{fig:Young}. The weights for level ($0,m$) antisymmetric fusion
represented by (\ref{zeronnfu}) are given as (\ref{zeron})
\be
\wf {W_{(0,m)}}{u+2n\la}bcd{\alpha}{\vspace*{-3mm}}{\beta}{}
 \sim \left(\prod_{k=1}^{m}(-r^1_k)\right)\;
\wf {W_{(m,0)}}{u+2n\la+\la}bcd{\alpha}{\vspace*{-3mm}}{\beta}{}\;.
\ee
Using the remaining faces with the other spectral parameter shifts in
Fig~\ref{fig:Young}, we can construct the fusion
$\wf {W_{(n,0)}}{u}bcd{\alpha}{\vspace*{-3mm}}{\beta}{}$. The one-row
symmetric $(n,m)$ fusion or the symmetric $(1,0)\times(n,m)$ fusion
can be constructed by studying the product
\be
\nmfusion \label{proj(n,m)}
\ee
The projector in (\ref{proj(n,m)}) is given by (\ref{(m,0)(n,0)})
with the spectral parameter $u=-2n\lambda-\lambda$ and we
sum over $\alpha$, $\beta$ and $c'$. Following the discussion in previous
sections, the fused face weight is found by introducing the
coefficients $\phi_{(n,m)}[a,b](j|\mu)$ and then splitting the
projector from the fused faces. To do so  we again need to know
the decomposition of paths into independent paths with respect to
the projector in (\ref{proj(n,m)}). For the  case of  level ($1,1$)
fusion this is gained by studying the single face $X^{(b,a)}(-3\la)$
in (\ref{blocks}) and thus we obtain the coefficients (\ref{coeff11}).
\begin{figure}[t]
\centerline{\Young}
\caption{\label{fig:Young} The sequence of the spectral parameter shifts
         for $Y=(n,m)$.}
\end{figure}

Let $(a,b;n,m)$ be the set of paths from $a$ to $b$ through $\alpha,c,\beta$
such that
\be
 \mndim \alpha\beta{a}bc\;\;:=\;
\wfa {W^{(m,0)}_{(n,0)}}{-2n\la\-\la}bca{\!\!\!\rho\!\!\!}{
  \beta}{\!\!\!\alpha\!\!\!}{\gamma} \ne 0 \; .
\ee
This means that
\be
(a,b;n,m)=\{(a,\alpha,c,\beta,b)|1\le\alpha\le A^{(n,0)}_{a,c},
1\le\beta\le A^{(m,0)}_{c,b}\; and \;A^{(n,0)}_{a,c}A^{(m,0)}_{c,b}\ne 0\}\;.
\ee
It is obvious that the number $|(a,b;n,m)|$ of paths in the set $(a,b;n,m)$
is $[A^{(n,0)}A^{(m,0)}]_{a,b}$. Similar to the elementary fusion
procedure at level ($2,0$) the fusion coefficients $\phi_{(n,m)}[a,b](i|\mu)$
are introduced by
\be
\mndim {\alpha^i}{\beta^i}{a}b{c^i}\;=
  \sum_{\mu=1}^{|[a,b;(n,m)]|}\phi_{(n,m)}[a,b](i|\mu)
    \h \mndim {\alpha^\mu}{\beta^\mu}{a}b{c^\mu}
\label{nmide}\ee
where the independent paths
$(a,\alpha^\mu,c^\mu,\beta^\mu,b)=[a,b;(n,m)|\mu]
\in\mbox{indpath}_{(n,m)}[a,b]$
and $|[a,b;(n,m)]|$ is the number of the independent paths. Obviously,
the number of independent paths,
$|\mbox{indpath}_{(n,m)}[a,b]|=|[a,b;(n,m)]|=A^{(n,m)}_{a,b}$,
  is equal to $|(a,b;n,m)|$ minus  the number of  independent
equations in (\ref{nmide}).
Notice that (\ref{nmide}) is over determined if $n+m\ge 2L$. This means
that there is no projector for the fusion when $n+m\ge 2L$ and so
no independent paths. Therefore  $A^{(n,m)}=0$ for $n+m\ge 2L$.

To calculate the level ($n,m$) fused face
weights it is necessary to find the coefficients
$\phi_{(n,m)}[a,b](j|\mu)$ and the set $\mbox{indpath}_{(n,m)}[a,b]$.
This can be done from the explicit weights
$W^{(m,0)}_{(n,0)}$. The projector is given by the fused face weights of
($m,0$)$\times$($n,0$).
The decomposition of  paths with respect to this projector follows from
(\ref{eq:mbynfusionprop1})--(\ref{eq:mbynfusionprop2})
and the face $X(-3\la)$ which sits in the right-hand corner of the projector.
One does not have a simple or unified form for the coefficients.
It is tedious but straightforward to calculate them. Instead of doing so we
formulate the fused weights of symmetric $1\times (n,m)$ fusion in terms of
the coefficients as follows
\be
\lefteqn{
\wf {W_{(n,m)}}{u}bcd{\mu}{\vspace*{-3mm}}{\nu}{}} \no\\
&=&\prod_{k=1}^{m}\(-s^1_{n+k}s^1_{n+k-5/2}\)\;\sum_{i=1}^{|(a,b;n,m)|}
                     \phi_{(n,m)}[a,b](i|\mu) \no \\
& &\times\;
\wf{W_{(n,0)}}{u}{e^i}{e^\nu}d{\gamma^i}{\vspace*{-3mm}}{\gamma^\nu}{}\;
\wb{W_{(m,0)}}{u+2n\lambda+\lambda}bc{e^\nu}{\eta^i}{
\vspace*{-5mm}}{\eta^\nu}{}
\label{W(n,m)}
\ee
where
\begin{eqnarray*}
(d,\gamma^\nu,e^\nu,\eta^\nu,c) & = & [d,c;(n,m)|\nu]
     \in\mbox{indpath}_{(n,m)}[d,c] \\
(a,\gamma^i,e^i,\eta^i,b) & = & (a,b;(n,m)|i)\in(a,b;n,m)
\end{eqnarray*}
and the notation $(a,b;(n,m)|i)$ denotes the $i$-th path in the set
$(a,b;n,m)$. These  bond variables $\nu,\mu$ are restricted by
$1\le\nu\le A^{(n,m)}_{(d,c)}$ and $1\le\mu\le A^{(n,m)}_{(a,b)}$.
The push-through property follows from (\ref{proj(n,m)}).
To see this we push the projector from the bottom through to the top
by using the Yang-Baxter equation (\ref{fuYBR}). It is easy to see that
the path along the top from $d$ to $c$ satisfies the same property
as the projector. So from (\ref{nmide}) we have the push-through property
for $W_{(n,m)}$.

Based on one row fusion $W_{(n,m)}$ we can easily build up multi-row
fusion. Let $m,n,\ol{m},\ol{n}$ be positive integers and define
\be
\lefteqn{
\wf {W^{(n,m)}_{(\ol{n},\ol{m})}}{u}bcd{\alpha}{\nu}{\beta}{\mu} =
\mnface } \no\\
& = & \prod_{k=1}^{\ol{n}}(-s^m_{k+1/2}s^m_{k-2})\;\prod_{k=1}^{\ol{m}}
   (-s^m_{k\!+\!\ol{n}\!+\!1}s^m_{k\!+\!\ol{n}\!-\!3/2})
   \sum_{i=1}^{|(a,d;n,m)|}\!\phi_{(n,m)}[a,d](i|\mu) \no\\
& & \hs{0.7} \sum_{\rho}\;\wdd {W^{(m,0)}_{(\ol{n},\ol{m})}}{u}{
  e^\nu}{\vs{-0.1}c}d{\rho}{\vs{-0.1}\eta^\nu}{\beta}{\eta^i}\;
   \wf {W^{(n,0)}_{(\ol{n},\ol{m})}}{u\-2m\lambda-\lambda}b{
     \vs{0.1}e^\nu}{e^i}{\alpha}{\vs{0.1}\gamma^\nu}{\rho}{\gamma^i} \;,
\label{eq:twomnfusion}\ee
where
\begin{eqnarray*}
(b,\gamma^\nu,e^\nu,\eta^\nu,c) & = & [b,c;(n,m)|\nu]
     \in\mbox{indpath}_{(n,m)}[b,c] \\
(a,\gamma^i,e^i,\eta^i,d) & = & (a,d;(n,m)|i)\in(a,d;n,m) \;.
\end{eqnarray*}
The sum over $\rho$ is over $1,2,\ldots,|[e^i ,e^\nu;(\ol{n},\ol{m})]|$.
The push-through property of each row carries over to the  multi-row
fusion. Therefore we have the following Lemma.
\begin{lemma}
Decomposing the path $(d,c;\ol{n},\ol{m}|j)$ in terms of the
independent paths $[d,c;(\ol{n},\ol{m}))|\beta]$ gives
\be
\wf {W^{(n,m)}_{(\ol{n},\ol{m})}}{u}bcd{\alpha}{\nu}{j}{\mu}=
    \sum_{\beta=1}^{A^{(\ol{n},\ol{m})}_{(d,c)}}
\phi_{(n,0)}{[d,c](j|\beta)}\;
\wf {W^{(n,m)}_{(\ol{n},\ol{m})}}{u}bcd{\alpha}{\nu}{\beta}{\mu}
\label{eq:mbyn1}
\ee
Similarly, ecomposing the path $(b,c;n,m|j)$ in terms of the
independent paths \newline $[b,c;(n,m)|\nu]$ gives
\be
\wf {W^{(n,m)}_{(\ol{n},\ol{m})}}{u}bcd{\alpha}{j}{\beta}{\mu}=
  \sum_{\nu=1}^{A^{(n,m)}_{(b,c)}}
\phi_{(m,0)}[b,c](j|\nu)\;
\wf {W^{(n,m)}_{(\ol{n},\ol{m})}}{u}bcd{\alpha}{\nu}{\beta}{\mu}
\label{eq:mbyn2}.\ee
\end{lemma}
Then the lemma implies that the fused faces satisfy the
Yang-Baxter relation (\ref{fuYBR1}).
\begin{theorem}
For a group of positive inters $m,n,l,\ol{m},\ol{n},\ol{l}$,
the fused face weights (\ref{eq:mbyn1}) satisfy the
Yang-Baxter equation
\be
\lefteqn{\hs{-0.5}\sum_{(\eta_1,\eta_2,\eta_3)}\!\sum_g \mbox{\small $
   \WWb fugde{\mh\eta_1\mh}{\eta_3}{\mh\mu\mh}{\vs{0.1}\nu}
   \WWbml g{u\-v}bcd{\mh\eta_2\mh}{\beta}{\mh\gamma\mh}{\vs{0.1}\eta_3}\times
\WWbln a{v}bgf{\mh\alpha\mh}{\eta_2}{\mh\eta_1\mh}{\vs{0.1}\rho} $}} \no\\
             \label{fuYBR1} \\
&=&\sum_{(\eta_1,\eta_2,\eta_3)}\!\sum_g\mbox{\small $
    \WWb aubcg{\mh\alpha\mh}{\beta}{\mh\eta_1\mh}{\vs{0.1}\eta_3}\times
    \WWbml f{u\-v}age{\mh\rho\mh}{\eta_3}{\mh\eta_2\mh}{\vs{0.1}\nu}
    \WWbln gvcde{\mh\eta_1\mh}{\gamma}{\mh\mu\mh}{\vs{0.1}\eta_2}$}\; .
\no\ee
\end{theorem}
\section{$su$(3) Fusion Hierarchy}
\label{Fusionhierarchy}
\subsection{Functional Equations }
\setcounter{equation}{0}

The $su(3)$ fusion rule (\ref{adjfusion}) relates the adjacency matrices of
fused models. We will see in this section that the similar relations carry
over to row transfer matrices.

Suppose that \mbox{\boldmath $a$} (\mbox{\boldmath $\alpha$}) and
\mbox{\boldmath $b$} (\mbox{\boldmath  $\beta$}) are allowed spin (bond)
configurations of two consecutive rows of an $N$ (even) column lattice
with periodic boundary conditions. The elements of the fused row
transfer matrix ${\bf T}(u)$ are given by
\be
\langle\mbox{\boldmath $a,\alpha$}|{\bf T}^{(n,m)}_{(\ol{n},
     \ol{m})}(u)|\mbox{\boldmath $b,\beta$}\rangle\; =\;
\rule[-16mm]{0mm}{34mm}
\prod_{j=1}^N\sum_{\eta_j}\; \WF {a_j}u{b_j}{b_{j\+1}}{a_{j\+1}}{\mh\eta_j
\mh}{\beta_j}{\mh\eta_{j\+1}\mh}{\alpha_j}\; =
\setlength{\unitlength}{0.0125in}
\begin{picture}(60,55)(-10,47)
\multiput(10,30)(40,0){2}{\line(0,1){40}}
\multiput(10,30)(0,40){2}{\line(1,0){40}}
\multiput(10,-2.5)(40,0){2}{\multiput(0,0)(0,10){11}{\line(0,1){5}}}
\put(27,47){$u$}
\put( 0,26){$\scriptstyle a_j$}
\put( 0,48){$\scriptstyle \alpha_j$}
\put(-9,69){$\scriptstyle a_{j+1}$}
\put(51,26){$\scriptstyle b_j$}
\put(51,48){$\scriptstyle \beta_j$}
\put(51,69){$\scriptstyle b_{j+1}$}
\end{picture}
\ee
where the periodic boundary conditions  $a_{N+1}=a_1$, $b_{N+1}=b_1$ and
$\eta_{N+1}=\eta_1$ are imposed. The Yang-Baxter equations (\ref{fuYBR1})
imply that
\begin{equation}
[{\bf T}^{(n,m)}_{(\ol{n},\ol{m})}(u),
     {\bf T}^{(n,m)}_{(\ol{n'},\ol{m'})}(v)]=0 \;.
\label{eq:rowcommute}
\end{equation}
Thus for each fixed  pair $(n,m)$  we obtain a hierarchy of
commuting  transfer matrices. Moreover, the fusion
procedure implies various relations among these
transfer matrices. Let us define
\be
\T^{(n,m)}_k&=&{\T}_{(n,m)}^{(n',m')}(u+2k\lambda) \no \\
\T^{(n,m)}&=&0 \h \mbox{if $n<0$ or $m<0$}        \\
\T^{(0,0)}&=&{\one}    \;  \no
\ee
and
\be
s_n^Y&=&\prod_{j\in Y} {\vartheta_1(u-u_j+2n\lambda)
        \over \sqrt{\vartheta_1(2\lambda)\vartheta_1(3\lambda)}} \\
f^Y_n&=&(-1)^{nN}\left[s_{n-3/2}^Ys_{n-1}^Ys_{n-1/2}^Ys_{n+1}^Y
s_{n+3/2}^Ys_{n+2}^Y \right]^N \label{f}
\ee
where the shifts $u_i$ are listed in  Fig~\ref{fig:Young} with
$u=0$. We summarize the functional relations in the following theorems.
\begin{theorem}[\protect\mbox{\boldmath $su$}(3) Fusion Hierarchy]
\label{theorem1}
The fused transfer matrices of the dilute $A_L$ lattice models
satisfy the functional relations
\be
\T^{(n,0)}_0\,\T^{(1,0)}_n&=&\T^{(n-1,1)}_0+\T^{(n+1,0)}_0 \label{fhT1}\\
\T^{(n,0)}_0\,\T^{(0,1)}_n&=&\T^{(n,1)}_0+f^Y_{n-1}\,\T^{(n-1,0)}_0 \;
\label{fhT2}
\ee
where $Y=(n',m')$ refers to the fixed vertical fusion level.
This hierarchy closes and the fused transfer matrices vanish for
$m+n\ge 2L$
\be
 {\bf T}^{(n,m)}=0\h  \mbox{if \ $m+n\ge 2L$.}
\label{close}
\ee
\end{theorem}
\begin{theorem}[Symmetry] \label{theorem2}
The fused transfer matrices $\T^{(n,m)}_{0}$ are given by
\be
\T^{(n,0)}_0\,\T^{(0,m)}_{n}&=&\T^{(n,m)}_{0}
+f^Y_{n-1}\,\T^{(n-1,0)}_0\,\T^{(0,m-1)}_{n+1}
\label{fhT3}
\ee
and satisfy the symmetry
\be
\lefteqn{\left[\prod_{k=0}^{n-1} s^Y_{k-2}s^Y_{k+1/2}\right]^N
     \T^{(n,m)}(u)\; = } \no\\
&&(-1)^{Nn(m-n)}\left[\prod_{k=0}^{m-1} s^Y_{n+k-3/2}s^Y_{n+k+1}\right]^N
   \left[\T^{(m,n)}(\!-u\!-\!2(n\!+\!m\!-\!2)\lambda)\right]^T
\ee
where the superscript $T$ denotes transpose and $Y=(n',m')$.
\end{theorem}
These theorems reduce to (\ref{fh1})--(\ref{fh3}) for $Y=(n',m')=(1,0)$.

\subsection{Bethe Ansatz Equations }
In section~\ref{FH} we discussed the connection of the fusion
hierarchy (\ref{fh1})--(\ref{fh3})  with the Bethe ansatz
equations for $(n',m')=(1,0)$. We now generalize the Bethe ansatz
equations to the fused transfer matrices with $Y=(n',m')$.

First let us consider the transfer matrix
${\T}^{(n',m')}_{(1,0)}(u)={\T}^{Y}_{(1,0)}(u)$.
The eigenvalues $\Lambda^Y_{(1,0)}(u)$ and the Bethe ansatz equations
of this transfer matrix are given by\addtolength{\jot}{4mm}
\be
T^{Y}_{(1,0)}(u) & = & \omega\,[s^Y_{-1}(u)s^Y_{-3/2}(u)]^N\frac{
          Q^{Y}(u+\lambda)}{Q^{Y}(u-\lambda)}
     \; +\; \omega^{-1}\,[s^Y_0(u)s^Y_{-1/2}(u)]^N
           \frac{Q^{Y}(u-4\lambda)}{Q^{Y}(u-2\lambda)}  \no\\*
     &   & \qquad +\; [(-1)^{n'}s^Y_0(u)s^Y_{-3/2}(u)]^N
         \frac{Q^{Y}(u)Q^{Y}(u-3\lambda)}{Q^{Y}(u-\lambda)Q^{Y}(u-2\lambda)}
\ee
where\addtolength{\jot}{-4mm}
\be
Q^{Y}(u)\;\; =\prod_{j=1}^{(n\!+\!2m)N} \vartheta_1(u-iu_j)
\ee
and the zeros $\{u_j\}$ satisfy the Bethe ansatz equations
\be
\omega \Biggr[\frac{s^{Y}_{1/2}(iu_j)}{s^{Y}_{-1/2}(iu_j)}\Biggr]^N
\!\!& = & (-1)^{n'N+1} \prod_{k=1}^{(n\!+\!2m)N}
\frac{\vartheta_1(iu_j-iu_k+2\lambda)\vartheta_1(iu_j-iu_k-\lambda)}
     {\vartheta_1(iu_j-iu_k-2\lambda)\vartheta_1(iu_j-iu_k+\lambda)}
\label{BAE1}
\ee
with $j=1,\ldots,N$ and $\omega=\exp(i\pi\ell/(L+1))$, $\ell=1,\ldots,L$.

As in section~\ref{FH} we now set\addtolength{\jot}{4mm}
\be
\sqbox{0}{-7}{1}{k} & = &
\omega\,[s^Y_{k-1}(u)s^Y_{k-3/2}(u)]^N\frac{Q^{Y}(u+k\lambda_2+\lambda)}{
       Q^{Y}(u+k\lambda_2-\lambda)}  \no\\
\sqbox{0}{-7}{2}{k}
    & = & [-s^Y_k(u)s^Y_{k-3/2}(u)]^N
   \frac{Q^{Y}(u+k\lambda_2)Q^{Y}(u+k\lambda_2-3\lambda)}{
   Q^{Y}(u+k\lambda_2-\lambda)Q^{Y}(u+k\lambda_2-2\lambda)}\label{boxes1}\\
\sqbox{0}{-7}{3}{k}
     & = & \omega^{-1}\,[s^Y_k(u)s^Y_{k-1/2}(u)]^N
      \frac{Q^{Y}(u+k\lambda_2-4\lambda)}{Q^{Y}(u+k\lambda_2-2\lambda)}\no
\ee
so that\addtolength{\jot}{-4mm}
\be
T^{Y}_{(1,0)}(u) & = & \sqbox{0}{-7}{1}{0} \; +\; \sqbox{0}{-7}{2}{0} \; +\;
    \sqbox{0}{-7}{3}{0} \; = \; \sum \sqbox{0}{-7}{}{0}
\ee
The $su(3)$ functional equations described in Theorem~\ref{theorem1} imply
 the eigenvalues of the fused row transfer
matrix at level $(n,m)$ can be written as
\be
T^{Y}_{(n,m)}(u) & = & \sum \; \genbox{n}{m}{0}
\ee
where the number of terms in the sum is given by the dimension
of the irreducible representations of $su(3)$
\be
 (n+1)(m+1)(n+m+2)/2.
\ee
Such a Young tableau denotes the product of the $(n+2m)$ labeled boxes as
given by (\ref{boxes1}) where it is understood that the relative shifts
in the arguments are given by Figure~4. These zeros $\{u_j\}$ satisfy
the Bethe ansatz equations (\ref{BAE1}). Thus $T^{Y}_{(n,m)}(u)$ can also
be represented in the determinantal form (\ref{determ}) where $f$ is
given by (\ref{f}).
\section{Concluding Remarks}
\setcounter{equation}{0}

We applied the fusion procedure to the dilute $A_L$ lattice models.
For these models, there are two types of fusion, related to the
values $u=\pm 2\lambda$ and $u=\pm 3\lambda$ of the spectral parameter $u$
where the local face operator $X_j(u)$ (\ref{localfaceop}) becomes singular.
These two types of fusion are very different in nature, one showing
an $su(2)$ structure, the other to an $su(3)$ structure. Here we concentrated
on the $su(3)$ type fusion, the $su(2)$ fusion hierarchies of the
dilute $A_L$ models are studied in \cite{Zhou:94}.

The fused models which we construct are labeled by altogether four integers,
one pair denoting the horizontal and the other pair the vertical fusion level.
Even in simplest non-trivial case, the actual fused face weights become
rather cumbersome which is the reason why we avoid to show any explicit
fused weights throughout this paper.

As the main result of our investigation, we derived the $su(3)$
fusion hierarchy (Theorems~\ref{theorem1} and \ref{theorem2}).
We also discuss the connection to te Bethe ansatz equations for the
dilute $A_L$ models. The hierarchy closes (\ref{close}) and thereby
yields functional equations for the row transfer matrices of the dilute
$A_L$ models. These equations can be written in a determinantal form
(\ref{determ}) and in principle can be solved for the eigenvalue spectra.
The solution of these equations will be presented in a subsequent
paper \cite{ZhPe:95}.

\section*{Acknowledgements}
This research is supported by the Australian Research Council (ARC) and
Stichting voor Fundamenteel Onderzoek der Materie (FOM).


\begin{thebibliography}{99}
\bibitem{WNS:92}
  S.~O.~Warnaar, B.~Nienhuis and K.~A.~Seaton,
  \newblock {\em New Construction of Solvable Lattice Models Including
               an Ising Model in a Field},
  \newblock Phys.\ Rev.\ Lett.\ {\bf 69} (1992) 710; \newline
  \newblock {\em A Critical Ising Model in a Magnetic Field},
  \newblock Int.\ J.\ Mod.\ Phys.\ {\bf B7} (1993) 3727.
\bibitem{Roche:92}
  Ph.~Roche,
  \newblock {\em On the construction of integrable dilute
             \mbox{A--D--E} models},
  \newblock Phys.\ Lett.\ {\bf 285B} (1992) 49.
\bibitem{Baxter:82}
  R.~J.~Baxter,
  \newblock {\em Exactly Solved Models in Statistical Mechanics},
  \newblock Academic Press, London (1982).
\bibitem{Pasquier:87}
  V.~Pasquier,
  \newblock {\em Two-dimensional critical systems labelled by Dynkin
             diagrams},
  \newblock Nucl.\ Phys.\ {\bf B285} (1987) 162; \newline
  \newblock {\em Exact Solvability of the $D_n$ series},
  \newblock J.\ Phys.\ {\bf A20} (1987) L217; \newline
  \newblock {\em $D_n$ models: local densities},
  \newblock J.\ Phys.\ {\bf A20} (1987) L221; \newline
  \newblock {\em Operator content for $ADE$ lattice models},
  \newblock J.\ Phys.\ {\bf A20} (1987) 5707; \newline
  \newblock {\em Lattice derivation of modular invariant partition functions
             on the torus},
  \newblock J.\ Phys.\ {\bf A20} (1987) L1229.
\bibitem{ABF:84}
  G.~E.~Andrews, R.~J.~Baxter and P.~J.~Forrester,
  \newblock {\em Eight-Vertex SOS Model and Generalized
             Rogers-Ramanujan-Type Identities},
  \newblock J.\ Stat.\ Phys.\ {\bf 35} (1984) 193.
\bibitem{WPSN:94}
  S.~O.~Warnaar,  P.~A.~Pearce, K.~A.~Seaton and B.~Nienhuis,
  \newblock {\em Order Parameters of the Dilute A Models},
  \newblock J.\ Stat.\ Phys.\ {\bf 74} (1994) 469.
\bibitem{Zam:89}
  A.~B.~Zamolodchikov,
  \newblock {\em Integrable Field Theory from Conformal Field Theory},
  \newblock Adv.\ Stud.\ in Pure Math.\ {\bf 19} (1989) 641; \newline
  \newblock {\em Integrals of motion and $S$-matrix of the (scaled) $T=T_c$
             Ising model with magnetic field},
  \newblock Int.\ J.\ Mod.\ Phys.\ {\bf A4} (1989) 4235.
\bibitem{BNW:94}
  V.~V.~Bazhanov, B.~Nienhuis and S.~O.~Warnaar,
  \newblock {\em Lattice Ising model in a field: E$_{8}$ scattering theory},
  \newblock Phys.\ Lett.\ {\bf B322} (1994) 198.
\bibitem{WaPe:95} S.~O.~Warnaar and P.~A.~Pearce,
  \newblock {\em Exceptional structure of the dilute $A_{3}$ model:
            $E_{8}$ and $E_{7}$ Rogers-Ramanujan identities},
  \newblock J.\ Phys.\ {\bf A27} (1995) L891.
\bibitem{OBPe:95} D. L. O'Brien and P.~A.~Pearce,
   \newblock {\em Lattice Realizations of Unitary Minimal Modular Invariant
              Partition Functions},
   \newblock Melbourne University preprint (1995).
\bibitem{CIZ:87} A.~Cappelli, C.~Itzykson and J.-B.~Zuber,
   \newblock {\em Modular invariant partition functions in two dimensions},
   \newblock Nucl.\ Phys.\ {\bf B280} (1987) 445; \newline
   \newblock {\em The A-D-E Classification of Minimal and A$^{(1)}_1$
              Conformal Invariant Theories},
   \newblock Comm.\ Math.\ Phys.\ {\bf 113} (1987) 1.
\bibitem{DJMO:86}
  E.~Date, M.~Jimbo, T.~Miwa and M.~Okado,
  \newblock {\em Fusion of the Eight Vertex SOS Model},
  \newblock Lett.\ Math.\ Phys.\ {\bf 12} (1986) 209.
\bibitem{BaRe:89}
  V.~V.~Bazhanov and N.~Yu.~Reshetikhin,
  \newblock {\em Critical RSOS models and conformal field theory},
  \newblock Int.\ J.\ Mod. Phys.\ {\bf A4} (1989) 115; \newline
  \newblock {\em Restricted solid-on-solid models connected with simply
            laced algebras and conformal field theory},
  \newblock J.\ Phys.\ {\bf A23} (1990) 1477.
\bibitem{KlPe:92}
  A.~Kl{\"{u}}mper and P.~A.~Pearce,
  \newblock {\em Conformal weights of RSOS lattice models and their
             fusion hierarchies},
  \newblock Physica {\bf A183} (1992) 304.
\bibitem{ZhPe:95}
  Y.~K.~Zhou and P.~A.~Pearce,
  \newblock {\em Solutions of functional equations of dilute lattice models},
  \newblock in preparation.
\bibitem{KRS:81}
  P.~P.~Kulish, N.~Yu.~Reshetikhin and E.~K.~Sklyanin,
  \newblock {\em Yang-Baxter equation and representation theory: I},
  \newblock Lett.\ Math.\ Phys.\ {\bf 5} (1981) 393.
\bibitem{Cherednik:82}
  I.~V.~Cherednik,
  \newblock {\em On the properties of factorized $S$-matrices in elliptic
             functions},
  \newblock Sov.\ J.\ Nucl.\ Phys.\ {\bf 36} (1982) 105.
\bibitem{JKMO:88b}
  M.~Jimbo, A.~Kuniba, T.~Miwa and M.~Okado,
  \newblock {\em The $A^{(1)}_{n}$ Face Models},
  \newblock Commun.\ Math.\ Phys.\ {\bf 119} (1988) 543.
\bibitem{ZhHo:89}
  Y.~K.~Zhou and B.~Y.~Hou,
  \newblock {\em On the fusion of face and vertex models},
  \newblock J.\ Phys.\ {\bf A22} (1989) 5089.
\bibitem{ZhPe:94}
  Y.~K.~Zhou and P.~A.~Pearce,
  \newblock {\em Fusion of $A$--$D$--$E$ lattice models},
  \newblock Int.\ J.\ Mod.\ Phys.\ {\bf B8} (1994) 3531;
  \newblock reprinted in: {\em Perspectives on Solvable Models},
  \newblock eds.\ U.~Grimm and M.~Baake,
  \newblock World Scientific, Singapore (1994), p.~83.
\bibitem{IzKor81}
  A.~G.~Izergin and V.~E.~Korepin,
  \newblock {\em The Inverse Scattering Method Approach to the
             Quantum Shabat-Mikhailov Model},
  \newblock Common.\ Math.\ Phys.\ {\bf 79} (1981) 303.
\bibitem{VichResh83}
  V.~I.~Vichirko and N.~Yu.~Reshetikhin,
  \newblock {\em Excitation spectrum of the anisotropic
             generalization of an SU$_3$ magnet},
  \newblock Teor.\ i Mat.\ Fiz.\ {\bf 56} (1983) 260.
\bibitem{KiRe:87}
  A.~N.~Kirillov and N.~Yu.~Reshetikhin,
  \newblock {\em Exact solution of the integrable $XXZ$ Heisenberg model
             with arbitrary spin: I.~The ground state and the excitation
             spectrum},
  \newblock J.\ Phys.\ {\bf A20} (1987) 1565.
\bibitem{KuSu:94}
  A.~Kuniba and J.~Suzuki,
  \newblock {\em Analytic Bethe ansatz for fundamental
             representations of Yangians},
  \newblock preprint (1994).
\bibitem{Baxter78}
  R.~J.~Baxter,
  \newblock {\em Solvable eight-vertex model on an arbitrary planar graph},
  \newblock Phil.\ Trans.\ Roy.\ Soc.\ (London) {\bf 289A} (1978) 315.
\bibitem{Zhou:94}
  Y.~K.~Zhou,
  \newblock {\em $SU(2)$ hierarchies of dilute lattice models},
  \newblock preprint (1994); \newline
  \newblock {\em On \ade  lattice models: selected topics},
  \newblock preprint (1995).
\end{thebibliography}
\end{document}